\pdfoutput=1 
\documentclass{article} 
\usepackage[a4paper, total={6in, 10in}]{geometry}
\usepackage[percent]{overpic} 
\usepackage{amsmath}
\usepackage{multirow} 
\usepackage{fancyvrb} 
\usepackage{enumitem} 
\usepackage{easylist}
\usepackage{graphicx}							
\usepackage{amssymb}
\usepackage[usenames]{color}
\usepackage[dvipsnames]{xcolor}
\usepackage{epsf}
\usepackage[bottom]{footmisc}
\usepackage{epsfig}
\usepackage{url}
\usepackage{wrapfig}
\usepackage{float} 

\usepackage[plainpages=false,backref=page]{hyperref} 

\newenvironment{s-itemize}{
\begin{itemize}
  \setlength{\itemsep}{1pt}
  \setlength{\parskip}{0pt}
  \setlength{\parsep}{0pt}
}{\end{itemize}}

\newenvironment{s-enumerate}{
\begin{enumerate}
  \setlength{\itemsep}{1pt}
  \setlength{\parskip}{0pt}
  \setlength{\parsep}{0pt}
}{\end{enumerate}}

\title{\bf Interactive Graph Visualization in DDLab}
\author{Andrew Wuensche%
\thanks{andy@ddlab.org,  \url{http://www.ddlab.org}}}%

\date{\small June 2024, with additions in March 2025}

\begin{document}

\maketitle

\vspace{-5ex}
\begin{abstract}
  
\noindent Interactive visualization of the basin of attraction field,
the ``ibaf-graph'', is a new feature in DDLab with the same
interactive functions as the ``network-graph'' and ``jump-graph''.
These functions allow any node and its connected fragment to
be dragged/dropped with the pointer as a graphic animation with elastic links.
The fragment itself depends on the node's link setting by inputs, outputs,
or either, and a distance in link-steps. Further options include
graph geometry, rescaling, node display, link editing, and isolating the fragment.
This article describes the three graph types, network/ibaf/jump, their selection,
enhanced functions, and applications to cellular automata, discrete
dynamical networks and random maps.
  
\end{abstract}

\begin{center}
  {\it keywords: DDLab, cellular automata, random Boolean networks,
    discrete dynamical networks, directed graphs, random maps, jump-graphs,
    basins of attraction, state transition graphs, interactive visualization.}
\end{center}

\section{Introduction} 
\label{Introduction}

We consider finite, deterministic (noise free), discrete dynamical
systems, including Cellular Automata (CA)\cite{wolfram2002,Wuensche92}, Random
Boolean Networks (RBN)\cite{kauffman69,kauffman93,Wuensche94a}, the general case
of Discrete Dynamical Networks (DDN)\cite{EDD}, and perhaps the most
general of all ---
random maps\cite{wuensche97}\hspace{-.2ex}\cite[\hspace{-1ex}\footnotesize{\#29.8}]{EDD}\footnote{References
to the book ``Exploring Discrete Dynamics''(EDD)\cite{EDD} usually include the
relevant chapter or section. EDD and DDLab\cite{Wuensche-DDLab}
are kept updated online at \url{http://www.ddlab.org} and mirror sites.}
--- random directed graphs with outdegree=1.
A system's state at any given moment is a string of $n$ elements, each with a value
$x \in(0,1,\dots, v-1)$ where $v$ is the value-range (often binary)
giving a state-space size $S$=$v^n$.  A state can be identified and
depicted by its string pattern in 1, 2 or 3 dimensions, or by the
string's equivalent value in decimal or hexadecimal.  How then is
state-space connected by directed links?

\begin{figure}[b]
  \begin{center}
    \begin{minipage}[t]{.8\linewidth},
      \includegraphics[height=.45\linewidth]{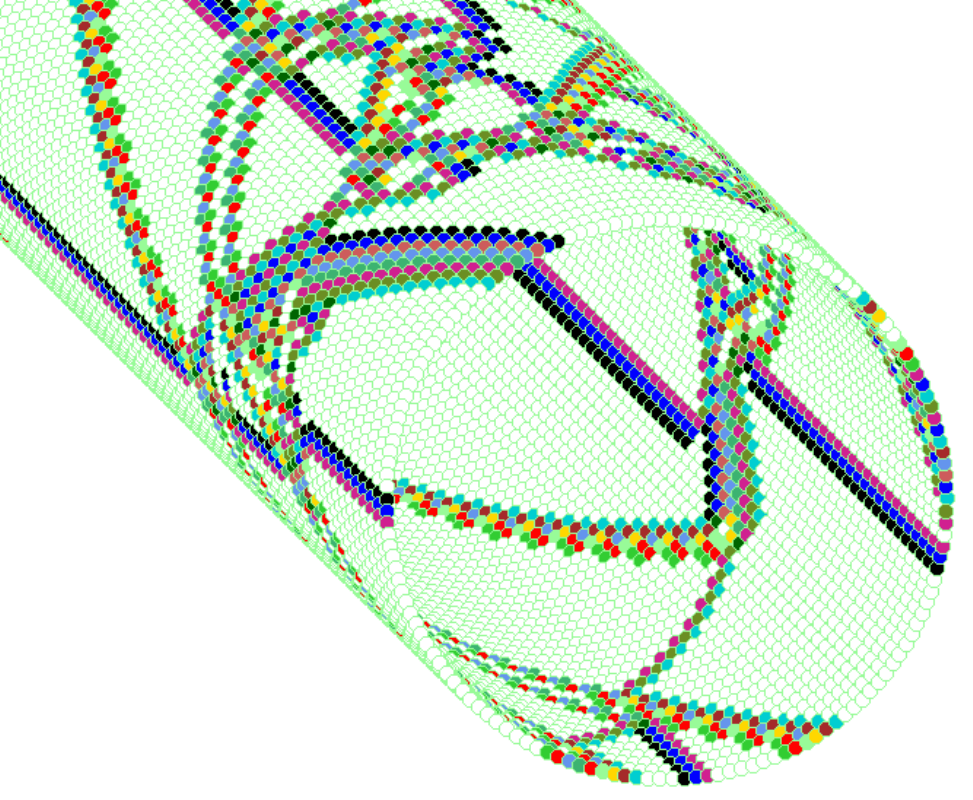}
      \includegraphics[height=.45\linewidth]{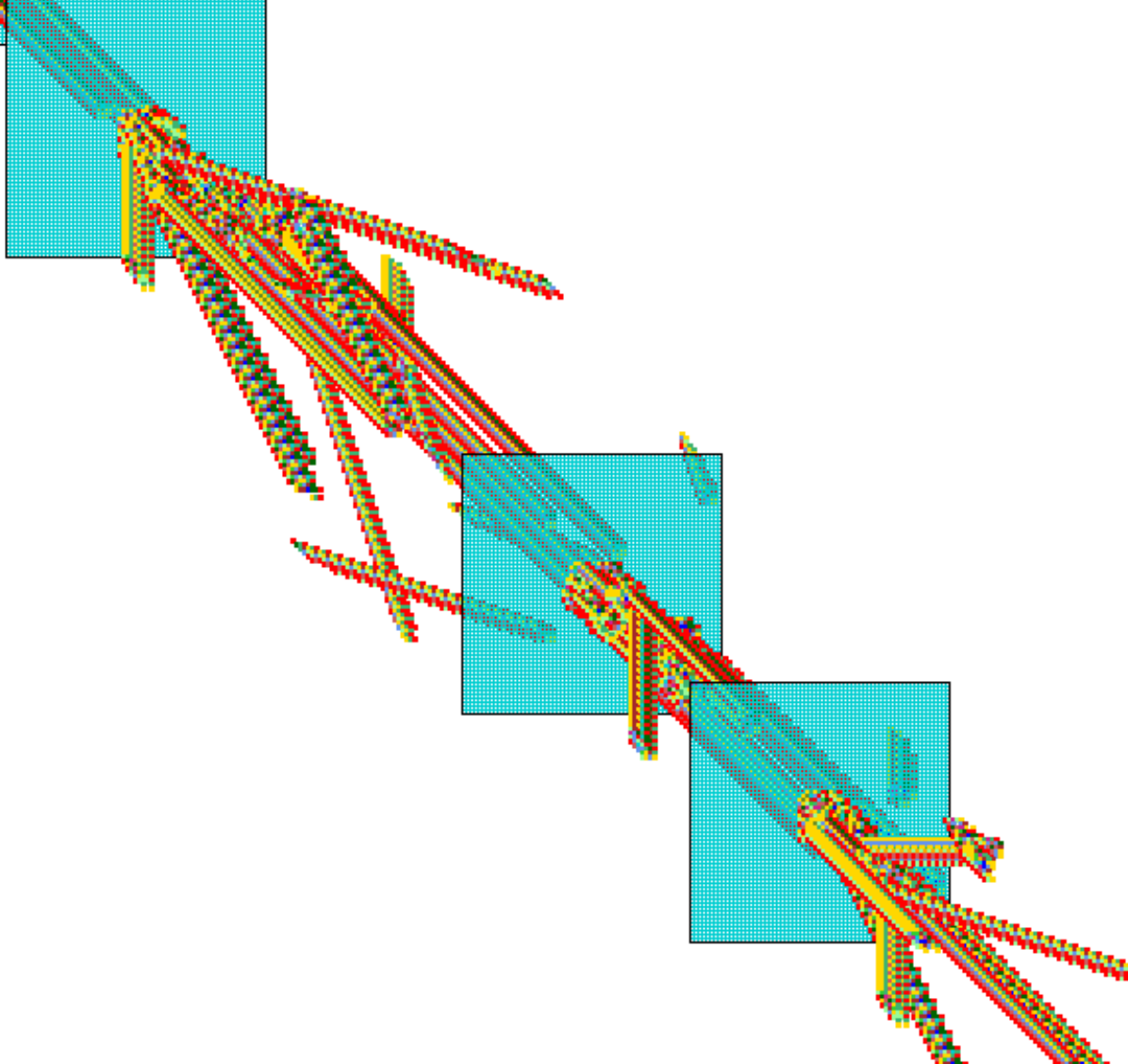}
   \end{minipage}
  \end{center}
   \vspace{-4ex}
   \caption[Scrolling network-graph space-time patterns diagonally]
           {\textsf{
               Snapshots of binary CA space-time patterns shown
               scrolling diagonally\cite[\hspace{-1ex}\footnotesize{\#4.8.2}]{EDD}
               with the present moment at the front, colored by neighborhood rather than value.
               Images were made with functions in the network-graph (section~\ref{The network-graph}).
             \underline{\it Left}: 1d ``circle'' layout and
             filtered\cite{wuensche98b}\hspace{-.2ex}\cite[\hspace{-1ex}\footnotesize{\#32.11.5}]{EDD},
             a ring of cells making a scrolling tube,
             $v2k5$, rcode (hex) e9f6a815, $n$=150.
             \underline{\it Right}: 
             the ``logically universal'' Variant rule\cite{Gomez2020} $n$=66$\times$66
             with random seeds 11$\times$11 set at intervals, and alternate time-steps skipped.
             }}
       \label{fig:quick_ring} 
\end{figure}

The system's dynamics iterate in discrete time-steps,
$t_0$, $t_1$, $t_2$, \dots, whereby each string element at $t_{+1}$
is updated, usually synchronously, applying a logical rule to its (size $k$)
neighbourhood at time~$t$. The input wiring scheme for all elements
is predefined but invariant over time.
Figure~\ref{fig:quick_ring} shows examples of resulting space-time patterns.
The rule and neighborhood are usually homogeneous for CA,
but heterogeneous with pseudo-neighborhoods\cite{wuensche96} for RBN and DDN,
and there can be intermediate hybrid
architectures\cite{wolfram2002,wuensche97}\hspace{-.2ex}\cite[\hspace{-1ex}\footnotesize{\#14.8}]{EDD}.
A random map transcends rules and wiring but can be tailored as equivalent to any CA,
RBN or DDN.

Iterating from initial states, the system falls along transients (if
any), then must eventually become trapped in one of its attractors,
repetitive state sequences of period one or more.  An attractor and
its transients comprise a basin of attraction, with a topology of
trees rooted on the attractor cycle. Each vertex has just one outgoing
edge --- its output --- and zero or more incoming edges --- inputs from
its predecessors called pre-images. A vertex without inputs --- zero
incoming edges --- is a leaf, or ``garden-of-Eden'' state in CA
terminology.  The global dynamics can be shown as a directed graph depicting how
states --- the vertices or nodes --- are linked to each other by
directed edges. This is the basin of attraction field (or state
transition graph) which depicts all the basins (one or more) showing how the
dynamics categorises state-space by attractors, basins, trees and
sub-trees, significant in many areas\cite{wuensche10} including
self-organisation\cite{langton90,kauffman93,wuensche97},
memory\cite{wuensche96,wuensche2005}, and gene regulation\cite{kauffman69,somogyi96,wuensche98a}.

\begin{figure}[b]
\begin{center}
  \includegraphics[bb=26 144 886 400,clip=,width=1\linewidth]{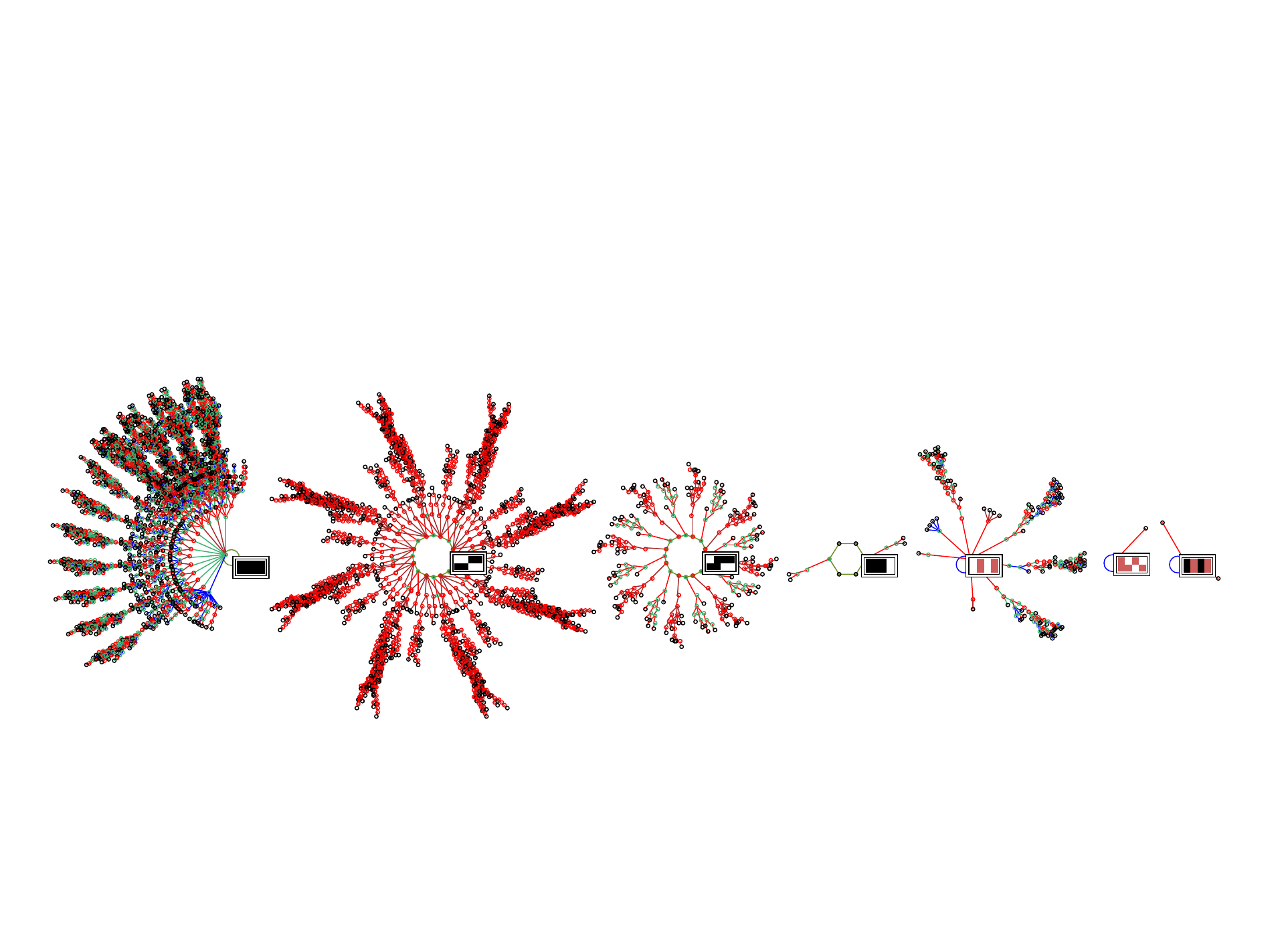}
\end{center} 
\vspace{-6ex}
\caption[The basin of attraction field of a 3-value Cellular Automaton]
        {\textsf{The non-interactive, classic basin of attraction field of a
            3-value Cellular Automaton, 
    $v$=3, $k$=3, $n$=8. All 17 basins could be drawn,
            but here ``compression'' is active so  equivalent basins have been
            suppressed leaving just the
              7 prototypes. One attractor state is shown for each basin.
              The rcode rule-table is 020120122021201201011011022.
              Note: time flows inward to attractors then clockwiss.}}
\label{fig:CA 3-value basin field preface}
\end{figure}

\begin{figure}[htb]
\begin{center}
\includegraphics[bb=42 140 1142 797,clip=,width=1\linewidth]{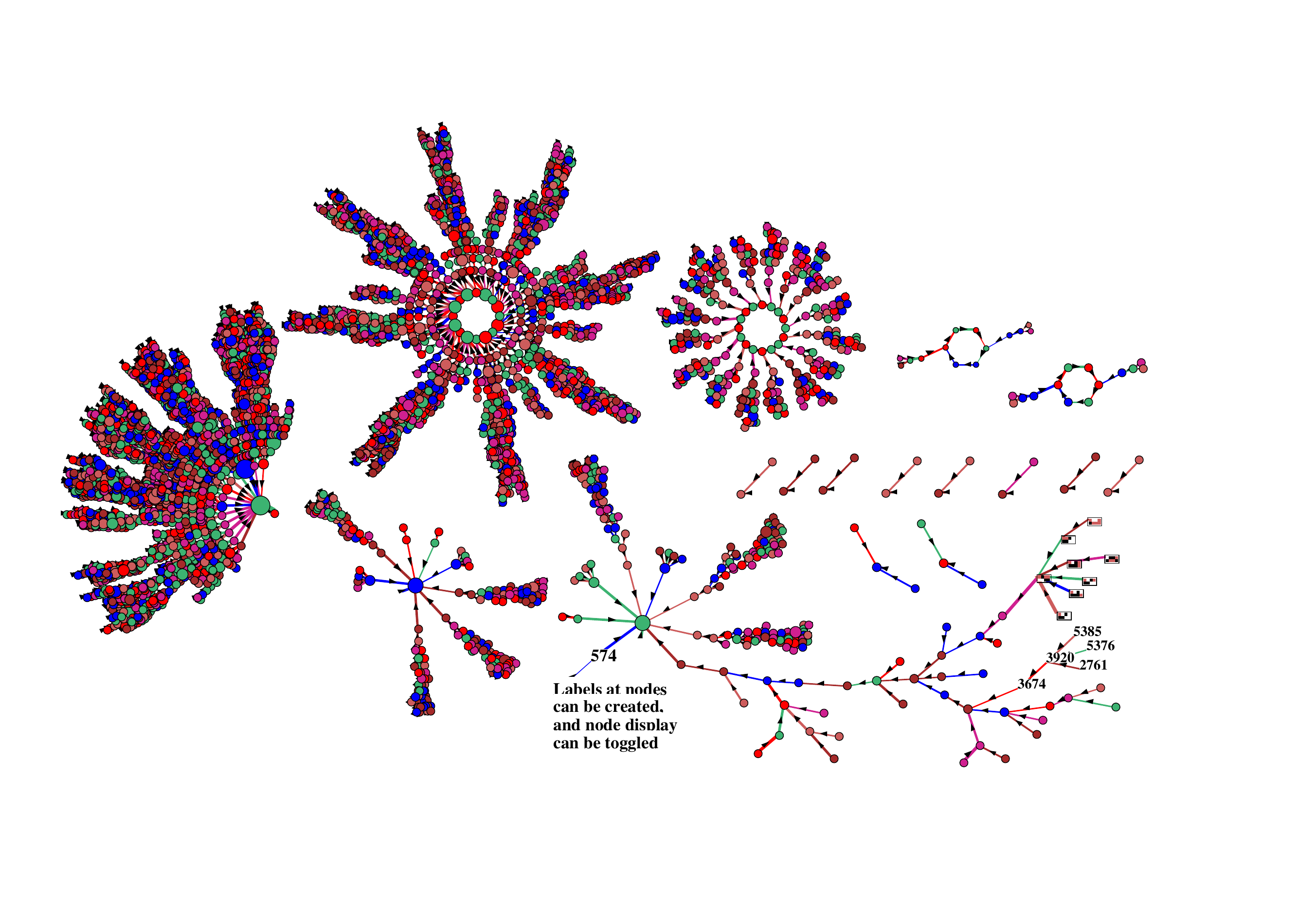}
\end{center}
\vspace{-5ex}
\caption[ibaf-graph v3k3 CA $n$=10]
        {\textsf{A snapshot of the ibaf-graph of the same CA
            in figure~\ref{fig:CA 3-value basin field preface}, based
            on an ``uncompressed'' classic graph of 17 basins, but now
            subject to interactive manipulation.
            Some basins (components) and fragments
            have been dragged to new positions. Nodes are scaled
            according to inputs. In the lower right basin
            some subtree node displays were toggled to 2d patterns and numbers, and a
            label was created upstream of node 574.}}             
\label{v33k3ibaf-2}
\end{figure}

\enlargethispage{6ex}
Visualizing basins of attraction computed by
reverse algorithms\cite{Wuensche92,Wuensche94a}\hspace{-.2ex}\cite[\hspace{-1ex}\footnotesize{\#2.19}]{EDD}
and drawn \mbox{according} to predefined graphic conventions
and parameters\cite{Wuensche92}\hspace{-.2ex}\cite[\hspace{-1ex}\footnotesize{\#24-26}]{EDD}
as in figure~\ref{fig:CA 3-value basin field preface},
has been the raison d'\^etre of the program
\mbox{\it Discrete Dynamics Lab} (DDLab)\cite{Wuensche-DDLab} since the book
``The Global Dynamics of Cellular Automata''\cite{Wuensche92} was publshed in 1992.
A graph of just subtrees or single basins,
or the complete basin of attraction field, the focus here, can be computed and drawn
with optional CA ``compression''\cite[\hspace{-1ex}\footnotesize{\#26.2}]{EDD}
by equivalent basins and subtrees as in figure~\ref{fig:CA 3-value basin field preface}.

The layout parameters are set and amended prior to (and possibly during)
drawing, but once complete the classic graph is static.  Given
that basin topology is inherently unpredictable (though deterministic)
pre-setting an acceptable layout usually requires multiple tries,
though an indirect workaround is to redraw basins at (adjustable)
jump-graph nodes (figures~\ref{r110-f-ibaf} and \ref{basin_r133_comp}).
However, DDLab always had the ambition to make the classic graph fully interactive.
This has now been accomplished with the new
{\it  interactive} basin of attraction field graph --- the ``ibaf-graph'' ---
as the snapshot in figure~\ref{v33k3ibaf-2} illustrates. The ibaf-graph
applies the exhaustive reverse algorithm treating any system as a random-map.
The method excludes CA ``compression'', but is able to
adopt methods for on-the-fly interactivity already present for the
network-graph and jump-graph.
The ibaf-graph with $v^n$ nodes typically has an order of magnitude
more nodes than the network-graph with just $n$ nodes, or the
jump-graph where the number of nodes is the number of basins, however
the three graph types share the coding and most functions for interactivity.

A directed graph is considered interactive in DDLab if any node,
singly or together with its connected ``fragment'', can be
dragged/dropped with the pointer as a graphic animation with elastic links
(or ``snap'' links --- figure~\ref{snap/elastic}). The fragment itself
depends on the node's link setting: just inputs, just outputs, or
``either'' (meaning irrespective of direction), and a distance in link
steps, which are time-steps in a basin of attraction.  A fragment
based on ``either'' and unlimited distance is a component in graph
theory, an independent subgraph which can be dragged away completely.
A fragment can also be defined as an arbitrary block of adjoining
nodes in 1, 2 or 3 dimensions.

On top of drag/drop, a node and its fragment can be manipulated
independently on-the-fly with a diversity of options including
node/link display and rescaling, spin, flip, link
editing, and the fragment can also be isolated (figures~\ref{frags-net}, \ref{frags-basin}).
Additional functions include: multi-line descriptive labels
at nodes, redefining graph geometry, the adjacency-matrix, launching a
Markov chain ``ant'', and PostScript filing of any image. A flexible
pre-layout of ibaf basin positions can be created from the node
layout of the jump-graph.

Sections \ref{The network-graph}, \ref{The ibaf-graph}, and
\ref{The jump-graph} describe the three interactive graphs types,
network/ibaf/jump, and their selection, noting there are two types of
jump-graph, {\it f-jump} from the basin of attraction field, and {\it
  h-jump} derived statistically from a sample of space-time patterns
--- finding attractors to assemble the
``attractor histogram''\cite{wuensche2002}\hspace{-.2ex}\cite[\hspace{-1ex}\footnotesize{\#31.7}]{EDD}.
Section\ref{Interactive graphs} describes and lists the functions for
interactive manipulation, common to the three graph types with just a
few exceptions.


\section{The network-graph}
\label{The network-graph}

In DDLab, the setup of CA, RBN or DDN (but not necessarily random
maps\cite[\hspace{-1ex}\footnotesize{\#29.8}]{EDD}) requires a
``wiring-scheme'' defining the origin of inputs to a
pseudo-neighborhood size $k$ for each of the system's $n$ nodes.  For
CA $k$ is homogeneous and there is no distinction between the pseudo
and actual neighborhood.  A rule, logic or
lookup-table\cite[\hspace{-1ex}\footnotesize{\#13-16}]{EDD}, is then
applied to the pseudo-neighborhood for the dynamics to unfold.
Default settings for the wiring-scheme are applied automatically, but
beyond that there are many ways of creating, depicting and amending a
purpose made or default 1d, 2d or 3d wiring-scheme, with homogeneous
or mixed-$k$\cite[\hspace{-1ex}\footnotesize{\#9}]{EDD}, with either
local or random wiring or some intermediate architecture, including
hybrid networks and networks of sub-networks\cite[\hspace{-1ex}\footnotesize{\#19.4}]{EDD}.

\begin{figure}[htb]  
  \footnotesize
\begin{minipage}[t]{.12\linewidth} 
  \fbox{\includegraphics[bb=172 82 325 385,clip=, width=1\linewidth]{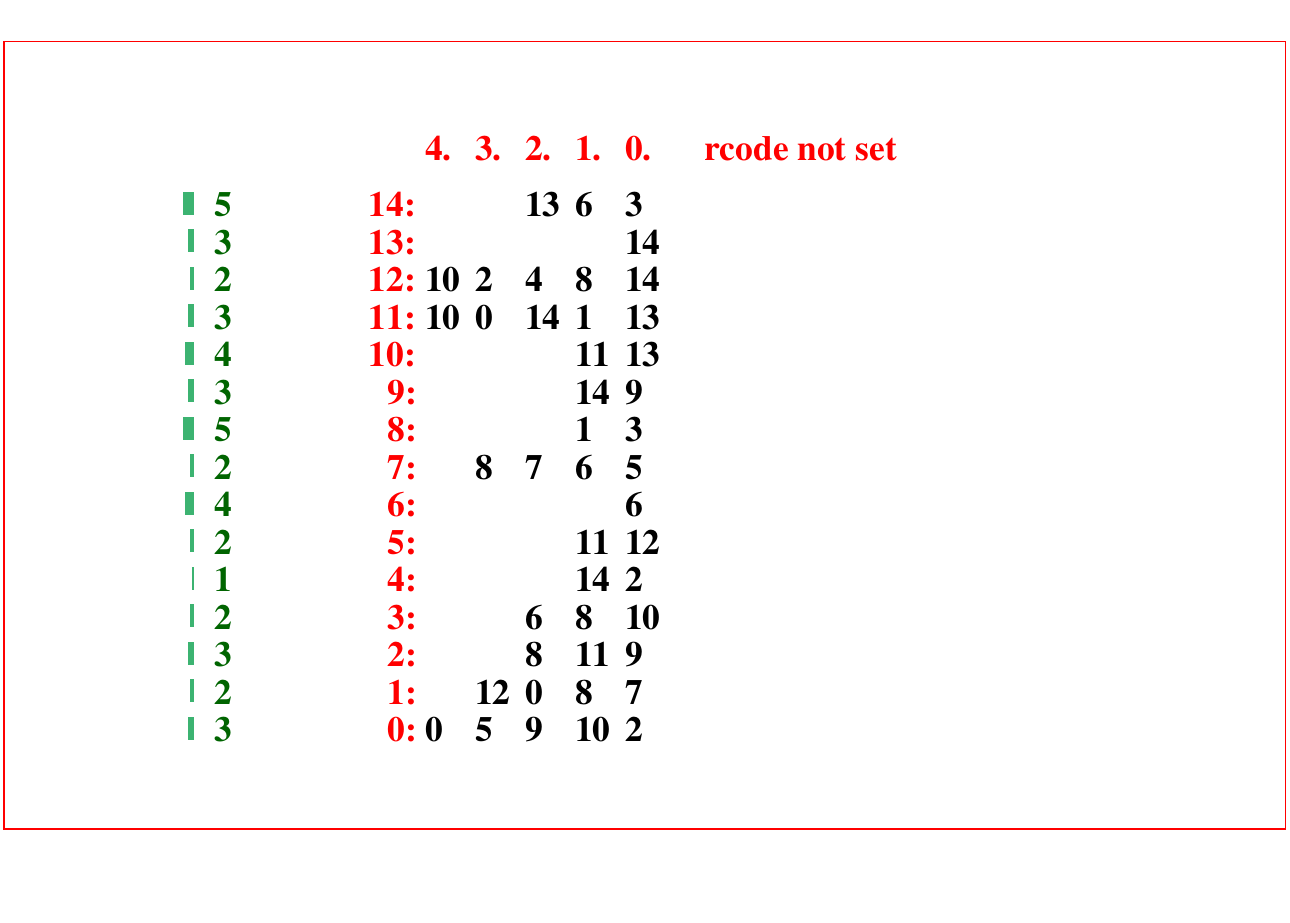}}
\begin{center}
  \vspace{-2ex}
\textsf{(a) spread-sheet}
\end{center}
\end{minipage}
\hfill
\begin{minipage}[t]{.16\linewidth}
\includegraphics[bb=101 130 278 407,clip=,width=1\linewidth]{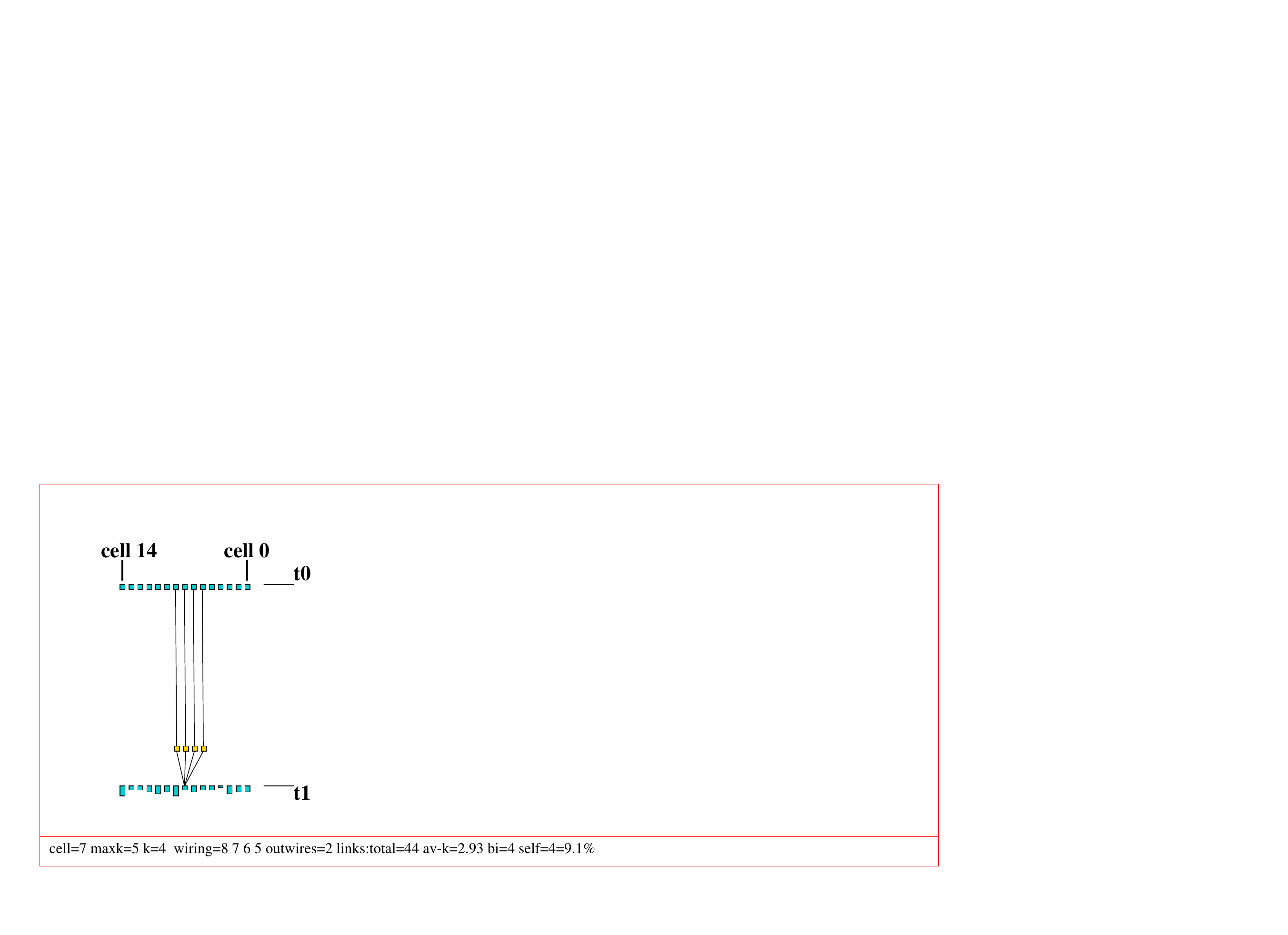}
\begin{center}
  \vspace{-2ex}
\textsf{(b) CA cell 7 with local wiring}
\end{center}
\end{minipage}
\hfill
\begin{minipage}[t]{.16\linewidth}
\includegraphics[bb=101 130 278 407,clip=,width=1\linewidth]{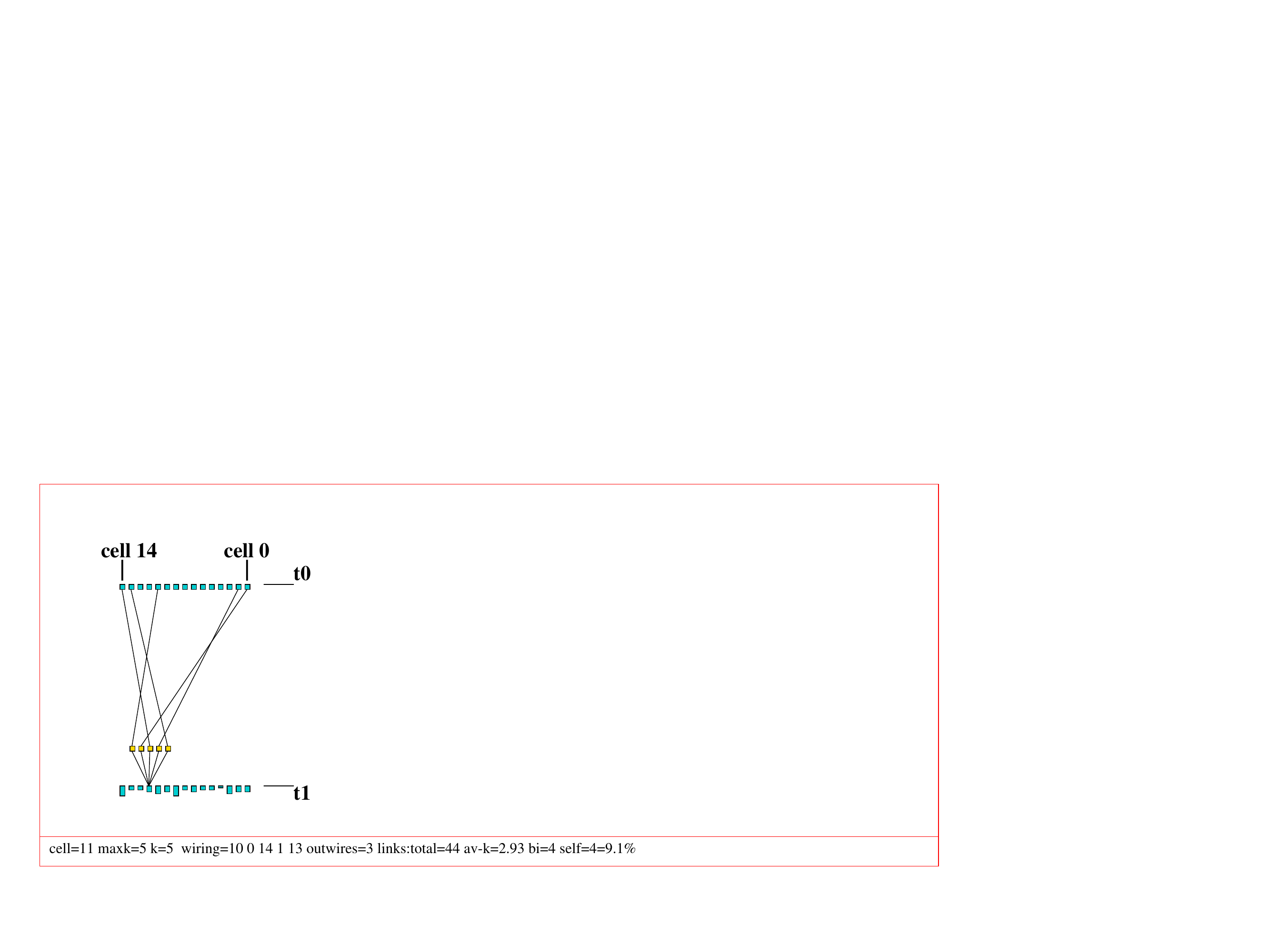}
\begin{center}
  \vspace{-2ex}
\textsf{(c) cell 11 with non-local wiring}
\end{center}
\end{minipage}
\hfill
\begin{minipage}[t]{.192\linewidth}
\includegraphics[bb=101 130 312 407,clip=,width=1\linewidth]{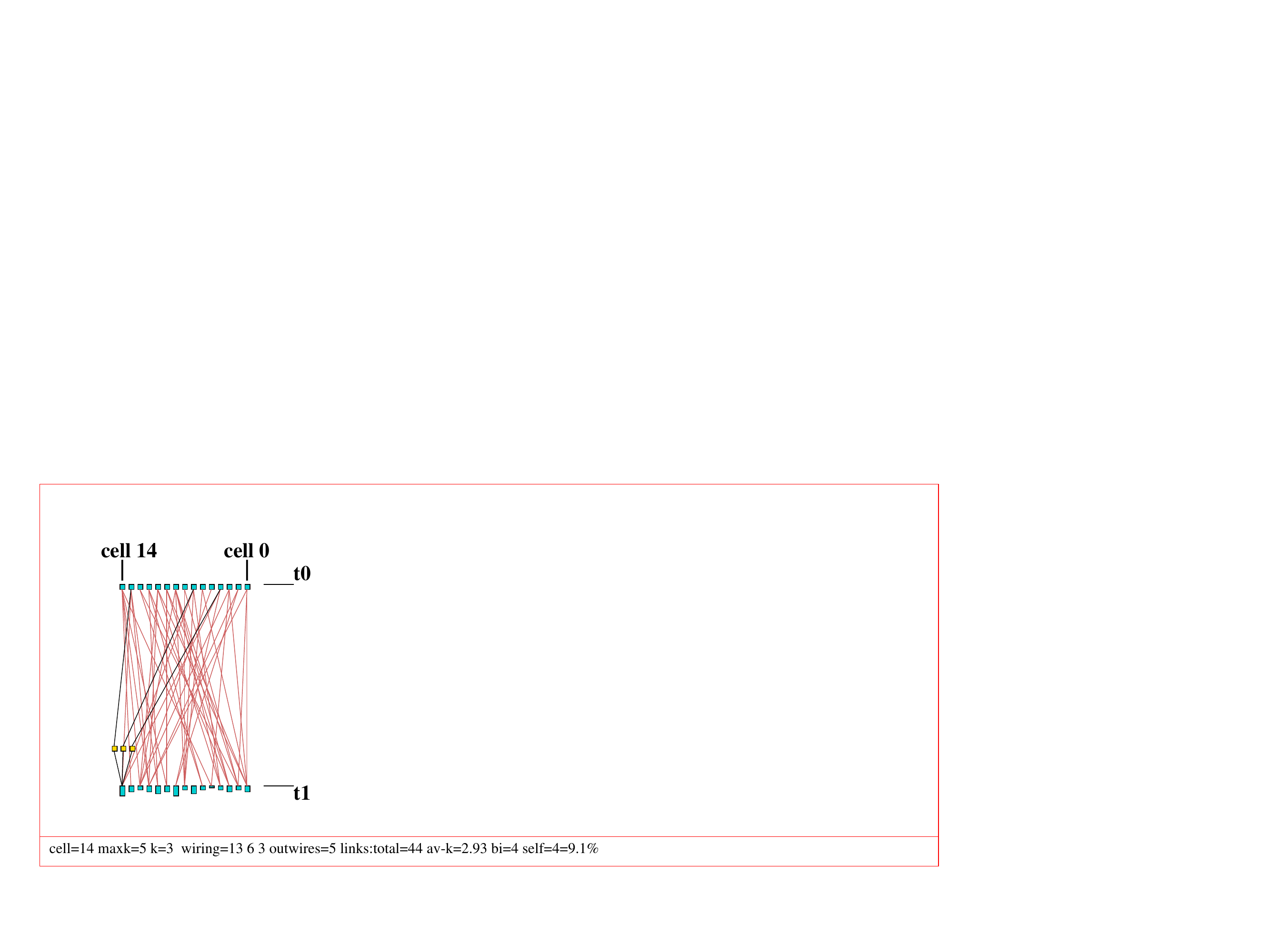}
\begin{center}
  \vspace{-2ex}
\textsf{(d) cell 14 with other links in red}
\end{center}
\end{minipage}
\hfill
\begin{minipage}[t]{.24\linewidth}
\includegraphics[bb=351 180 936 717,clip=,width=1\linewidth]{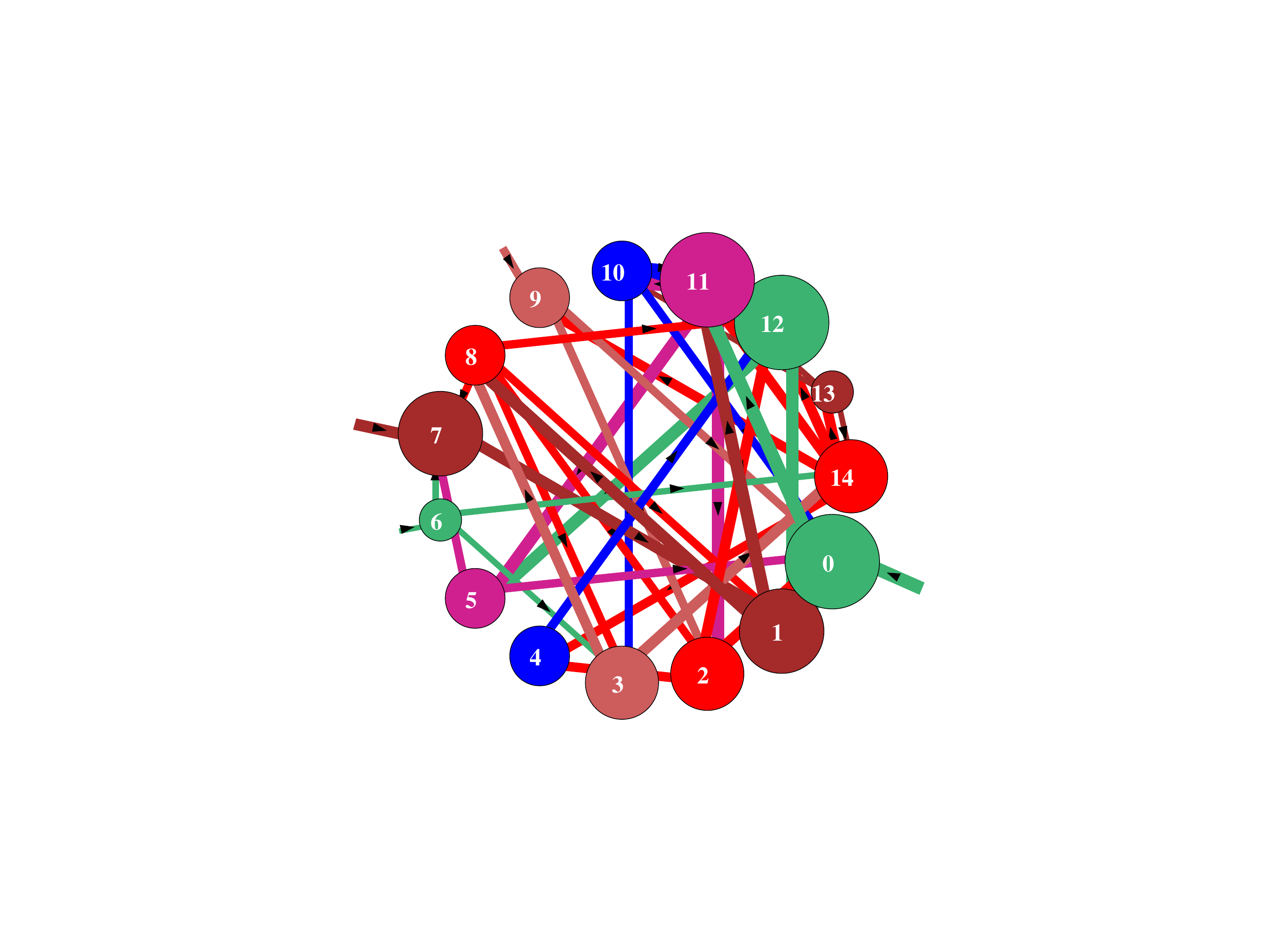}
\begin{center}
  \vspace{-2ex}
\textsf{(e) network-graph, nodes and links scaled by inputs}
\end{center}
\end{minipage}
\vspace{-2ex}
\caption[RBN wiring and network-graph)]{
  \textsf{A 1d wiring scheme where links can be set to the pseudo-neighborhoods.
    (a) in a wiring matrix spread-sheet\cite[\hspace{-1ex}\footnotesize{\#17.2.2}]{EDD},
    (b)-(d) in a ``wiring graphic''\cite[\hspace{-1ex}\footnotesize{\#17.3}]{EDD}
    according to time-steps $t_0$ to  $t_1$.
    (e) The corresponding network-graph with nodes and links scaled according to inputs $k$.
        Scaling can be toggled between inputs outputs.
}}
\label{n15rbn1}
\end{figure}

\begin{figure}[htb]  
\footnotesize
\begin{center}
\parbox[t]{0.5\linewidth}{
\begin{center}  
\includegraphics[bb=255 130 780 520, width=1\linewidth]{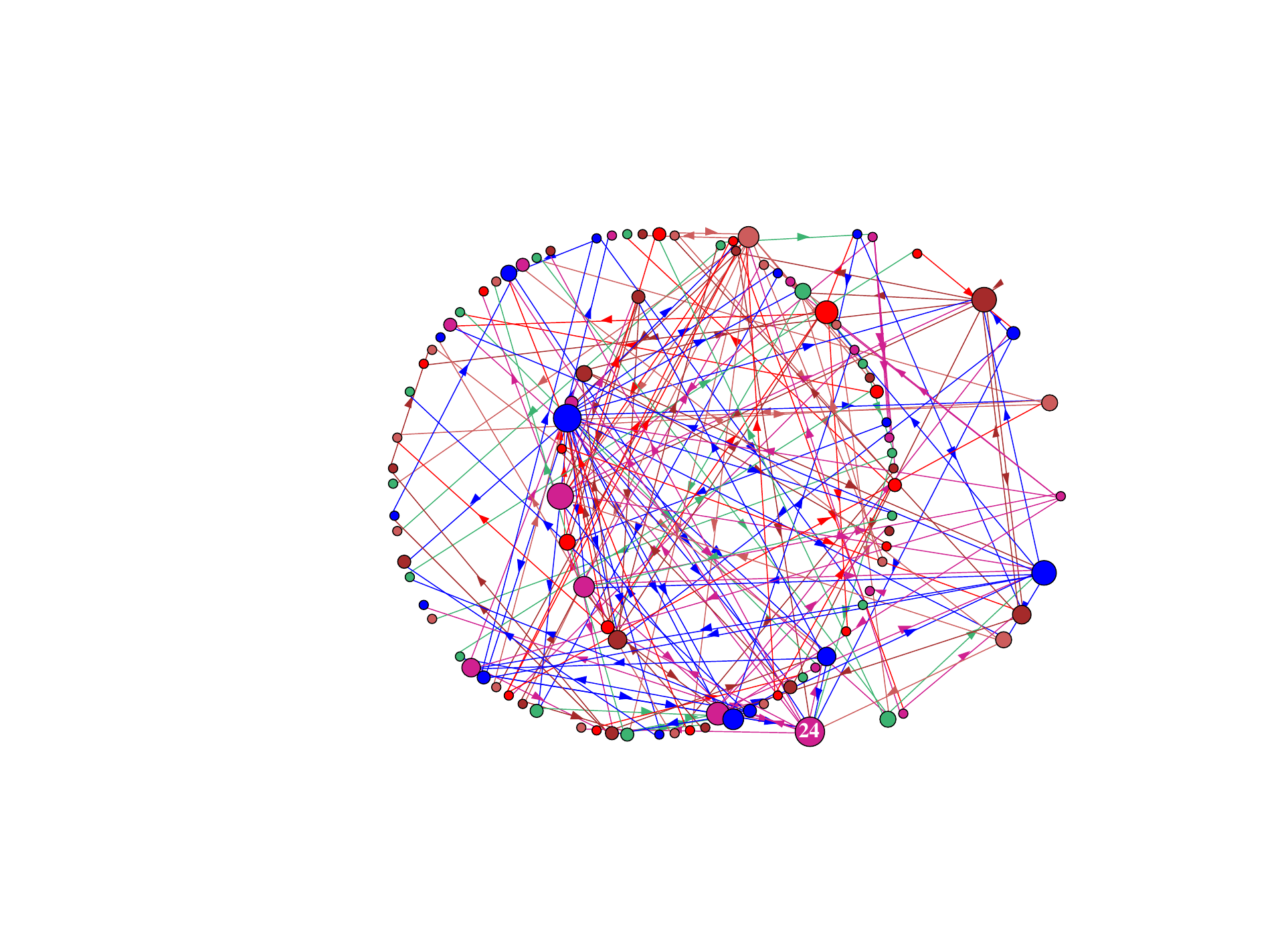}
\textsf{(a) circle layout, one node dragged + 2-step inputs}\end{center}}
\parbox[t]{0.47\linewidth}{
\begin{center}    
\includegraphics[bb=238 175 670 518, width=1\linewidth]{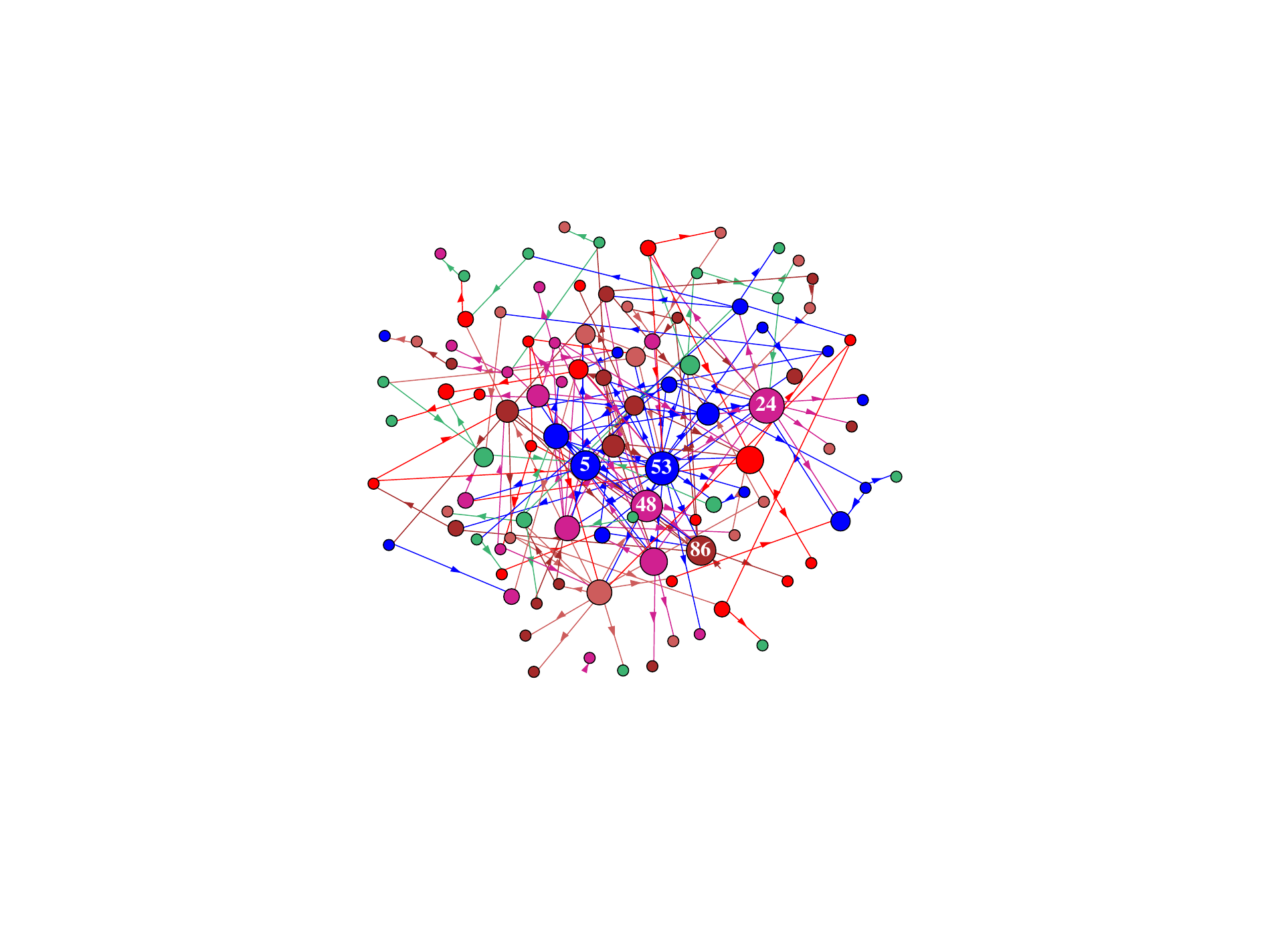}
\textsf{(b) spiral/rank/unscramble}
\end{center}}
\end{center}
\vspace{-6ex}
\caption[Power-law network-graphs)]{
\textsf{Two versions of the network-graph for the
same network with a power-law wiring distribution,
both inputs ($k$=1 to 10) and outputs, $n$=100.
Nodes are scaled according to inputs $k$.
}\label{power-law-network}}
\end{figure}

The wiring scheme can be invoked, reviewed and amended at various stages in DDLab,
including at ``quick wire settings''\cite[\hspace{-1ex}\footnotesize{\#11}]{EDD}, 
special wiring\cite[\hspace{-1ex}\footnotesize{\#12}]{EDD},
as a spread-sheet ``wiring matrix''\cite[\hspace{-1ex}\footnotesize{\#17.2}]{EDD},
as a 1d, 2d or 3d  ``wiring graphic''\cite[\hspace{-1ex}\footnotesize{\#17.4}]{EDD},
and loading a previously saved wiring-scheme\cite[\hspace{-1ex}\footnotesize{\#17.9.12}]{EDD}.
At any of these stages, the network-graph ---
figure~\ref{n15rbn1}(e) --- can be immediately activated with the prompt
{\bf graph-g} in the related wiring scheme prompt window.
Link changes within the network-graph do not
affect the current wiring-scheme.
When the network-graph is activated from the
wiring graphic, the option {\bf win-w} showing what lies below the
graph window provides a useful toggle between the two presentations.
For large networks where only the network itself is of interest, not
the rules, it is best to select totalistic (forwards-only) rules at
the first DDLab prompt. This allows a mixed-$k$ network with large
max-$k$, where a power-law $k$-distribution may be applied as
in figure~\ref{power-law-network}. 

The network-graph can be activated when interrupting space-time patterns
for a simultaneous presentation of space-time patterns within
the network-graph\cite[\hspace{-1ex}\footnotesize{\#32.19}]{EDD},
which can then be scrolled as in figure~\ref{fig:quick_ring}. In this case
links would usually be omitted.

\enlargethispage{3ex}
The network-graph can also be activated 
indirectly from the “attractor basin complete”
prompt\cite[\hspace{-1ex}\footnotesize{\#14-16}]{EDD} or when
basins are paused\cite[\hspace{-1ex}\footnotesize{\#30.5-25.3}]{EDD}.
\clearpage

In the network-graph the $n$ nodes are shown as discs (by default)
or numbers starting at cell 0 (zero). Each node has $k$$\geq$$1$
inputs, its neighborhood, and a variable number of outputs, but the wiring scheme
requirement that a node must have at least one input is
relaxed within the network-graph environment so inputs can be
cut.

\begin{figure}[H]
  \vspace*{-8ex}
\footnotesize  
\textsf{
   \begin{center}  
   \begin{minipage}[t]{.8\linewidth}
     \includegraphics[bb=85 200 836 500,clip=, width=.63\linewidth]{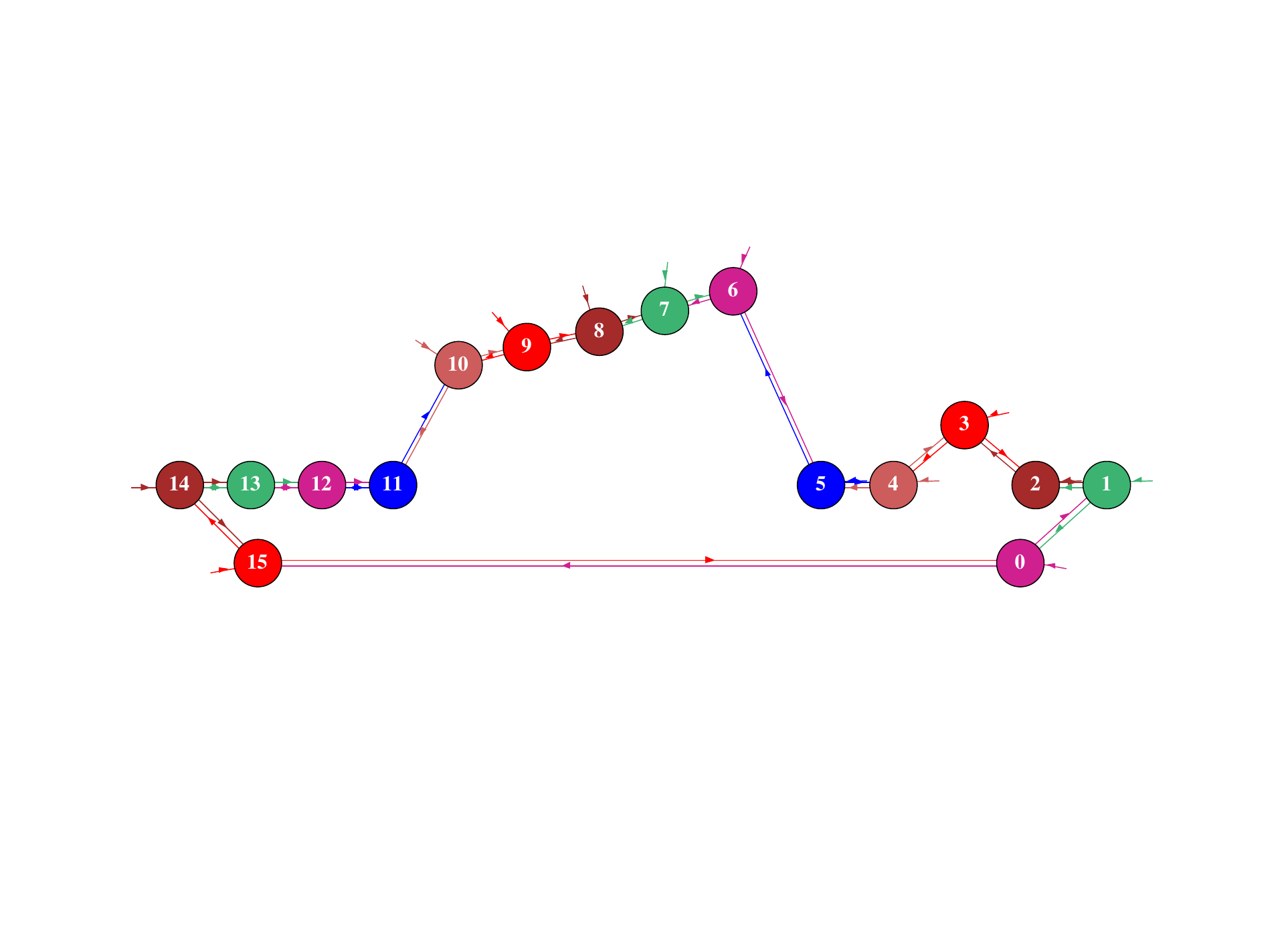}
   \hfill
   \includegraphics[bb=257 138 656 517,clip=, width=.36\linewidth]{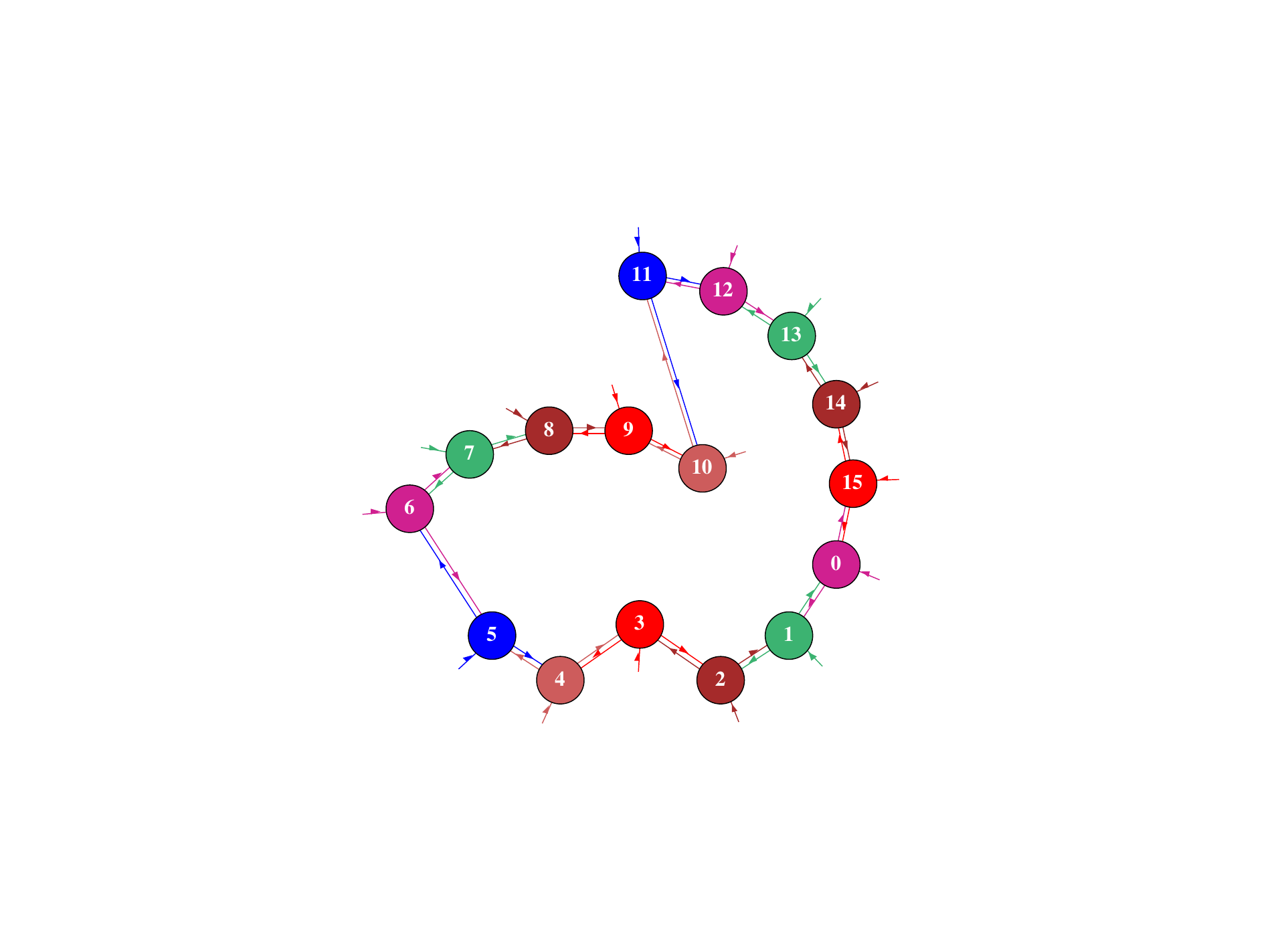}
   \end{minipage}
   \end{center}
   \vspace{-4ex}
       Starting with a 1d CA  $k$=3 $n$=16
       \underline{\it Left}:~with 1d layout 
       and \underline{\it Right}: with circle layout. 
       Single nodes have been dragged, and in both cases node 8 and its 
       2-step neighbors have been dragged and rotated.\\[-4ex]   
   \begin{center} 
     \begin{minipage}[t]{.9\linewidth}
       \includegraphics[ bb=209 152 641 506,clip=,height=.39\linewidth]{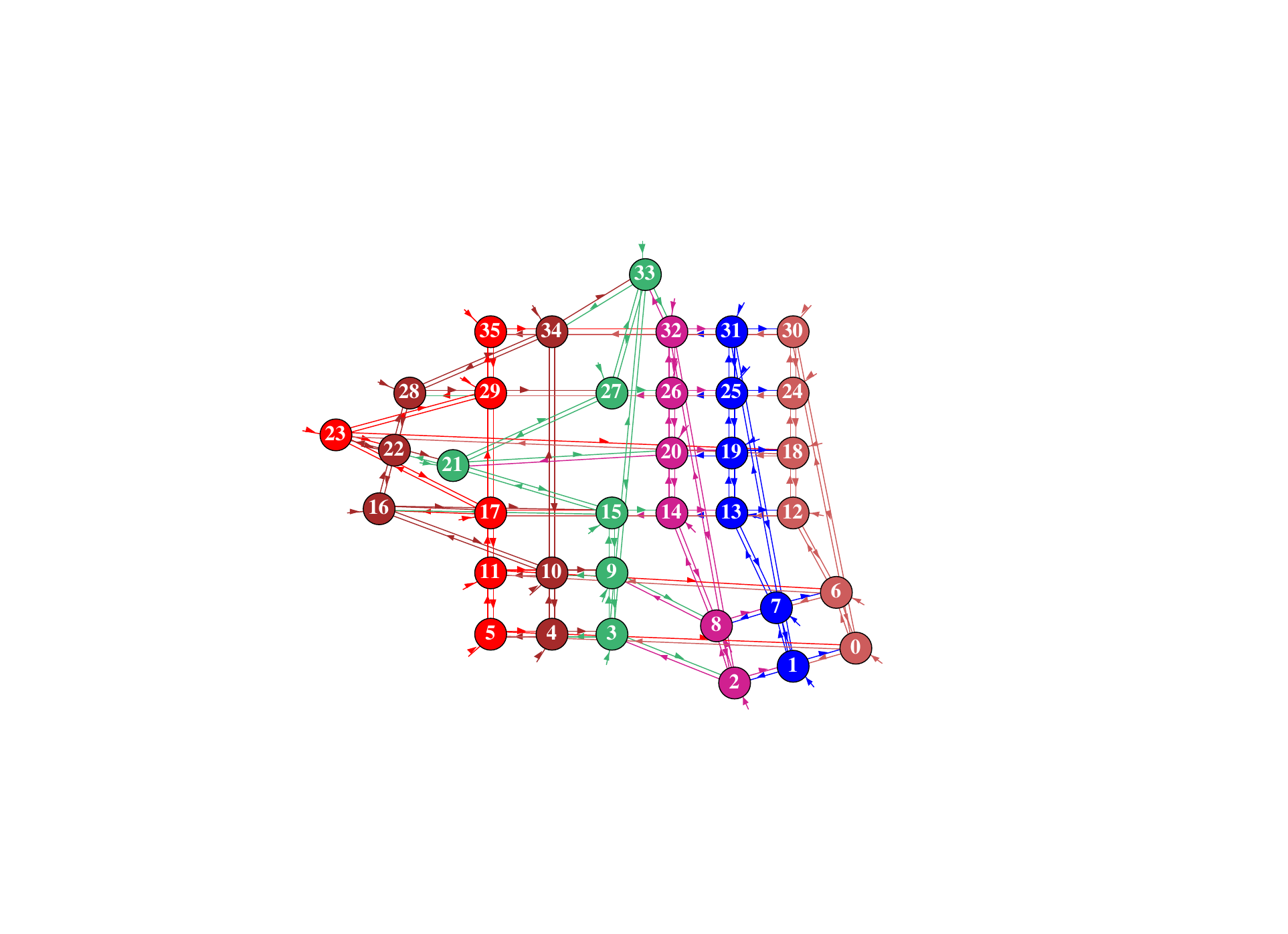}
     \hfill
     \includegraphics[bb=196 164 661 483,clip=,height=.37\linewidth]{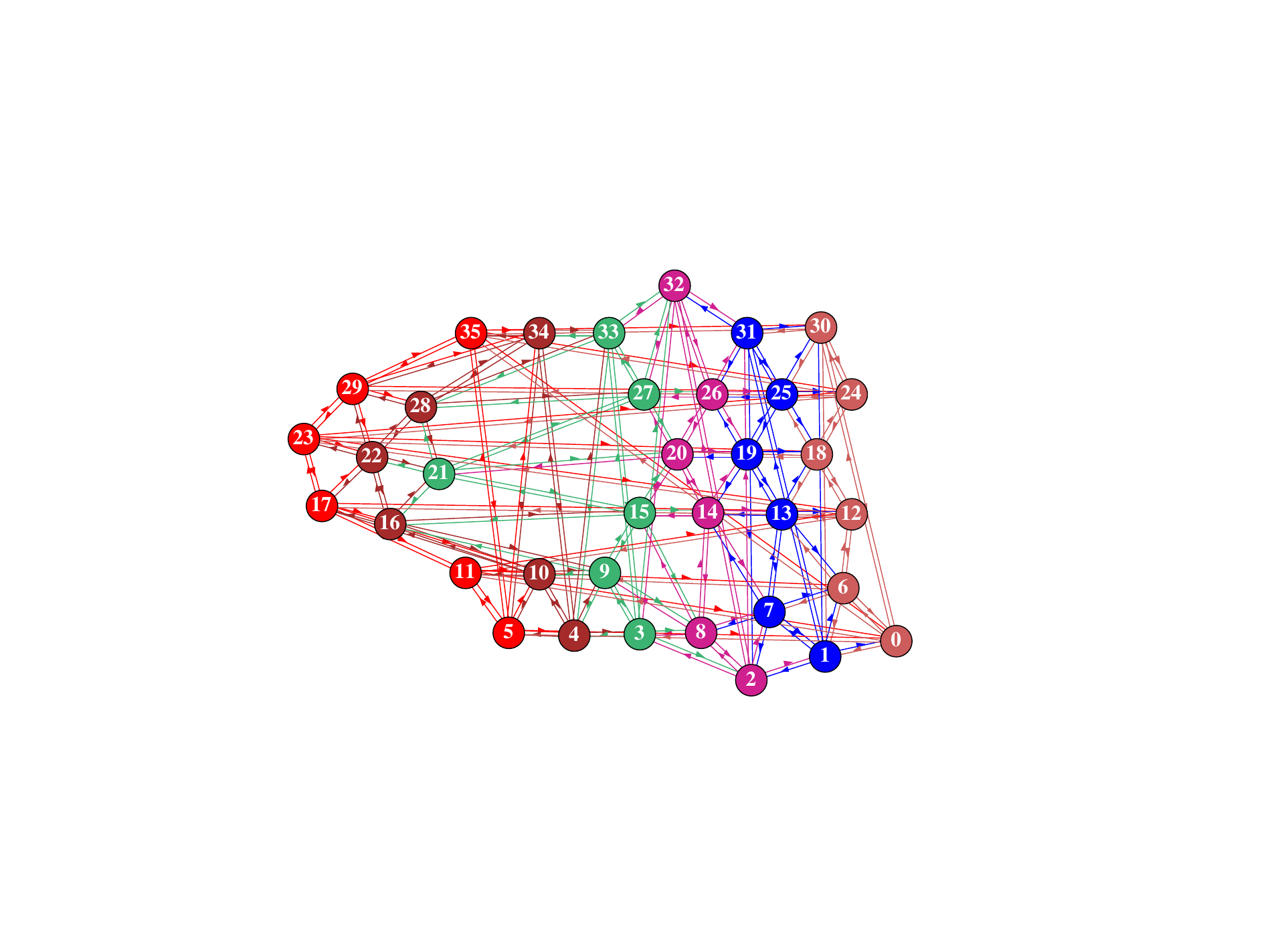}
   \end{minipage}
   \end{center}
   \vspace{-2ex}
   \begin{minipage}[b]{1\linewidth}
       A 2d CA 6$\times$6,
       \underline{\it Left}:  square $k$=5,
       and \underline{\it Right}: hexagonal $k$=6. 
       In both cases, a node has been dragged from the top row,
       node 22 and its 1-step neighbors have been dragged and rotated,
       and a 2d block (0 --- 8) has been 
       dragged and rotated from the lower right hand corner.
   \end{minipage}\\
   \includegraphics[bb=238 174 758 496,clip=,width=.5\linewidth]{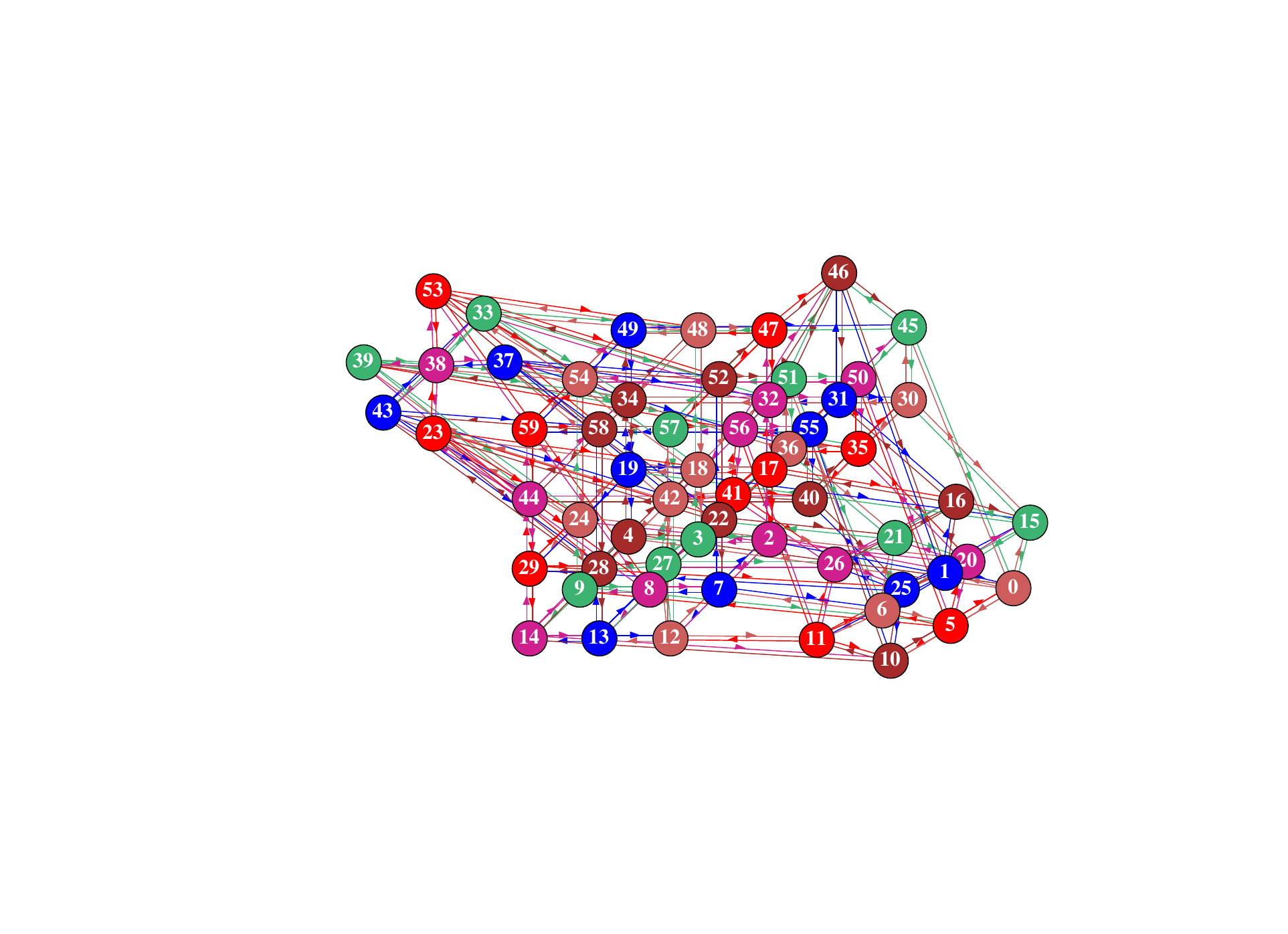}
   \hfill
   \begin{minipage}[b]{.45\linewidth}
     Starting with a regular
       3d network-graph 5$\times$3$\times$4, $k$=6. 
       Node 46 was dragged from the top level,
       node 38 and its 1-step neighbors were dragged and rotated,
       and a 3d block (0 --- 26) was
       dragged and rotated from the lower right hand corner.\\[7ex]
    \end{minipage}
}
\normalsize
   \vspace{-3ex}  
    \caption[Dragging network-graph nodes and fragments --- 1d, 2d, and 3d]
       {\textsf{Examples of dragging nodes and fragments
       for 1d, 2d, and 3d CA network-graphs, including
       single nodes, linked fragments and blocks.}}
     \label{net_drag123}
\end{figure}

CA wiring\cite[\hspace{-1ex}\footnotesize{\#10}]{EDD} in 1d, 2d
or 3d is presented in the same way by default in network-graphs (1d as circle layout).
Fragments  defined by link distance also follow the same dimensions, and
can be dragged as in figure~\ref{net_drag123}.  There may be a $k$-mix and random wiring.
Scaled discs/links can toggle between dependence on inputs or
outputs.  As with the other graph-types, links can be cut/added
allowing any arbitrary network-graph, but this is self-contained and does not affect the
underlying wiring-scheme outside the network-graph itself.

\enlargethispage{5ex}
\begin{figure}[H]
  \footnotesize
\begin{center}
  \begin{minipage}[t]{1\linewidth} 
\begin{minipage}[t]{.21\linewidth}
\includegraphics[bb=385 176 957 724,clip=,width=1\linewidth]{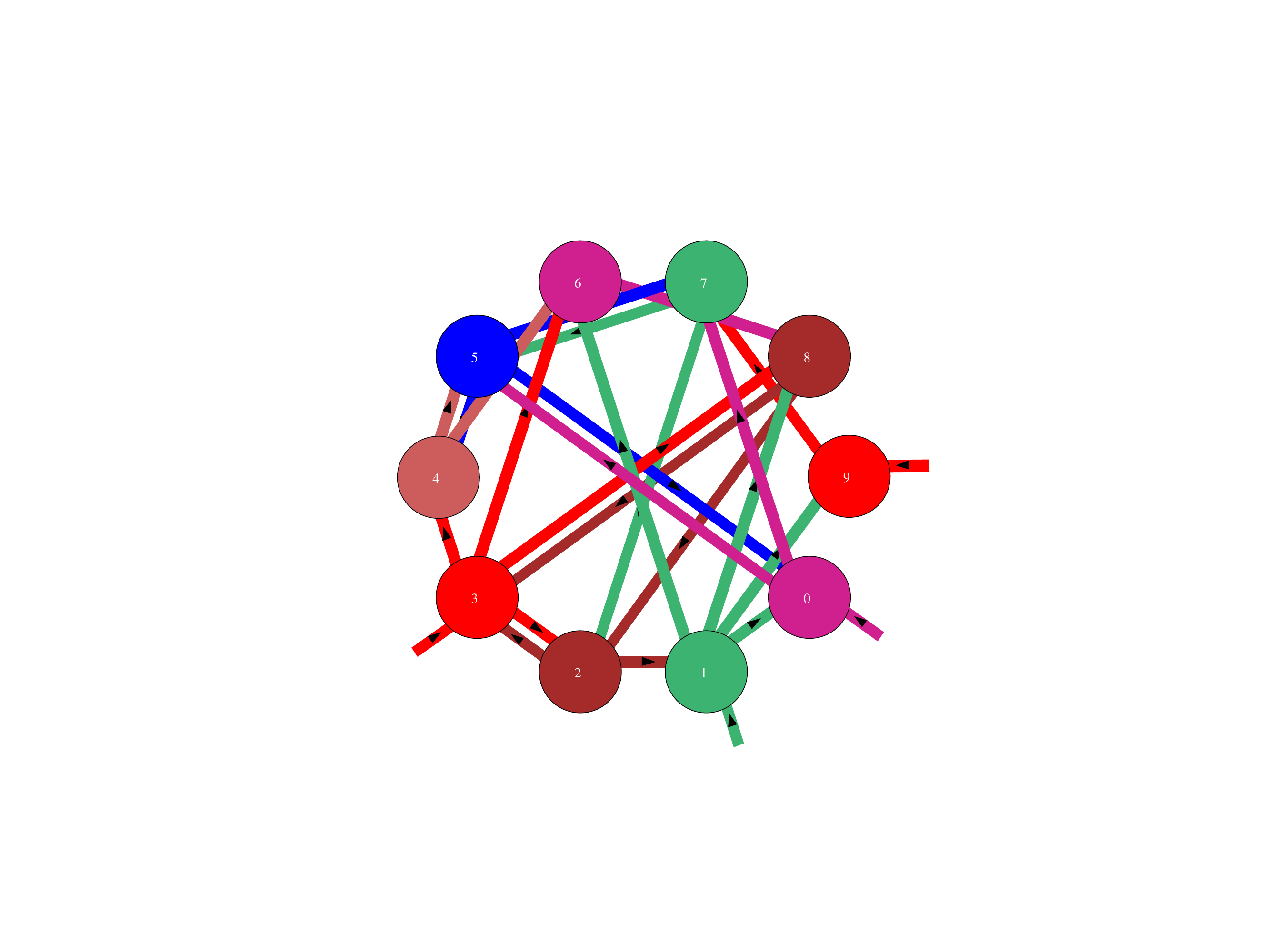}
\begin{center}
\vspace{-2ex}\textsf{(a) default circle layout, nodes/links scaled by (equal) inputs}
\end{center}
\end{minipage}
\hfill
\begin{minipage}[t]{.25\linewidth} 
\includegraphics[bb=402 218 917 679,clip=, width=1\linewidth]{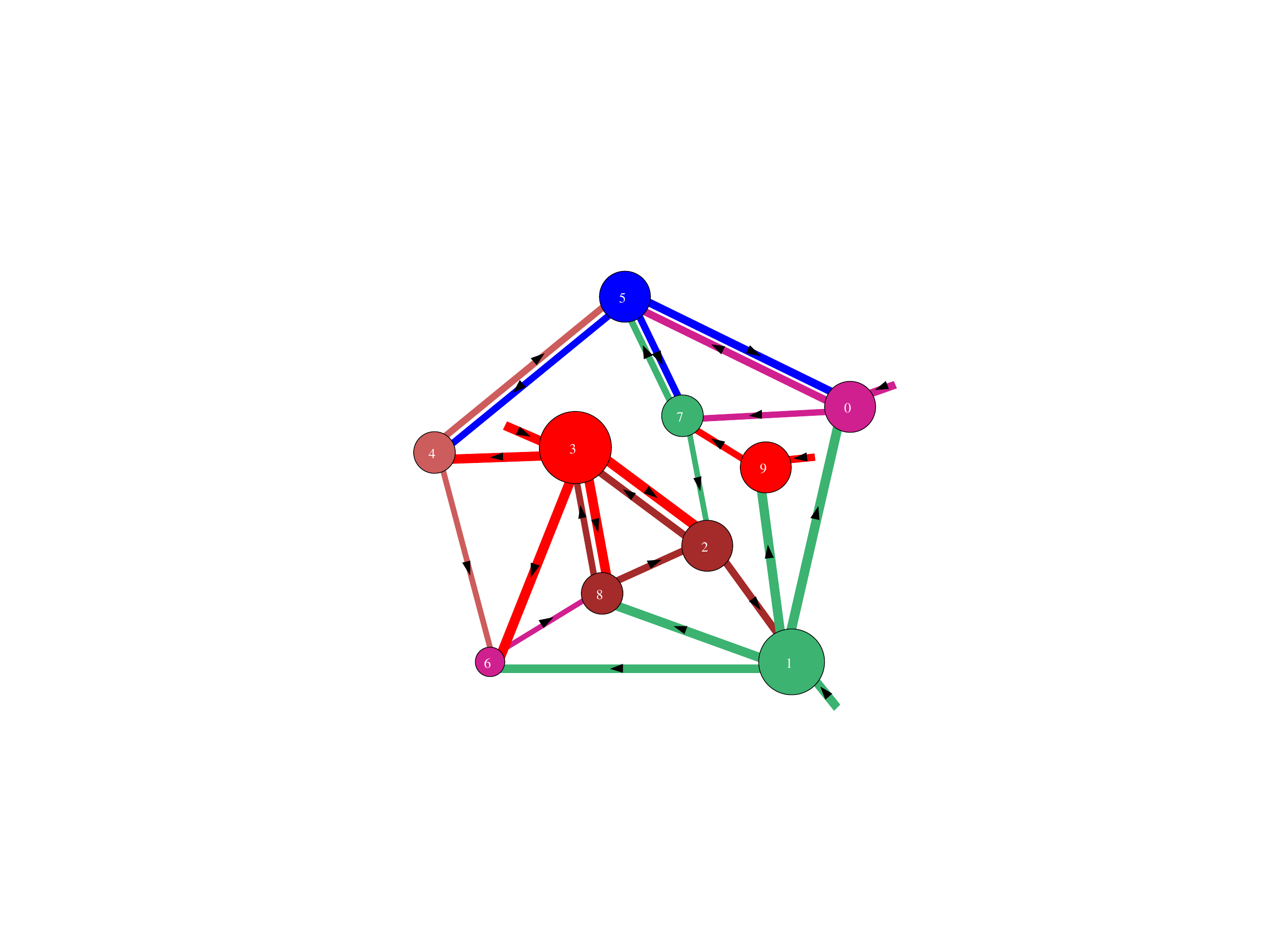}
\begin{center}
\vspace{-2ex}\textsf{(b) nodes/links scaled by outputs --- and rearranged by drag/drop}
\end{center}
\end{minipage}
\hfill
\begin{minipage}[t]{.25\linewidth} 
  \includegraphics[bb=402 218 917 679,clip=,width=1\linewidth]{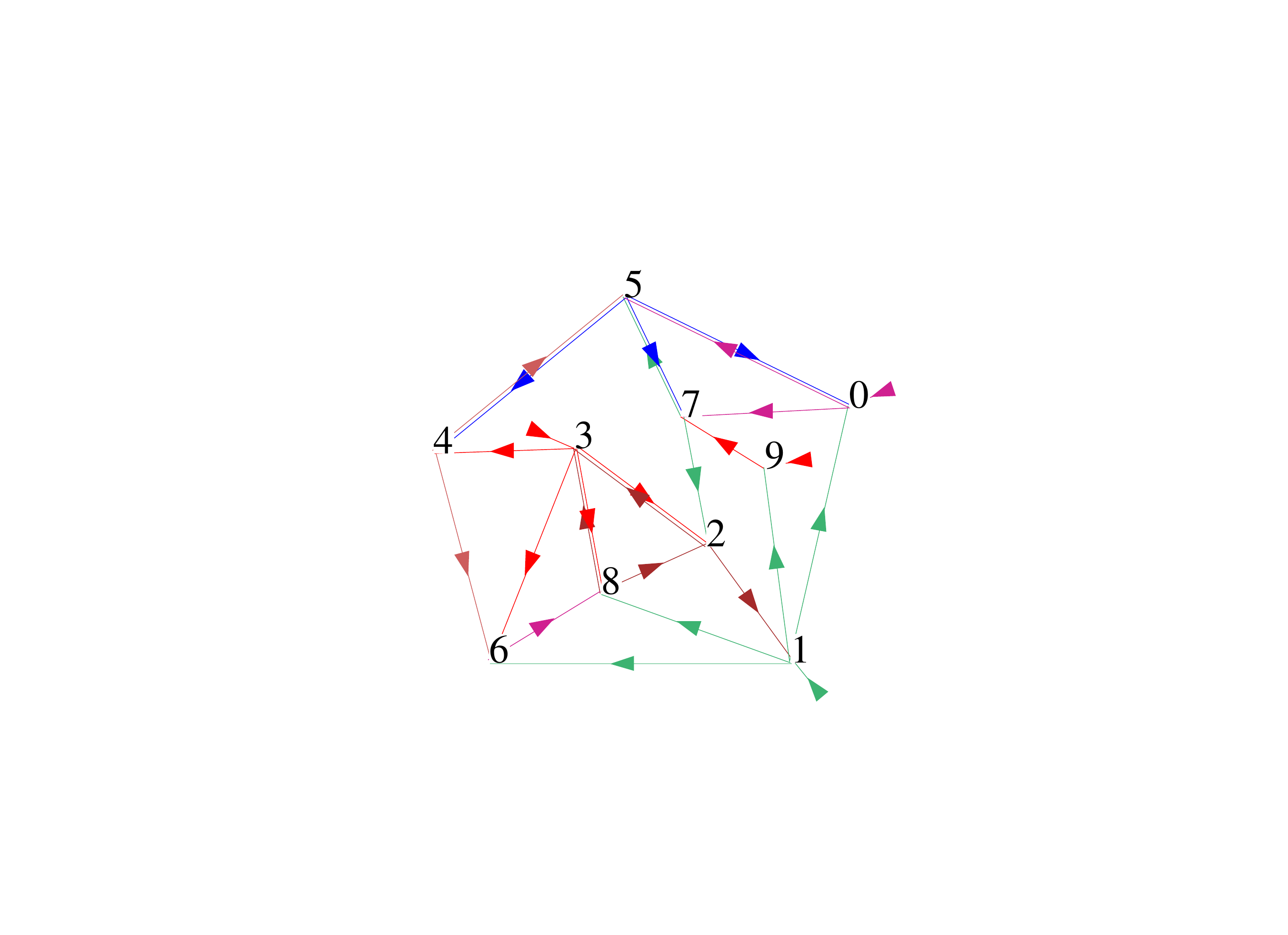}
\begin{center}
\vspace{-2ex}\textsf{(c) nodes as numbers, links as simple lines and arrows expanded}
\end{center}
\end{minipage}
\hfill 
\begin{minipage}[t]{.27\linewidth} 
  \fbox{\includegraphics[bb=52 69 300 244,clip=, width=1\linewidth]{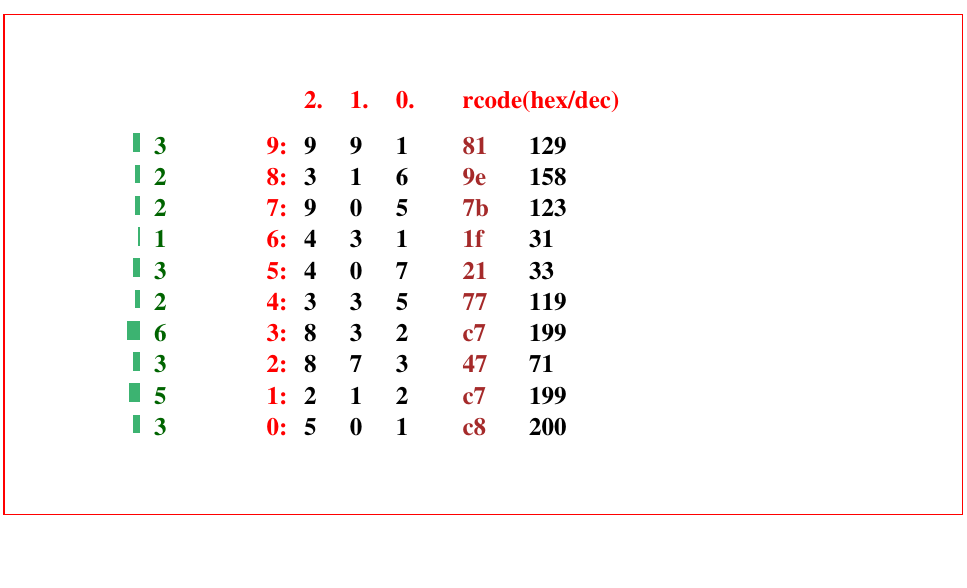}} 
\begin{center}
  \vspace{-2ex}\textsf{(d) The `wiring matrix'' and  rules defining the RBN, green column
    shows out-degree}
\end{center}
\end{minipage} 
\end{minipage}
\end{center}
\vspace{-4ex}
\caption[Simple network-graphs]{
  {\textsf{(a)-(c) network-graph alternative presentations of a small RBN defined in (d)
      with homogeneous $k$=3 inputs but variable outputs.
}}}
\label{rbn10k3_network_graphs}
\end{figure}

\clearpage

\section{The ibaf-graph}
\label{The ibaf-graph}

\begin{figure}[htb]
\begin{center}
  \mbox{\includegraphics[bb=9 487 1248 862,clip=,width=1\linewidth]{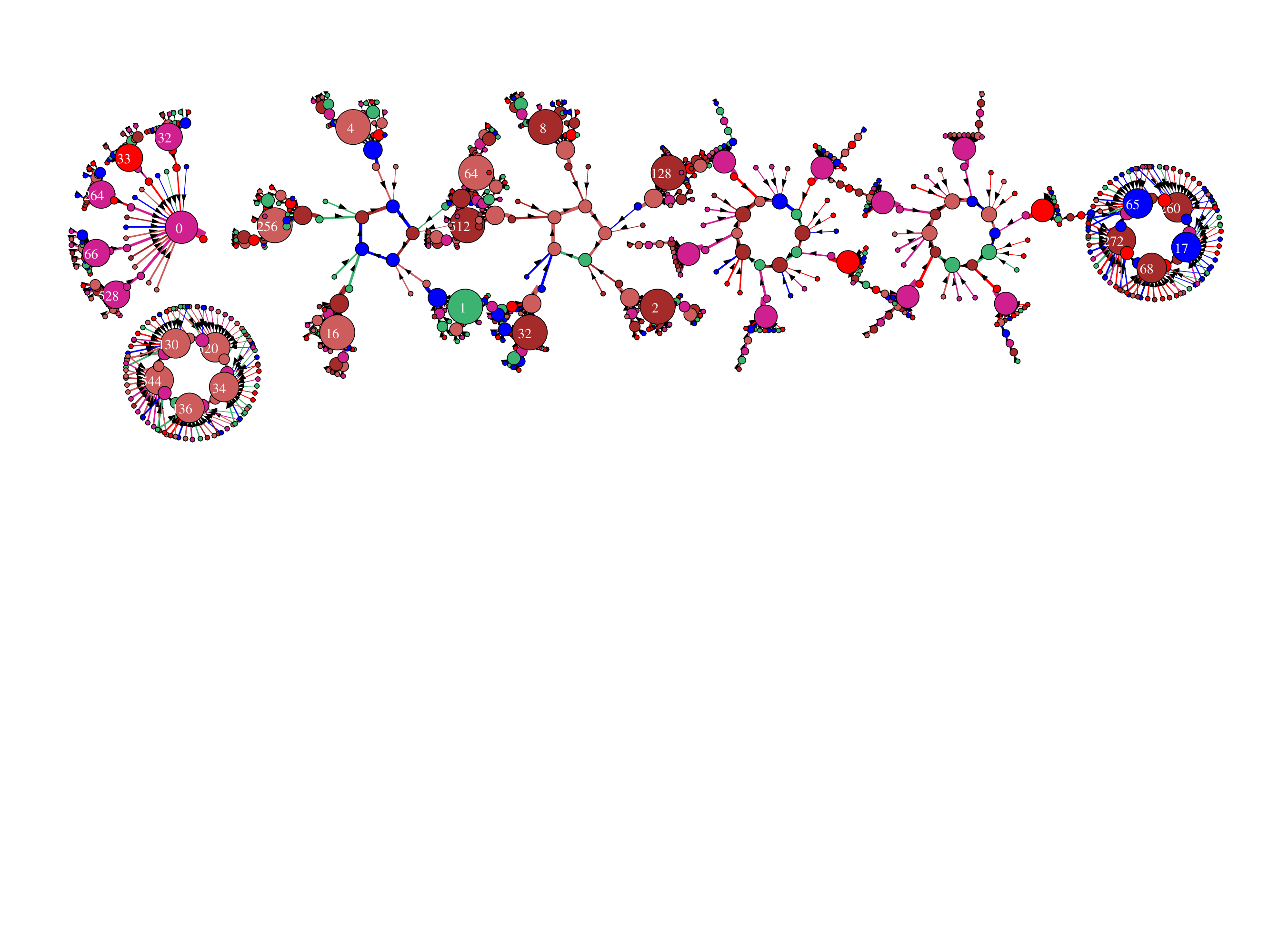}}
\end{center}
\vspace{-7ex}
\hspace{22ex}
\footnotesize{
\begin{minipage}[t]{.77\linewidth}
        \textsf{(a) The default presentation of the ibaf-graph follows\\
      the original uncompressed basin of attraction field layout.}
\end{minipage}
}
\begin{center}
  \includegraphics[width=1\linewidth,bb=71 78 1253 676,clip=]{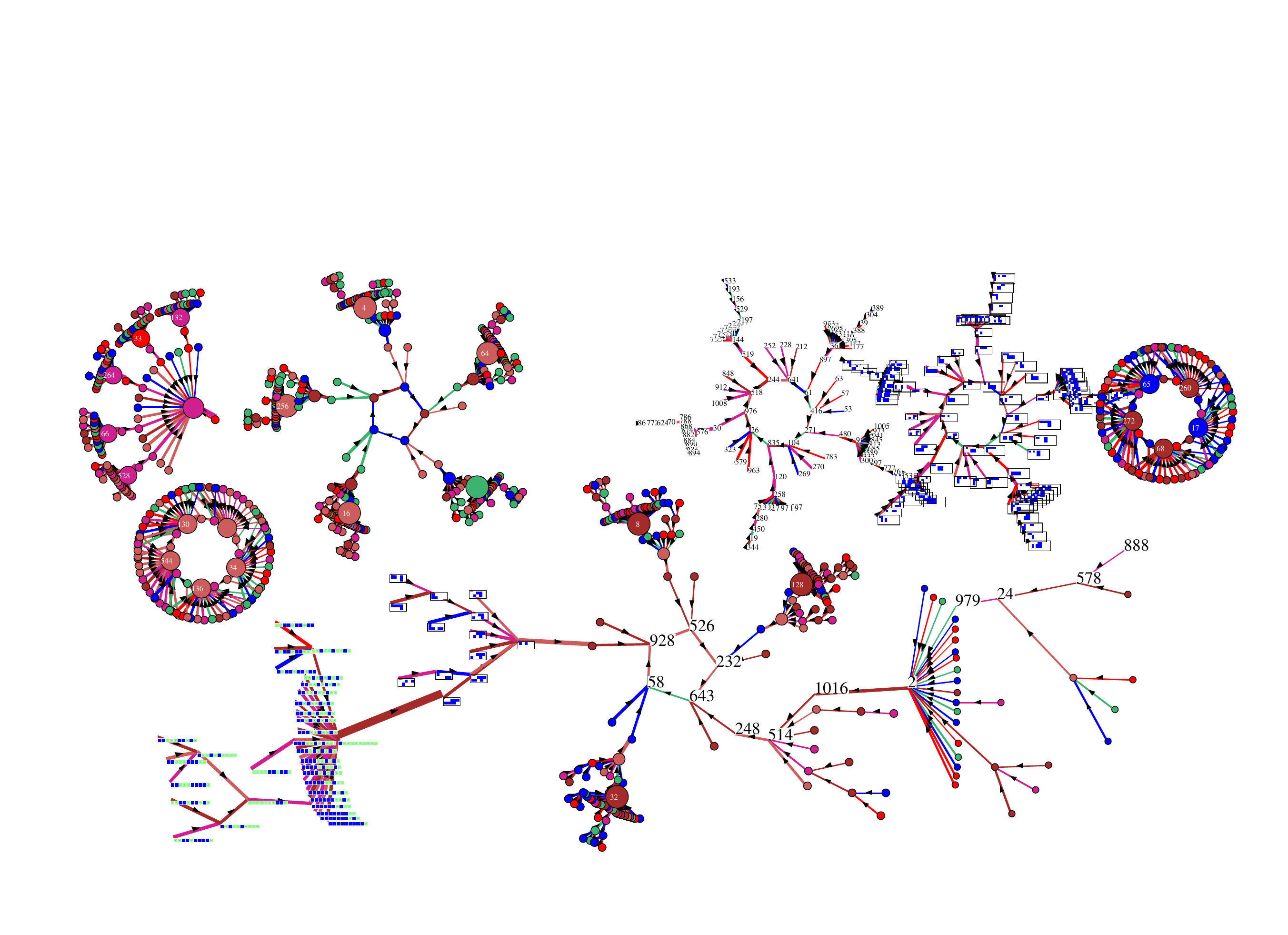}\\
\footnotesize{
\begin{minipage}[t]{.95\linewidth}
  \textsf{(b) The amended ibaf-graph:
            Basin 3 was dragged lower, expanded, and some fragments were
            dragged and relabelled in decimal, and as patterns in 1d and 2d.
            Note the sequence of decimal nodes
            starting at 888 (top right) which follow output arrows
            leading to the attractor. Nodes were relabelled to decimal in basins 4 and to 2d
            patterns in basin 5.}
\end{minipage}
}
\end{center} 
\vspace{-2ex}
\caption[A rearranged  ibaf-graph]
        {\textsf{
            The default ibaf-graph (a) was manipulated and amended as described in
            (b) which shows the current snapshot of the result.
            1d CA, $v2k3$, $n$=10, rcode 9, with 7 basins.  
        }}
        \label{r9-ibaf}
\end{figure}

\noindent The ibaf-graph, as the example in figure~\ref{r9-ibaf},
is an interactive image of the classic basin of
attraction field (state transition graph) as drawn in DDLab according
to the graphic conventions and settings for layout, scale and
presentation\cite[\hspace{-1ex}\footnotesize{\#24-\#26}]{EDD}.
The ibaf-graph has $v^n$ nodes, each with out-degree=1 and
variable in-degree$\geq$0 depending on the number of each state's pre-images.
The in-degree provides the weight of discs/links.
Just one basin (or fragment) can be isolated while the rest are rendered invisible.

\begin{figure}[htb]
 \footnotesize
 \begin{center}
   \begin{minipage}[t]{.49\linewidth}
     \includegraphics[bb=361 121 916 704,clip=, width=1\linewidth]{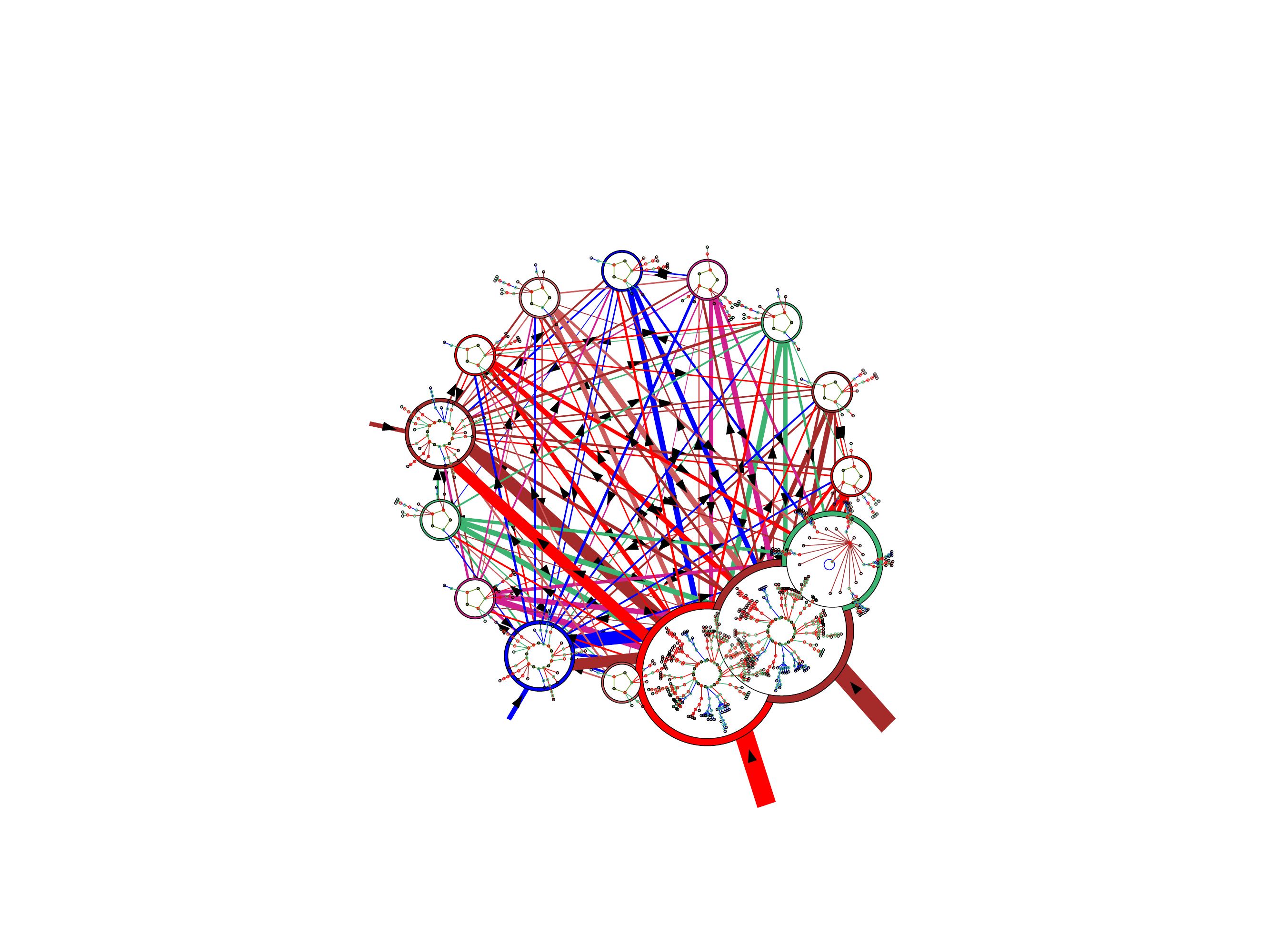}\\[-7.5ex]     
          {\textsf{(a) circle layout jump-graph\\ with classic basins drawn inside nodes.\\
          The basins coordinates data are saved.}}
   \end{minipage}
   \hfill
   \begin{minipage}[t]{.49\linewidth}
    \includegraphics[bb=361 121 916 704,clip=, width=1\linewidth]{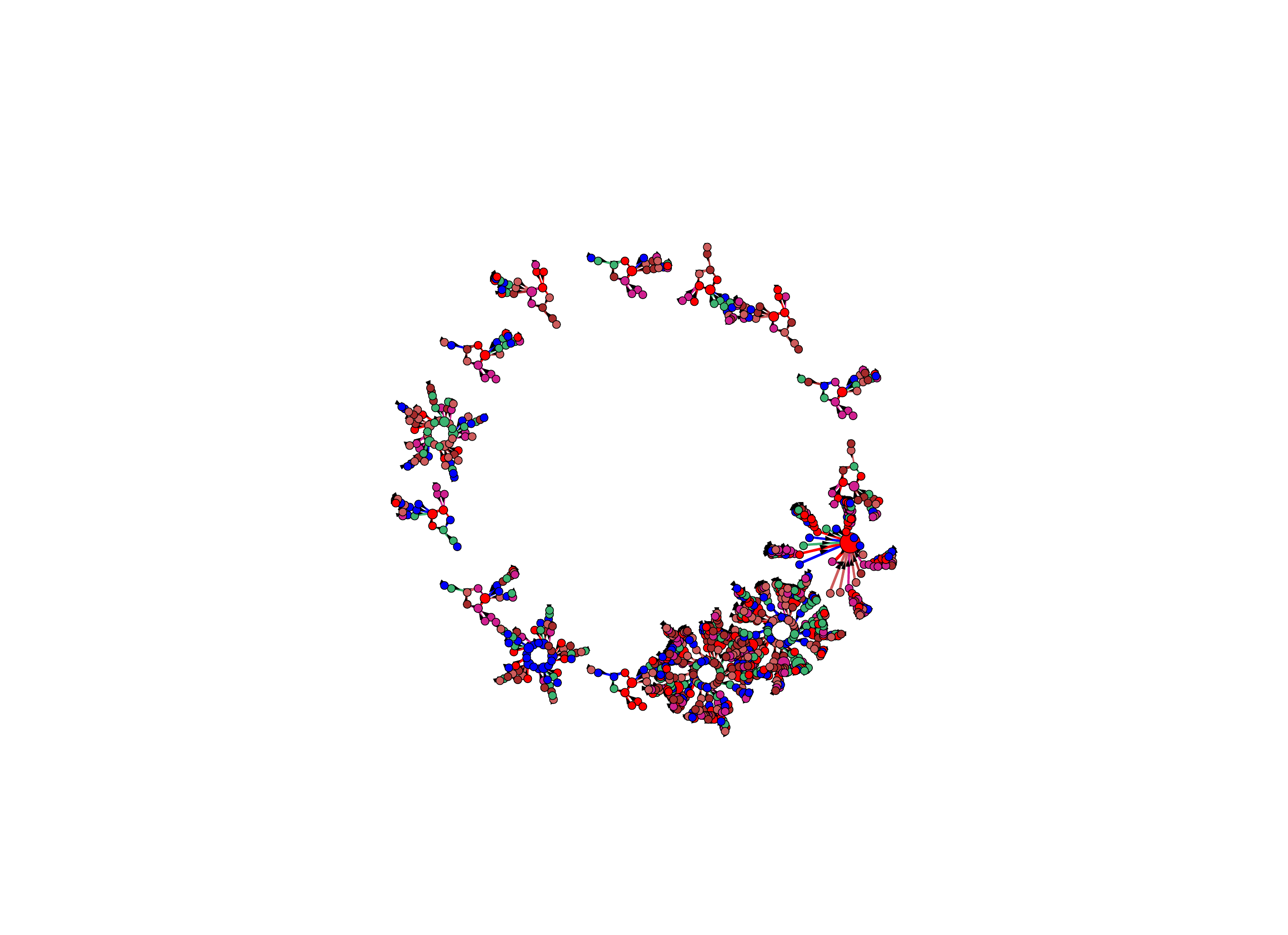}\\[-5ex]
          {\textsf{(b) ibaf-graph with jump-graph basin coordinates data loaded from (a)
              has exactly the same layout as (a).
               }}
    \end{minipage}
   \normalsize
       \caption [Reload jump-graph basins to ibaf-graph]       
                {\textsf{(a) Drawing basins of attraction (scale adjustable)
                    inside the nodes of the jump-graph.
           (b)~Reloading jump-graph basin coordinate data (\texttt{.grh} file) into
           the ibaf-graph for an exact copy of the state-space layout, in this case
           showing basins of attraction in circle layout, which can be manipulated as usual.
       1d $v2k3$ CA rcode 110, $n$=10}
          }
        \label{r110-f-ibaf}
 \end{center}
\vspace{-4ex}
\end{figure}

The layout of basins can also be flexibly preset by drawing the field
within the nodes of its jump-graph
(figures~\ref{r110-f-ibaf}, \ref{basin_r133_comp}), where the 
basin coordinates can be saved/loaded. This allows an ibaf-graph to have arbitrary
or special basin layouts including geometric --- circle/spiral/1d/2d/3d.

The ibaf-graph is generated in two stages starting with the
exhaustive reverse algorithm\cite[\hspace{-1ex}\footnotesize{\#29.7}]{EDD}
which computes ``exhaustive pairs'' --- the outputs of the entire
$v^n$ state-space providing the data for directed links between
states. The second stage computes and draws the ibaf-graph itself with
nodes/links scaled by inputs (pre-images), and nodes depicted as discs,
numbers starting at 0 (zero), or patterns. Numbers in decimal or hexadecimal
are the equivalents of the pattern string.

\enlargethispage{2ex}
The exhaustive algorithm limits network size but
small sizes are in any case preferable for practical
graphics and computation time of both exhaustive pairs and the
ibaf-graph --- the times for both stages are
independent of $k$ or the wiring/rule setup,
but very sensitive to the size of state-space.
The exhaustive data and the ibaf-graph data can
be saved/loaded to avoid re-computation.
As well as CA, RBN and DDN, the exhaustive algorithm makes the
ibaf-graph compatible for null boundaries, random maps and sequential
updating.
\clearpage

\begin{figure}[htb]
\begin{center}
\includegraphics[bb=59 252 1096 744,clip=,width=1\linewidth]{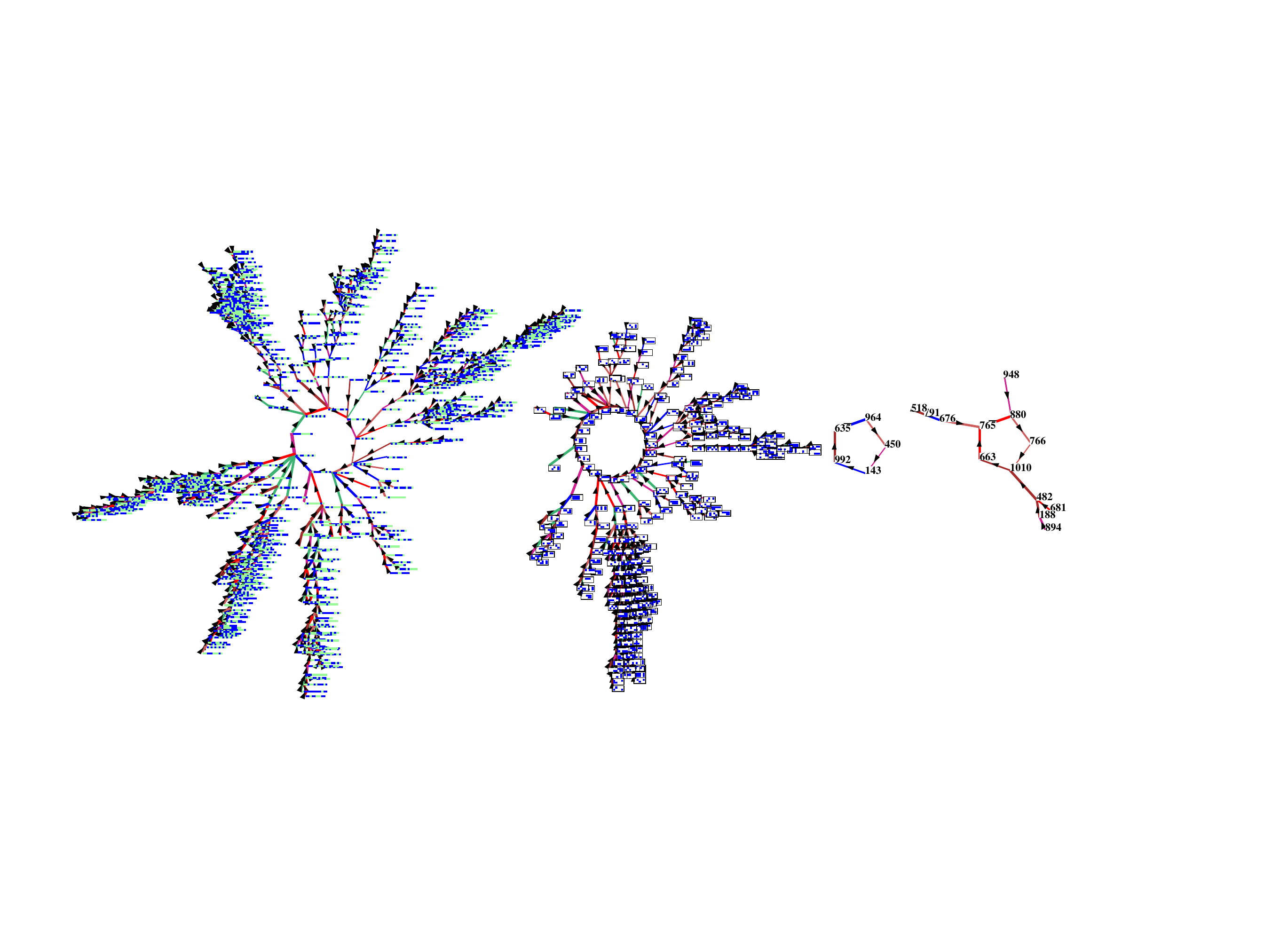}
\end{center}
\vspace{-5ex}
\caption[The ibaf graph of a random map]
        {\textsf{The ibaf graph of a binary random map where
            each state in state-space is assigned a successor at random (but optionally
            with some bias), $v$=2, $n$=10.
            Other CA, RBN or DDN parameters previously assigned are irrelevant.
            In this snapshot some basins were moved and rotated to
            avoid overlap, and nodes were changed to 1d and 2d patterns, and decimal equivalent
            numbers.}}
\label{fig:rnd-map}
\end{figure}
\clearpage

The ibaf-graph is selected from the basin parameter
prompts\cite[\hspace{-1ex}\footnotesize{\#24.3}]{EDD},
then the exhaustive algorithm is selected at the final ``basin field'' prompt.
Alternative options select sequential updating
of the current system, or a
random map\cite{wuensche97}\hspace{-.2ex}\cite[\hspace{-1ex}\footnotesize{\#29.8}]{EDD},
as the example in figure~\ref{fig:rnd-map}.
The classic basin of attraction field is drawn first,
then a subsequent option draws the ibaf-graph (toggle between the two with {\bf win-w}). 
A ``quick-start'' example for selecting an
ibaf-graph can be found in \cite[\hspace{-1ex}\footnotesize{\#4.2.3}]{EDD}.

\section{The jump-graph}
\label{The jump-graph}

\begin{figure}[htb]
   \begin{center}
   \begin{minipage}[t]{.92\linewidth}
     \includegraphics[bb=261 63 671 472, clip=, height=.4\linewidth]{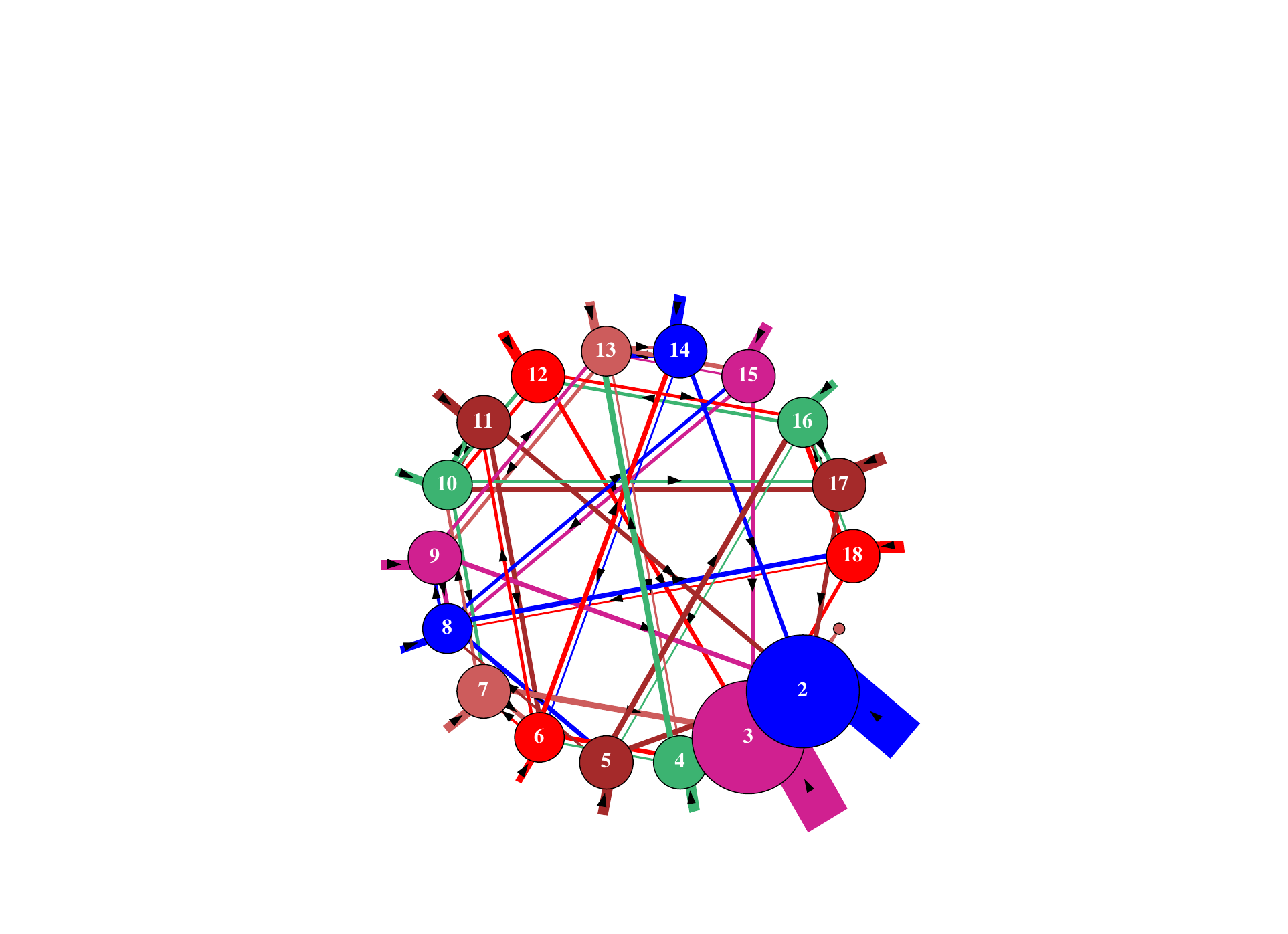}
   \hfill
   \includegraphics[bb=163 90 765 473, clip=, height=.4\linewidth]{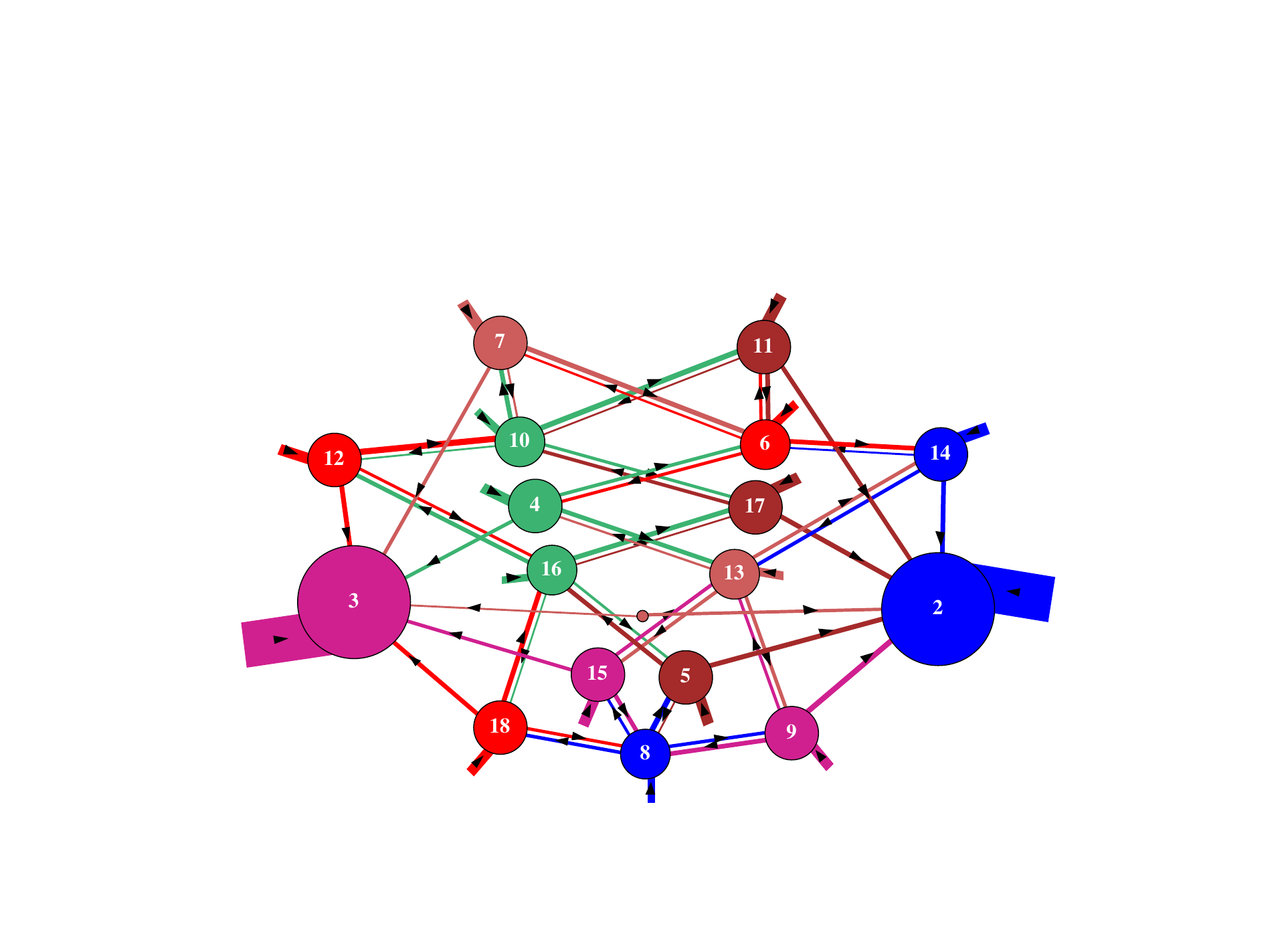}
   \end{minipage}
   \end{center} 
    \vspace{-4ex}   
    \caption [Dragging the jump-graph, further examples]
    {\textsf{The f-jump-graph for a 1d CA $v2k3$, $n$=10, rcode 141.
    \underline{\it Left}: the default circle layout, and
           \underline{\it Right}: rearranged with the mouse
           to clarify the jumps.}}
     \label{fig:jump_CA141_10}
\end{figure} 

\noindent Perturbations due to noise or external signals are most likely 
to take affect once a system has relaxed to its attractor because
that is where the dynamics spends the most time. Taking single-bit or single-value
perturbations to attractor states as the simplest case,
the jump-graph\cite{wuensche2002}\hspace{-.2ex}\cite[\hspace{-1ex}\footnotesize{\#20.7}]{EDD}
represents the probabilities of jumping between basins.
Apart from the philosophical/mathematical implications\cite{Voorhees},
insights into the stability and adaptability of dynamics is
relevant in attractor models of memory and learning\cite{Wuensche92,wuensche96,wuensche2005}
and of cell differentiation in genetic networks\cite{kauffman69,wuensche98a}.

To compute jump-graphs, algorithms track
where all possible single-bit/value flips to attractor states end up,
whether to the same or to which other attractor. The information is
presented as a graph with weighed vertices --- nodes shown as discs scaled by basin volume
--- the number of states in the basin, and the weight of links (edges) by jump probability
--- the actual jumps comprising a link divided by the total of possible
jumps from the attractor. The data is also available in the
adjacency-matrix\cite[\hspace{-1ex}\footnotesize{\#20.9}]{EDD}
or jump-table (figure~\ref{jump_matrix_RBN}).

\begin{figure}[htb]
\footnotesize{
\begin{center}
\includegraphics[height=.25\linewidth]{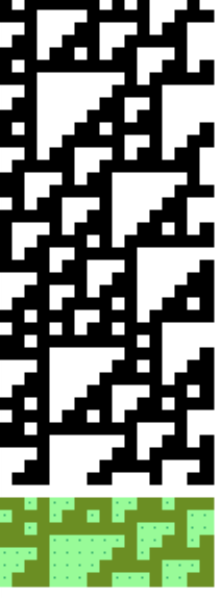}
\includegraphics[height=.25\linewidth]{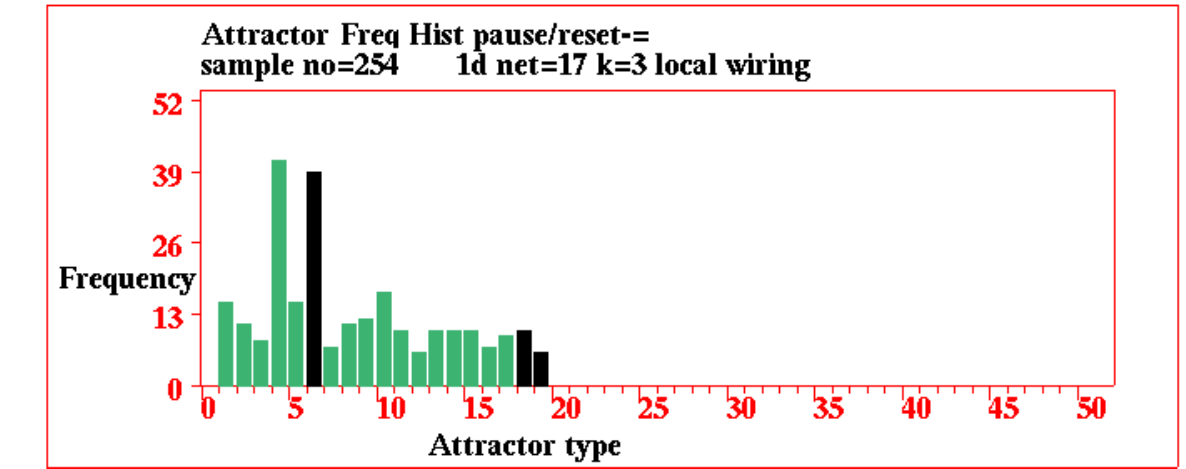}
\includegraphics[bb=352 166 884 714, clip=, height=.25\linewidth]{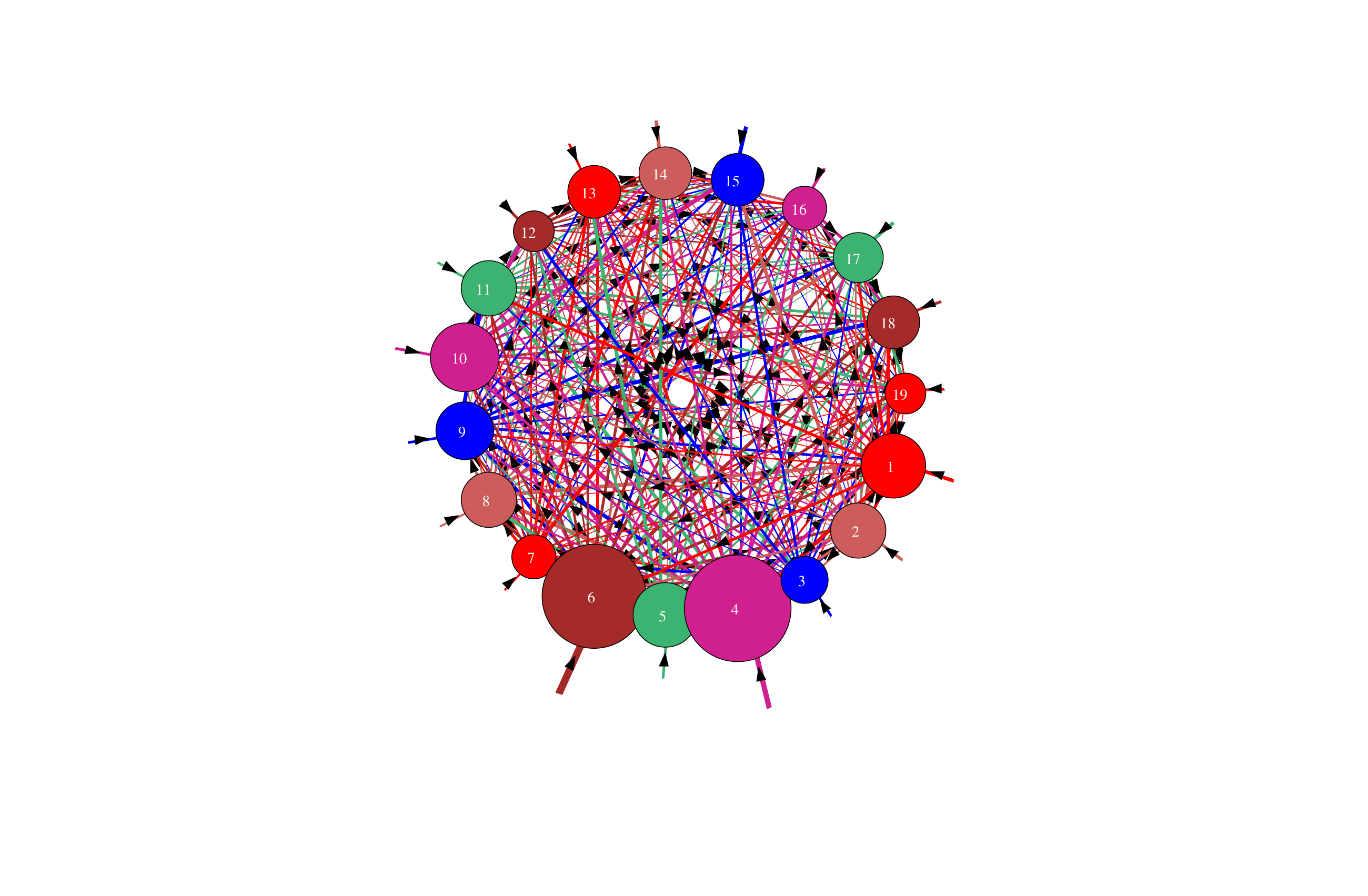}
\end{center}
\vspace{-2ex}
\textsf{(a) A sample of 254 random initial states captures 19 basins so far.
      {\it Left}: space-time patterns + green attractor,
      {\it Center}: unsorted attractor histogram continuously
     updates (in black) showing the type
($x$ axis) against the frequency of arriving at that type
($y$ axis), rescaling automatically (bars turn green) on reaching
the current scale limits.
     {\it Right}: h-jump-graph}
\begin{center}
  \includegraphics[height=.25\linewidth]{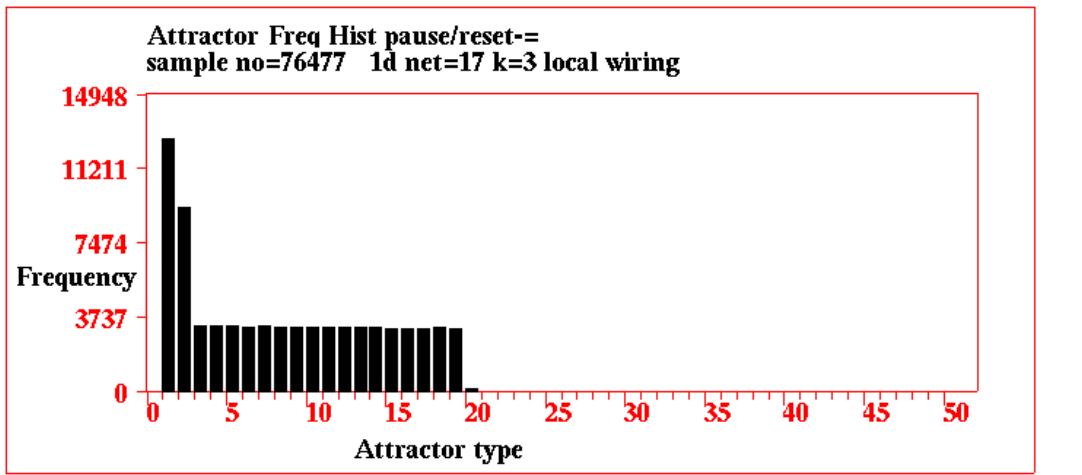}
  \includegraphics[bb=371 251 1037 795, clip=, height=.25\linewidth]{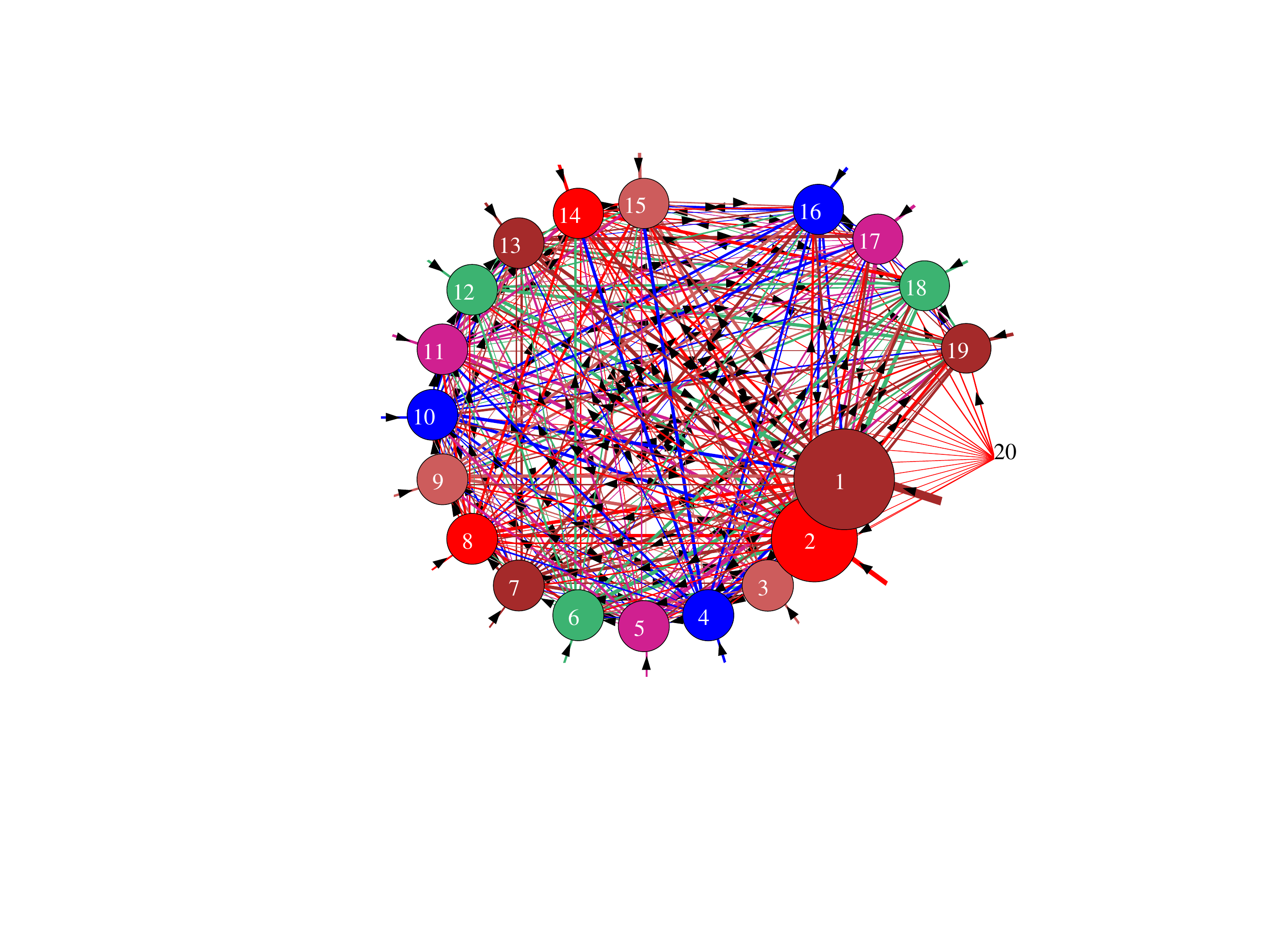}
\end{center}
\vspace{-2ex}
\textsf{(b) A sample of 76477 random initial states captures
  another basin, basin 20, with very small volume.
     {\it Left}: attractor histogram sorted by frequency,
     {\it Right}: h-jump-graph with block 16 to 20 dragged.}
}
\vspace{-1ex}
 \caption[Attractor histograms and h-jump-graphs, large sample]
         {\textsf{Attractor histogram and h-jump-graph:
             (a) interrupted early and unsorted, (b) interrupted
             after a large sample which has uncovered another small basin.
              1d CA rcode 110 n=17.}
           \label{att-hist-h-jump-large}}
\vspace{-4ex}         
\end{figure}

\begin{figure}[htb]
       \begin{minipage}[t]{.52\linewidth}
       \includegraphics[bb=321 180 1024 774, clip=, width=1\linewidth]{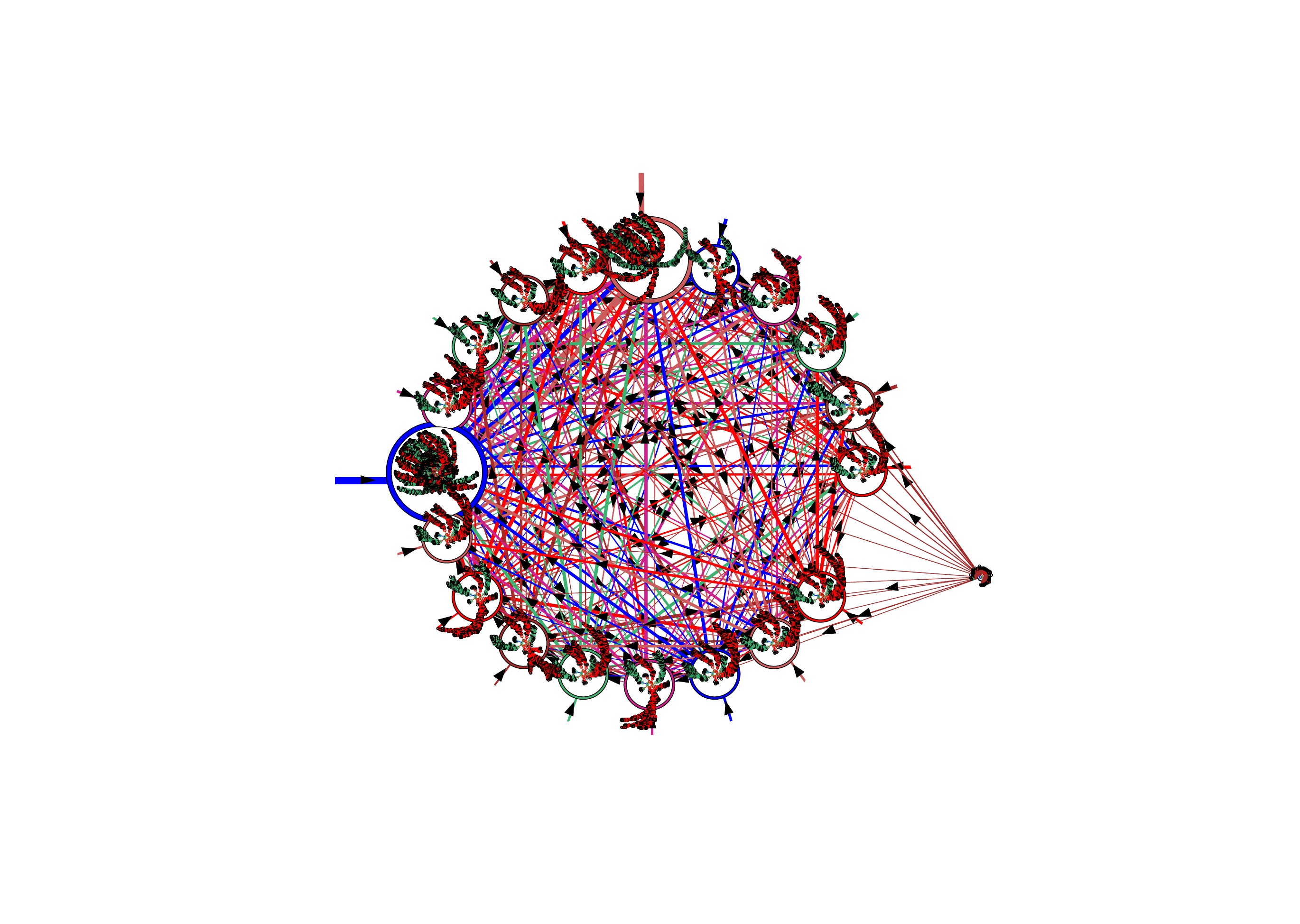}
       \end{minipage}
        \hfill
        \parbox[t]{.44\linewidth}{  \vspace{-32ex} 
       \caption [The f-jump-graph with basins inside]       
                {\textsf{The f-jump-graph (with basins drawn inside) for the same CA
                    (rcode 110 $n$=17)  as in figure~\ref{att-hist-h-jump-large} ---
                    as a reality check to compare
                    with the \mbox{h-jump-graph}. This is the default circle
                    layout but with the smallest basin dragged out on the right,
                    which has the lowest frequency/volume in the \mbox{h-jump-graph}
                    so the least likely to be found --- a large sample is required.
                    }}
                \label{jump+basins-r110n17}}
\vspace{-2ex}        
\end{figure}

Two types of jump-graph can be created,
the {\it f-jump}-graph as in
figures~\ref{fig:jump_CA141_10} and \ref{jump+basins-r110n17}
directly from the basin of attraction field, 
or the {\it h-jump}-graph from the
``attractor histogram''\cite{wuensche2002}\hspace{-.2ex}\cite[\hspace{-1ex}\footnotesize{\#29.8}]{EDD}
derived statistically from a sample of space-time patterns  as in
figure~\ref{att-hist-h-jump-large}.

The nodes in a jump-graph are numbered starting from 1, as opposed
to nodes in the network-graph or ibaf-graph where the start is node 0
(zero). For the \mbox{f-jump-graph} the number and order of nodes
follows the basin of attraction field.  For the h-jump-graph the
number of nodes and order follows attractor types discovered so far
--- bars in the attractor histogram --- and the order of discovery,
but the order can sorted by several measures: the frequency
of each type (which approximates basin volume), by attractor period,
or by average transient length.

Some differences between the two jump-graph types are worth noting.
The \mbox{f-jump-graph} allows basins to be redrawn at nodes, and there are
options for ``layout-only'' without links, either uncompressed, or
compressed for 1d CA.  The h-jump-graph depends on attractors
discovered so far (there may be more as the sample grows) but allows
much larger systems, though large systems should be biased towards
order--- otherwise transients will be too long for the effective
accumulation of data.  In other respects the two jump-graphs and their options
are very similar.

The f-jump-graph applies to any system where a basin of attraction
field can be generated with a suitable reverse algorithm, including CA
(for 1d with or without compression), RBN or DDN, with synchronous
updating. Sequential updating and random maps will use the
exhaustive algorithm by default.  The f-jump-graph is selected from
the basin parameter prompts\cite[\hspace{-1ex}\footnotesize{\#24.3}]{EDD}
and will be drawn automatically after the
basin of attraction field is complete. Mutants, by wiring or rules, can be
generated sequentially.  As noted in section~\ref{The ibaf-graph} and 
figures~\ref{r110-f-ibaf} and \ref{jump+basins-r110n17},
basins of attraction (with adjustable scale) can be drawn at (or
inside) f-jump-graph nodes, and this action also provides a layout method
for the ibaf-graph.

The h-jump-graph applies to any system running space-time patterns.
The attractor histogram (figure~\ref{att-hist-h-jump-large})
gathers data on attractors and their relative
sizes statistically,  by automatically running space-time patterns
from many random initial states, identifying the different 
attractors types, and continuously building a histogram of the frequency of falling into
each type which approximates the relative basin volume.
The method can be applied to large networks, especially RBN/DDN ---
too large to generate the basin of attraction field.
The h-jump-graph is generated while pausing the attractor histogram.
The selection of SEED-mode or TFO-mode (at the ``first prompt''\cite[\hspace{-1ex}\footnotesize{\#6}]{EDD})
and the attractor histogram in space-time parameters\cite[\hspace{-1ex}\footnotesize{\#24.3}]{EDD}
is required.

Space-time patterns will run
top-left in the DDLab screen, from successive random initial states, identifying each attractor ---
each type is either added as a new bar to the histogram (x-axis), or is added to
an existing  bar's height/frequency (y-axis).
As the histogram builds, rescaling automatically in a lower center window,
information is updated simultaneously top-right and at the top of the of the histogram window.
Pausing the histogram\cite[\hspace{-1ex}\footnotesize{\#31.7.3}]{EDD}
then allows various options including sorting and
drawing the h-jump-graph computed at the pause,
or continuing the search --- a larger sample can capture more attractors.
Toggle showing the graphics of space-time patterns {\it off} to drastically increase histogram sample speed,
which can be done on-the-fly\cite[\hspace{-1ex}\footnotesize{\#32.2}]{EDD}.


\section{Interactive graphs}
\label{Interactive graphs}

The ibaf-graph adapts pre-existing coding for interactive functions
of the network-graph and jump-graph\footnote{The network graph and jump-graph were part of
DDLab's May 2002 update\cite{DDLab2002update}, where the jump-graph was
originally named the ``meta-graph''.}.
With the introduction of the ibaf-graph the commands and options have been
expanded, updated and improved. They are the same or very similar for the three graph types
--- this section gives a summery\footnote{The full
  operating instructions for interactive graphs, too lengthy for this article,
  can be found in the latest online update of ``Exploring Discrete
  Dynamics'' (EDD)\cite[\hspace{-1ex}\footnotesize{\#20}]{EDD},
  Check also for updates to DDLab\cite{Wuensche-DDLab}.},
noting the few exceptions.

Once selected and activated, an interactive graph (network, ibaf or jump)
is presented in a large central window in the DDLab screen, together with
the initial top-right reminder of the various on-the-fly key-press
options, and showing the current graph type.

The graphs and options have two interchangeable stages with distinct
reminders: the {\it initial}-graph/reminder where options
apply to all nodes simultaneously, and the {\it drag}-graph/reminder
where options generally apply to a single node, a geometric range of adjacent
nodes (blocks), or most significantly to the selected node and its
linked fragment.  Examples below are for the ibaf-graph --- the
network-graph and jump-graph reminders are similar --- the
differences are noted. Keyhits usually take effect without entering return,
but in some cases there are suboptions that are indicated but not
fully described in this article.

\subsection{initial-reminder}
\label{initial-reminder}

The initial-reminder options apply to the whole graph, 
described in sections~\ref{initial and drag graphs --- shared options} and \ref{initial-graph --- exclusive options}.

\begin{quote}
  \underline{\it this example for the ibaf-graph}\\[1.5ex]    
     {\color{BrickRed}{\bf IBAF-graph:}} {\bf drag-(def) PScript-P net-\# ant-a unscram-u win-w rank0-k}\\
     {\bf settings-S rot-x/X flip-h/v nodes-(/)/= links-\{/\} both-[/]}\\
     {\bf Unreach-U matrix-t/T nodes-n/N links-l Labels-+ arrows-A/{$\boldsymbol <$}/{$\boldsymbol >$}}\\
     {\bf layout: file-f graph-g circle/spiral-o/O 1d/2d(tog)/3d-1/2/3 rnd-r/R quit-q:}\\
     \underline{\it note:}\\
     option {\bf graph-g} applies to the ibaf-graph only\\
     there are extra options: {\bf in/out-z} for network-graph, and {\bf basins-i/I/s} for f-jump-graph         
\end{quote}

Click the left or right mouse button with the pointer anywhere on the
screen (or press {\bf return}) to change to the drag-graph and the
context dependent drag-reminder.

\subsection{drag-reminder}
\label{drag-reminder}

The drag-reminder options apply to a selected node --- just a single node,
or the node and its linked fragment by {\it inputs}, {\it outputs}, or {\it either}, or potentially
a ``block'' of nodes, or ``allnodes''.
The initial node is 1 for the jump-graph with ``single'' active,
otherwise node 0 with ``either, step=nolimit'' active.
Sections~\ref{initial and drag graphs --- shared options} and \ref{drag-graph --- exclusive options}
describe the drag-options.

\enlargethispage{3ex}

\begin{quote}
     \underline{\it this example for the ibaf-graph with ``either'' active}\\[1.5ex]
     {\color{BrickRed}{\bf node 0, either, step=nolimit:}} {\bf leftb-drag  PScript-P elstc/snap-d gap-g}\\
     \mbox{\bf inactive?-rightb first, rot-x/X flip-h/v nodes-(/)/=/E links-\{/\} both-[/] just-j/J}\\
     {\bf Lnk0:cut/restore-c/r Lnks0-0:cut/add/restore-C/A/R net-\#}\\
     {\bf step-(1-9) nolimit-0 single-s in/out/either-i/o/e all-a exit-q:}\\
     \underline{\it note:} \\
     for {\color{BrickRed}{\bf single}} status,
         options {\bf rot-x/X flip-h/v links-\{/\} both-[/]} are absent,\\
               and the 3rd line changes to {\bf block-B Label-L/+ net-\#}
\end{quote}

The drag-reminder is context dependent on the current activation of
{\color{BrickRed}{\bf single}},
{\color{BrickRed}{\bf either}}, {\color{BrickRed}{\bf inputs}}, {\color{BrickRed}{\bf outputs}},
{\color{BrickRed}{\bf block}} or {\color{BrickRed}{\bf allnodes}}, which appear in the title.
Click the left mouse button on a node to activate it and hold down to
drag the node or fragment --- then release (drop) in a new
position. Initially, to activate a node it may be necessary to click
the right button first, then the left, possibly a few times.  Hence
the reminder,
                       \begin{quote} 
                        {\bf inactive?-rightb button-first}
                       \end{quote}

Enter ``{\bf q}'' to return to the initial-graph/reminder.

\subsection{initial and drag graphs --- shared options}
\label{initial and drag graphs --- shared options}

\noindent By default, nodes (vertices) are displayed as colored
discs cycling through four colors, with a central number if the disc is big enough,
Links (edges)  have centrally placed directional arrows.
Links are colored according to the parent node and aligned
asymmetrically clockwise relative to the parent to separate outgoing
and incoming links by a central gap to avoid mutual link overlap.
Self-links are show as short stubs projecting from a node.

The following options feature in both the
initial and drag graphs.  They apply to the whole initial-graph, or in
the drag-graph just to the active (single) node or the active
fragment.

\enlargethispage{4ex}
\begin{list}{$\Box$}{\parsep 0ex \itemsep .0ex  
 \leftmargin 6ex  \rightmargin 0ex \labelwidth 5ex \labelsep 1ex}
\item[\underline{\it options} $\dots$]   \underline{\it what they mean}

\item[{\bf rot-x/X} $\dots$] lower-case ``{\bf x}'' rotates clockwise, upper-case ``{\bf X}'' rotates anti-clockwise.
\item[{\bf flip-h/v} $\dots$] ``{\bf h}'' flips horizontally, ``{\bf v}'' flips vertically, 
\item[{\bf nodes-(/)} $\dots$] simple brackets: ``{\bf (}'' to contract, ``{\bf )}''
  to expand the node display as discs, numbers, or patterns.
\item[{\bf nodes-= $\dots$}] (equals sign) to cycle (toggle) though successive
  node displays (figure~\ref{node-dis}).

\begin{figure}[H]
   \begin{center}
     \fbox{\includegraphics[bb=266 213 1116 575, clip=, width=.8\linewidth]{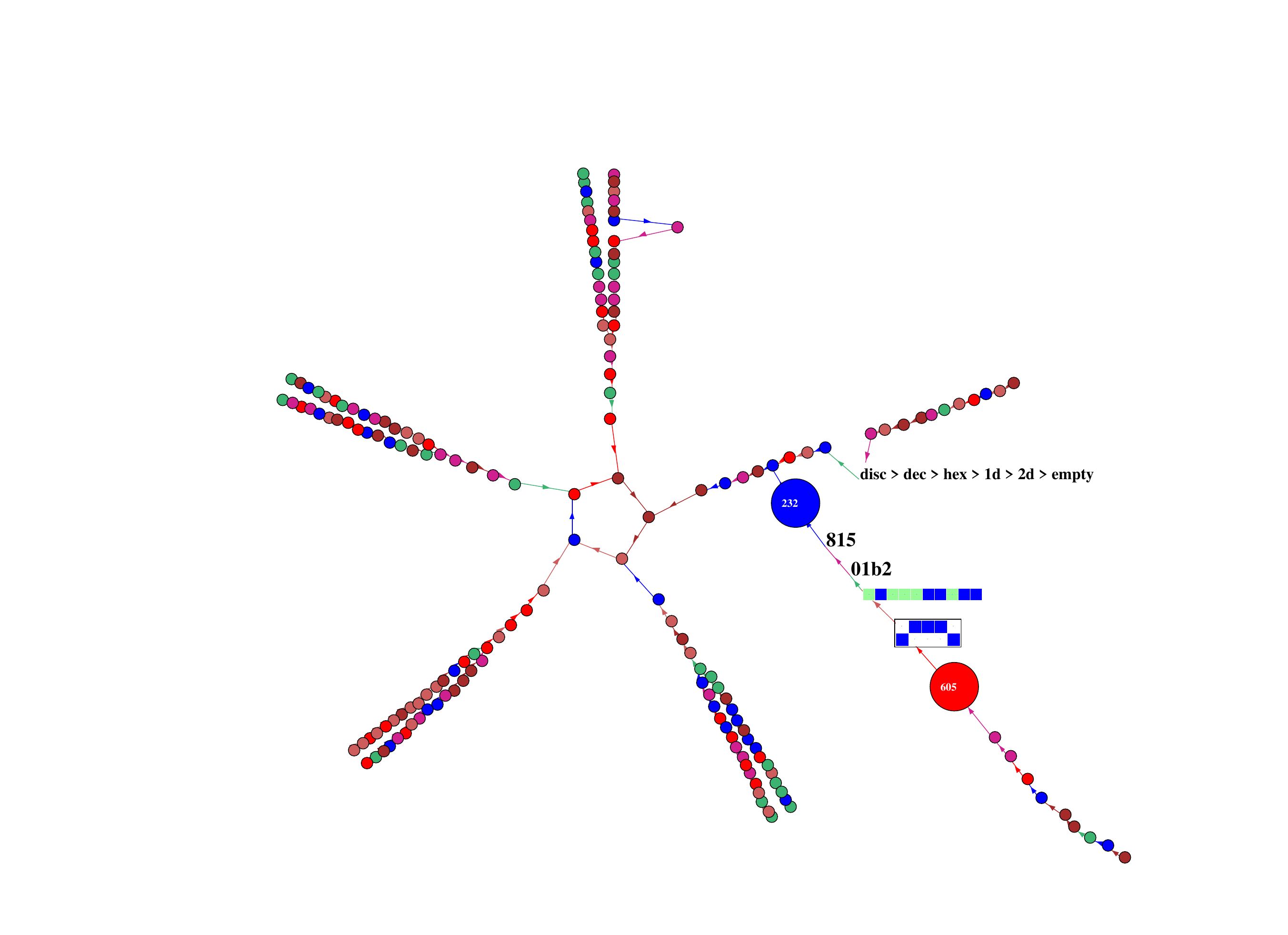}}
   \end{center} 
    \vspace{-4ex}   
    \caption [Node displays --- ibaf-graph]
    {\textsf{Toggling between 6 successive node displays in the ibaf-graph ---
    \underline{\it Lower Right}: Examples of 5 displays (``empty'' is skipped).
   \underline{\it Upper Right}: a created ``label''
             noting the 6 alternatives. Part of a basin for 1d CA $v2k3$, $n$=10, rule 30.}}
     \label{node-dis}
\end{figure}
\clearpage

\item[{\bf links-\{/\}} $\dots$] curly brackets: ``{\bf \{}'' to contract,
           ``{\bf \}}'' to expand the length of links, or the distance between nodes.
                 Scaled link thickness follows disc size.
\item[{\bf both-[/]} $\dots$] square brackets:  ``{\bf [}'' to contract,
  ``{\bf ]}'' to expand, both nodes and links together.
\item[{\bf PScript-P} $\dots$] to save the current graph image as a vector PostScript file.
\item[{\bf net-\#} $\dots$] to redraw the graph, restoring any link cuts or additions
                             made in the drag-graph. 
\end{list}

The  operations for  rotations, flips, and contract/expand link length, are made with reference to
the window center for the initial-graph, and the pointer position for the drag-graph.

\subsection{initial-graph --- exclusive options}
\label{initial-graph --- exclusive options}

\noindent In addition to the shared options in section~\ref{initial and drag graphs --- shared options},
options exclusive to the initial-graph which apply to all nodes/links simultaneously are listed below
in roughly the order they appear in the initial-reminder,

\begin{list}{$\Box$}{\parsep 0ex \itemsep .0ex  
    \leftmargin 6ex  \rightmargin 0ex \labelwidth 5ex \labelsep 1ex}
\item[\underline{\it options} $\dots$]   \underline{\it what they mean}   

\item[{\bf drag-(def)} $\dots$] enter {\bf return}, or click the left or right
                             mouse button, for the ``drag''options,
                             then enter ``{\bf q}'' to revert.
                             Click on a node to activate it in the drag-graph.
\item[{\bf ant-a} $\dots$]   to launch a projectile --- a probabilistic 
                             ``ant'' --- in the graph to trace a (Markov chain) path 
                             according to link probabilities, keeping track 
                             of the frequency of visiting vertices (figure~\ref{jump_RBN_ant_hits}).
                             Top-right suboptions appear\cite[\hspace{-1ex}\footnotesize{\#20.13.2}]{EDD}.
\begin{figure}[H]
\textsf{\footnotesize
     \begin{minipage}[t]{.48\linewidth}
       \begin{center}
         \includegraphics[width=.8\linewidth]{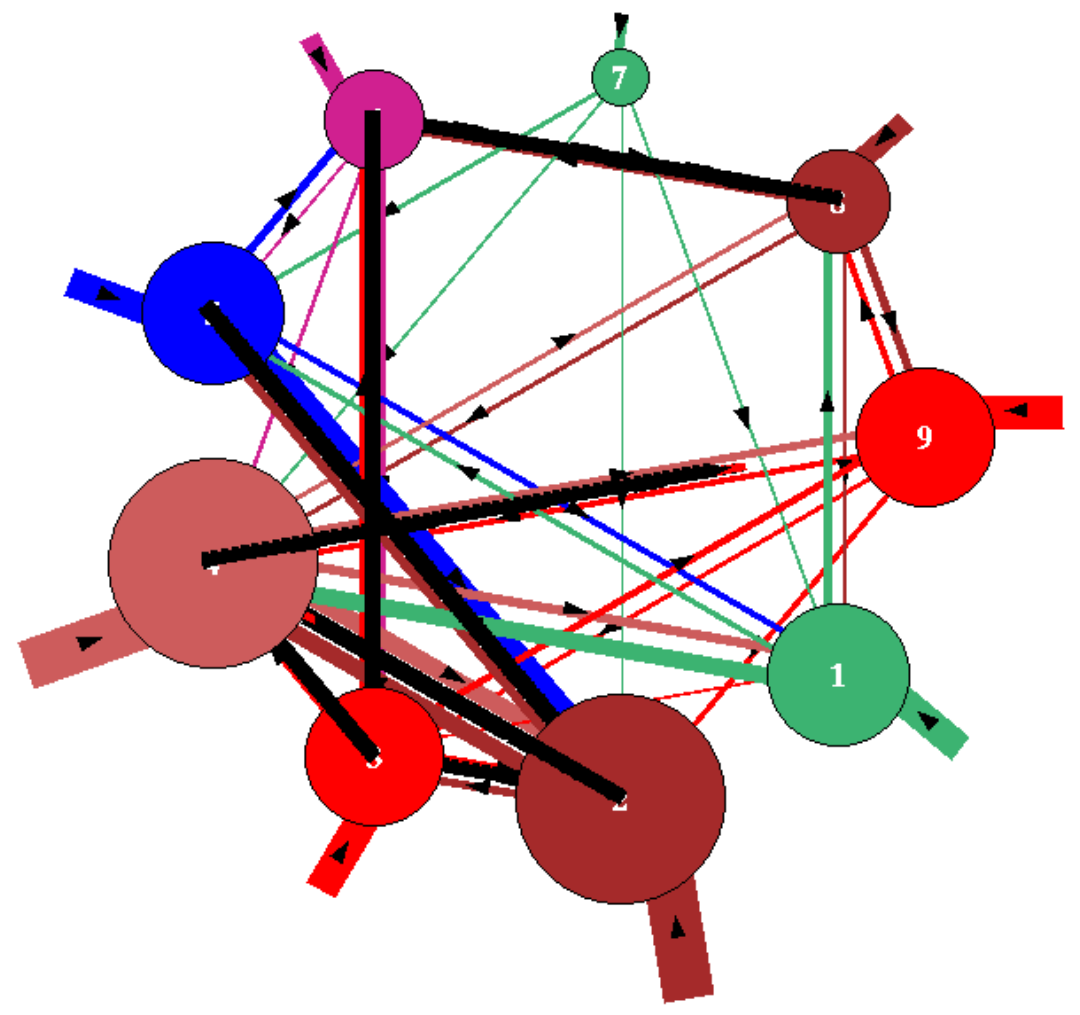}
       \end{center}
       \vspace{-3ex}
             (a) The ant has a red body and leaves a black trail. Soon after launch the ant
             was paused while moving (in slow motion) from node 4 to 9. The default animation
             speed is very fast
             but can be controlled with ``{$\boldsymbol <$}{\bf /}{$\boldsymbol >$}''. 
        \end{minipage}
             \hfill
        \begin{minipage}[t]{.48\linewidth}
          \begin{center}
           \includegraphics[bb=0 14 507 484, clip=, width=.8\linewidth]{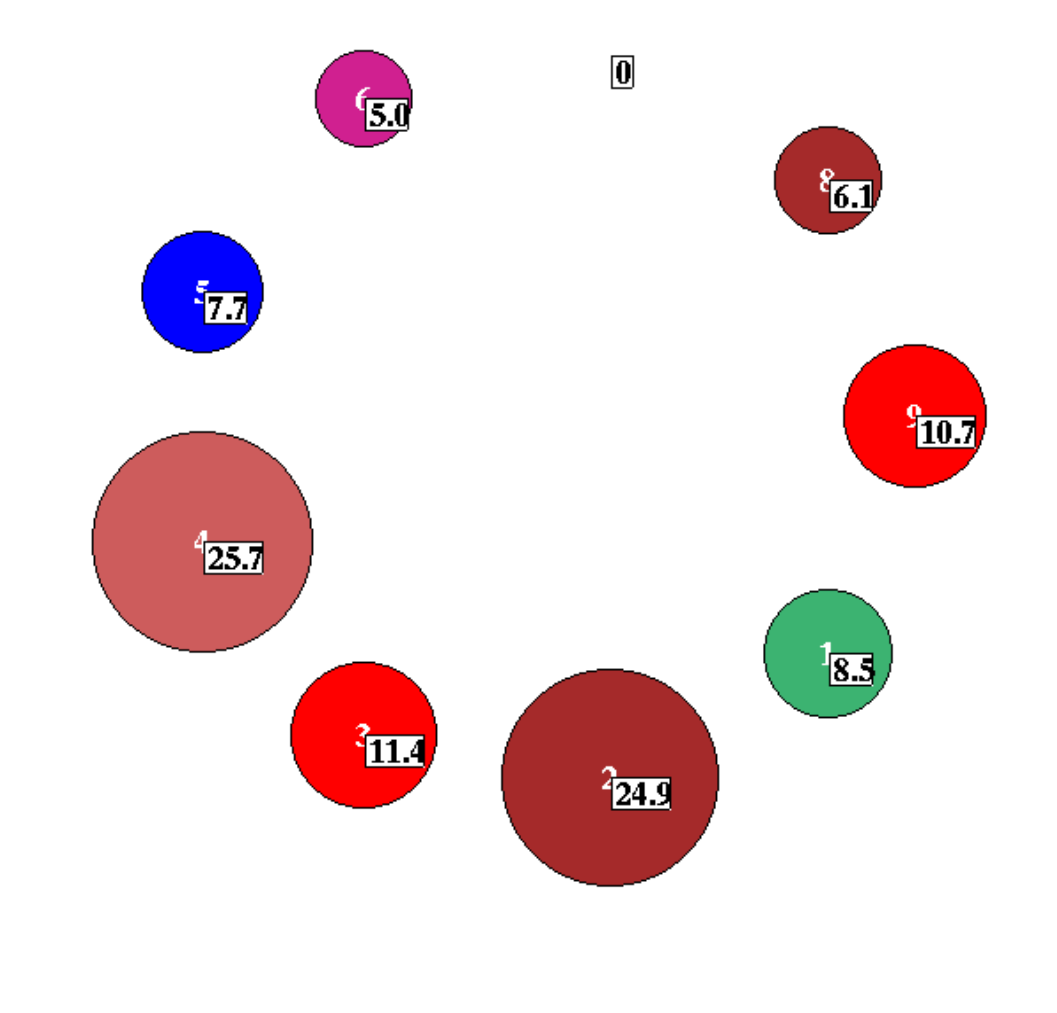}
          \end{center}
          \vspace{-3ex}
             (b) The percentage of ant hits at each node after 5 million moves ---
             with ant animation disabled this took a few seconds.             
             Node~7 indicted by%
               {\bf\setlength{\fboxsep}{.25ex}
                  \fbox{%
                   \parbox{1.2ex}{0}%
               }}
               is unreachable from other nodes so has zero hits.
        \end{minipage}
        } \vspace{-1ex}     
       \caption [Probabilistic ant-hits]
       {\textsf{(a) A probabilistic ant moving along directed links between
               nodes. (b) Ant hits, or node visits (enter ``{\bf h}''
               followed ``{\bf n}''). Ant hits can be compared with basin (node) volume in (a)
               and the jump-table.
               This example is for the jump-graph of the RBN $v2k3$, $n$=10, defined in
               figure~\ref{rbn10k3_network_graphs}.               
               }} \label{jump_RBN_ant_hits}
\end{figure}

\item[{\bf unscram-u} $\dots$]  to automatically unscramble the graph,
                             placing weakly connected nodes on the outer edges,
                             nearby their connections.
                             This works best in the circle or spiral layout
                             (figure~\ref{fig:netgraph_RBN_power150_s}).
                             For the ibaf-graph the option is not so useful --- it will scramble
                             the basin layout (figure~\ref{ibaf-r110rank})
                             ~--- but can nicely demonstrate dragging components.

\begin{figure}[H]
 \footnotesize
   \begin{center}
    \begin{minipage}[t]{1\linewidth}
      \begin{minipage}[t]{.328\linewidth}
        \includegraphics[bb=236 104 689 550, clip=,width=1\linewidth]{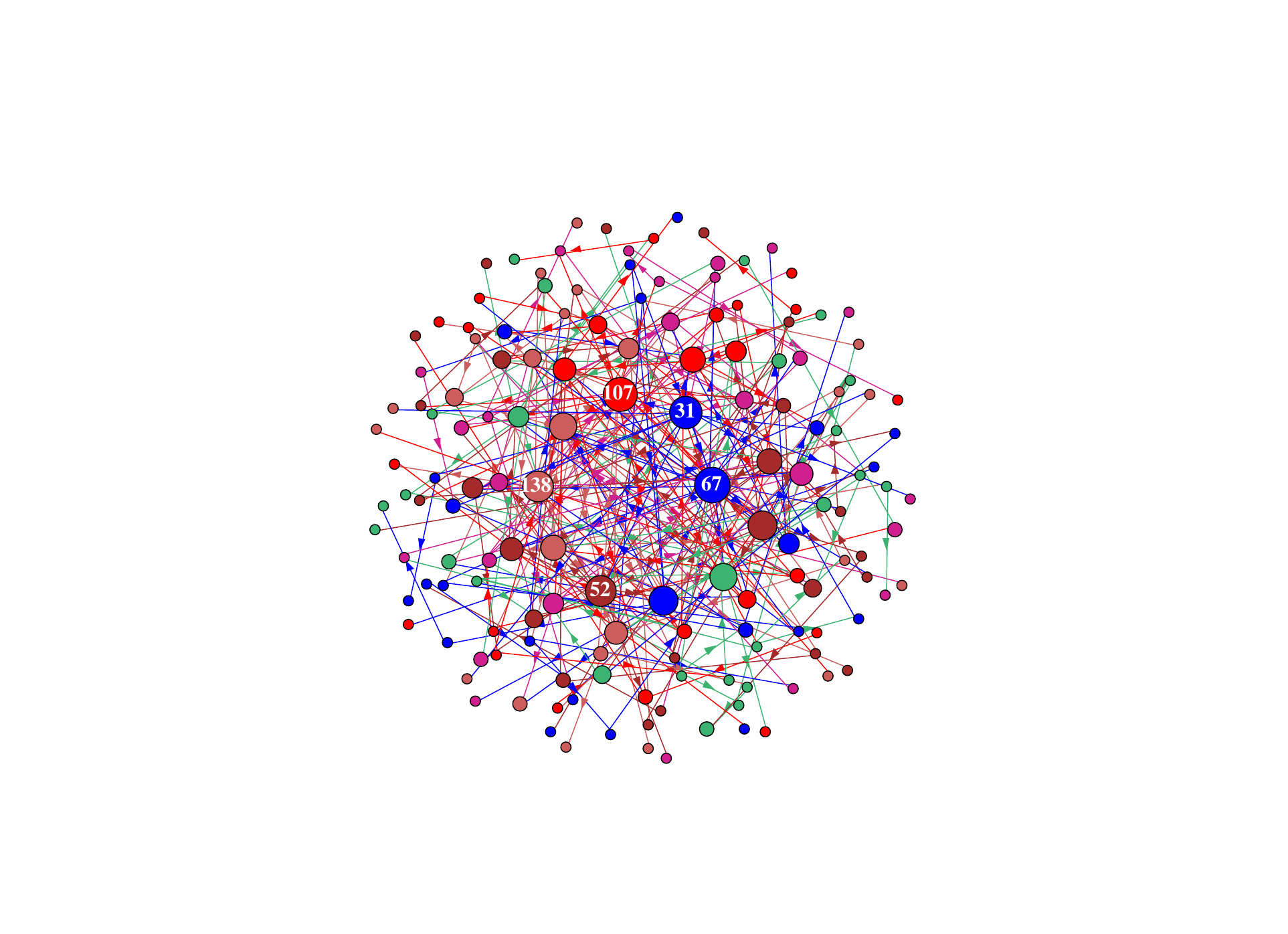}
     \begin{center} \vspace{-2ex} \textsf{(a) show as a spiral, and ranked} \end{center}
    \end{minipage} 
   \hfill
     \begin{minipage}[t]{.328\linewidth}
        \includegraphics[bb=236 104 689 550, clip=,width=1\linewidth]{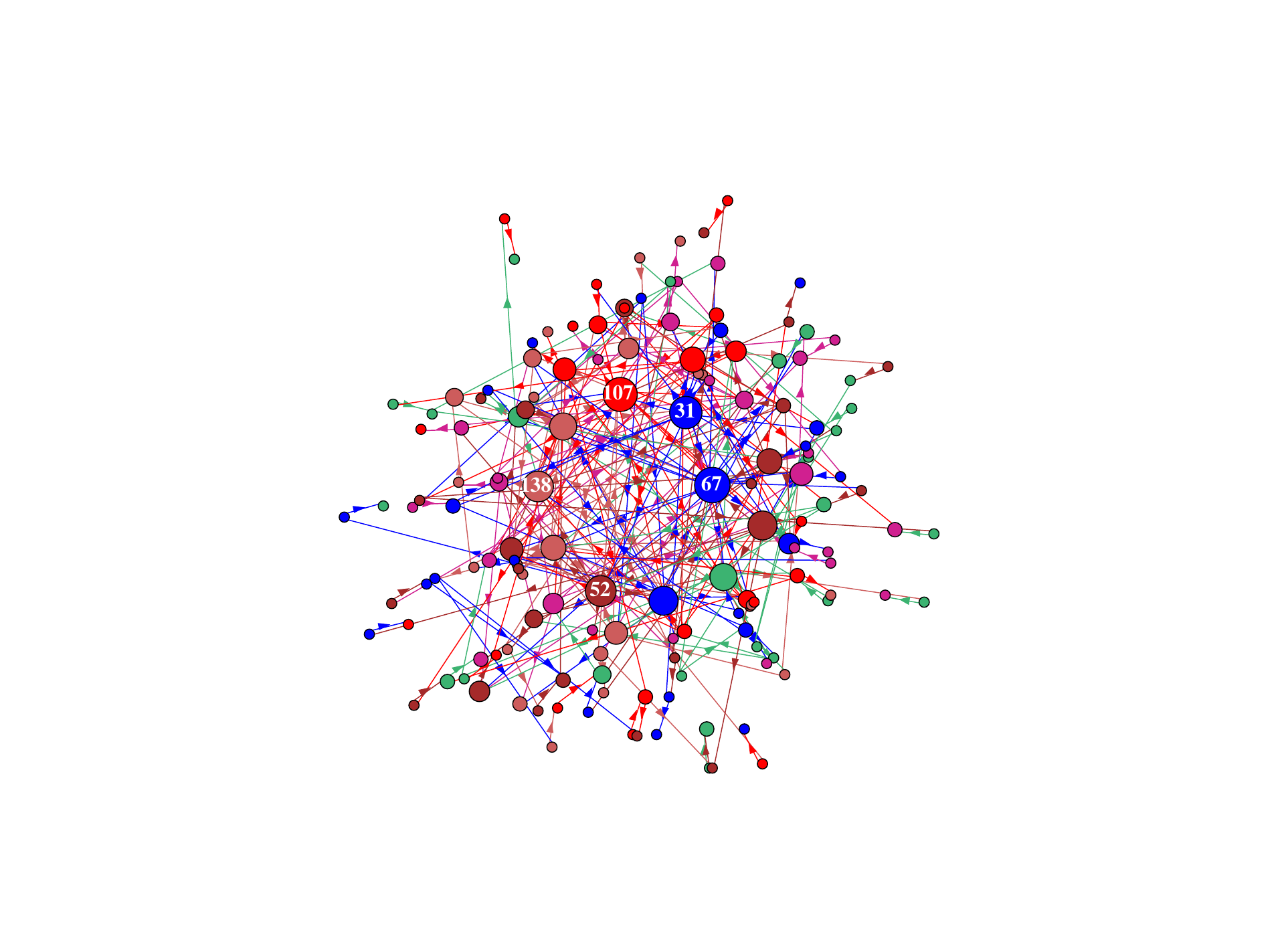}
    \begin{center}  \vspace{-2ex} \textsf{(b) automatically unscramble}  \end{center}
    \end{minipage} 
    \hfill
       \begin{minipage}[t]{.328\linewidth}
         \includegraphics[bb=236 104 689 550, clip=,width=1\linewidth]{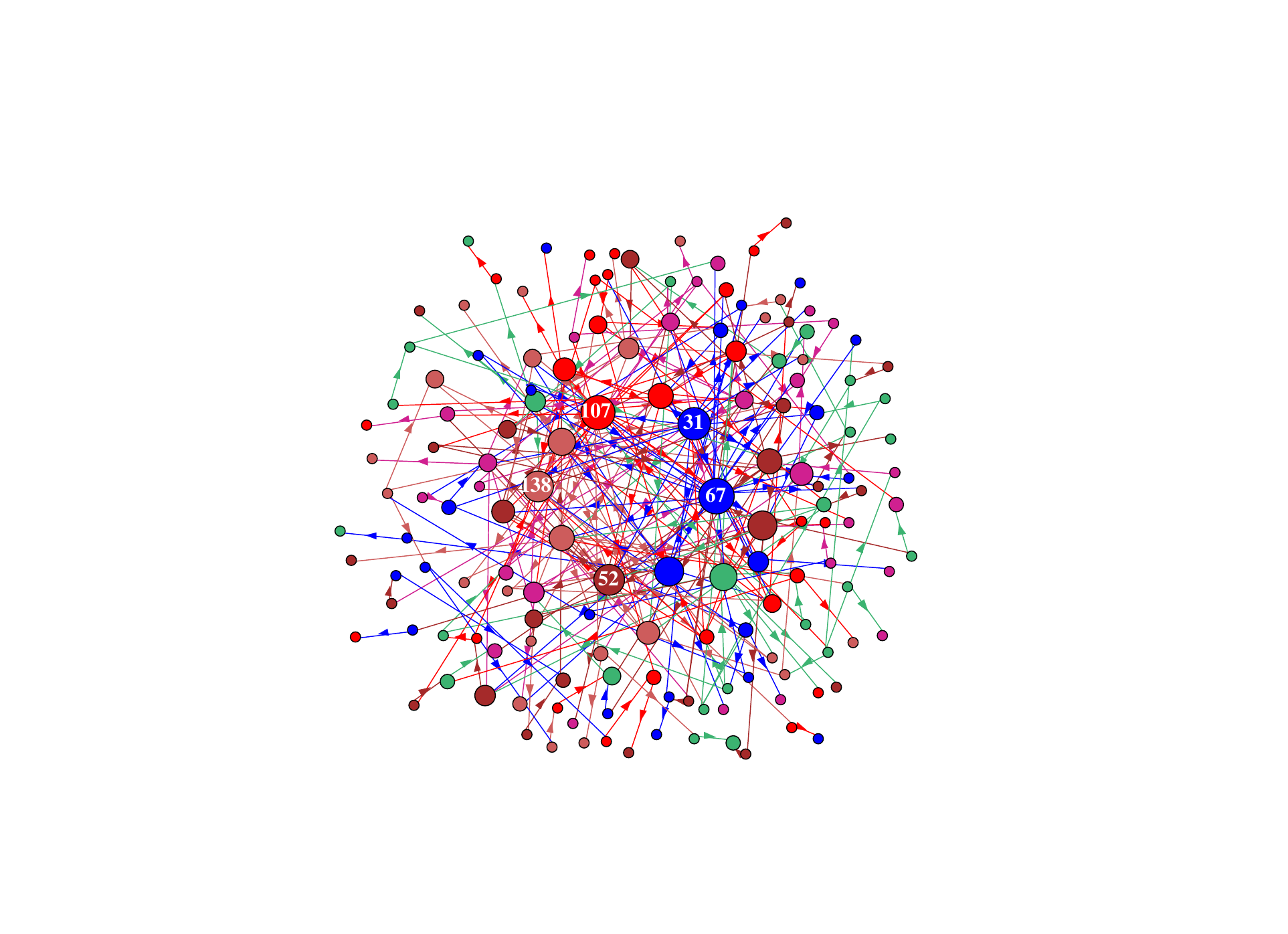}
     \begin{center} \vspace{-2ex} \textsf{(c) shake, then drag nodes} \end{center}
    \end{minipage} 
   \end{minipage}
   \end{center} 
    \vspace{-3ex}   
    \caption [Unscrambling the network-graph of a scale-free RBN]
    {\textsf{One possible method of unscrambling the network-graph of a scale-free RBN, $n$=150.}}
     \label{fig:netgraph_RBN_power150_s}
\end{figure}                              

\item[{\bf win-w} $\dots$]  to show what lies below the graph window ---
                            press any key to restore. This is useful
                            for both ibaf and jump graphs to see the original basin graphic
                            and its top-right data window, and for the network-graph to see the
                            underlying wiring graphic (figure~\ref{n15rbn1}).

\item[{\bf rank0-k $\dots$}] to toggle between default and ranked node positions.
                             {\bf rank0}/{\bf rank1} shows which is current.
                             The nodes are ranked (reordered) according to node scaled size (weight),
                             which differs for graph type as follows,
                             
                             {\bf network-graph:} node size based on inputs or outputs, whichever is active.
                             
                             {\bf ibaf-graph:} node size based on inputs (in-degree or number of pre-images).
                                
                             {\bf jump-graph:} node size based on basin volume.               
                          
\vspace{-2ex}
\begin{figure}[H]
   \begin{center}
        \begin{minipage}[t]{.45\linewidth}
          \includegraphics[bb=46 140 1245 767, clip=, height=1\linewidth]{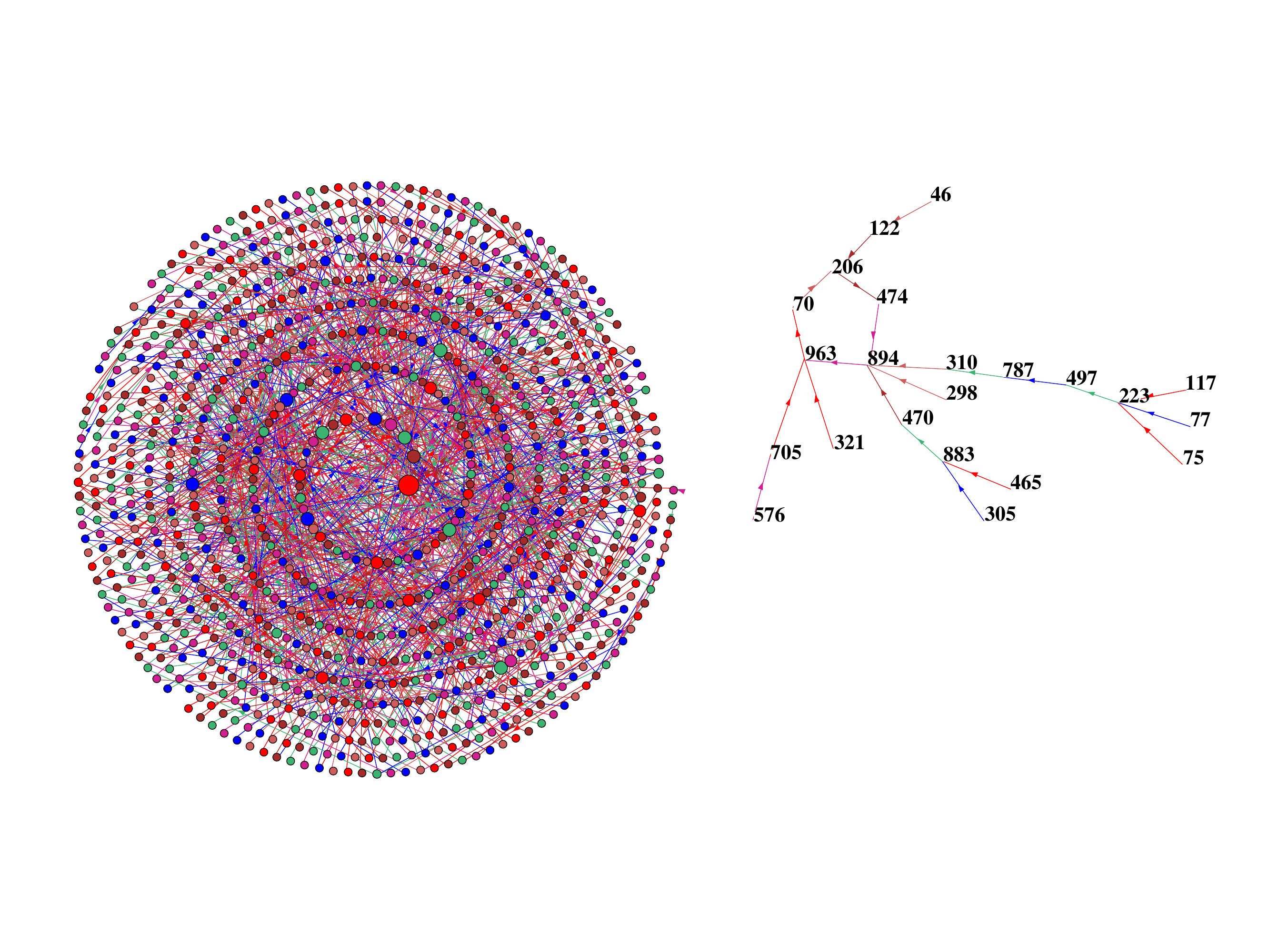}
        \end{minipage}
       \hfill
       \parbox[t]{.52\linewidth}{ \vspace{-20ex}   
    \caption [ibaf-graph: ranking a spiral layout]
    {\textsf{\underline{\it Left}: Ranked spiral layout of the ibaf-graph in figure~\ref{r110-f-ibaf}(b)
    with $2^{10}$=1024 nodes, ranked from the center in descending
    size based on in-degree --- garden-of-Eden states with in-degree zero form the outer layers.
    \underline{\it Above}: A component (basin of attraction) dragged out, rearranged, with numbered nodes. 
    \label{ibaf-r110rank}}}}
 \end{center}
\end{figure} 
\vspace{-4ex}                                  

\item[{\bf settings-S} $\dots$] to change the default settings for the rotation 
angle, the expand/contract factors for
both nodes and links, and the arrow size on links --- sub-options apply.                             
                            
\item[{\bf Unreach-U $\dots$}]  to identify and
  separate unreachable or hard to reach nodes in the graph by
  disconnecting and/or isolating those node from the rest of the
  graph.  A~node is unreachable if it has no input links from other
  nodes in spite of possible output links.  Sub-options apply to set
  the unreachability threshold, so that a node is considered
  ``unreachable'' if it has $x$ or fewer input links. To restore links
  reenter ``{\it Unreach-}{\bf U}'', then {\bf return}.
  
\vspace{-2ex}
\begin{figure}[H]
 \begin{center}
   \includegraphics[bb=42 77 674 464, clip=, width=.7\linewidth]{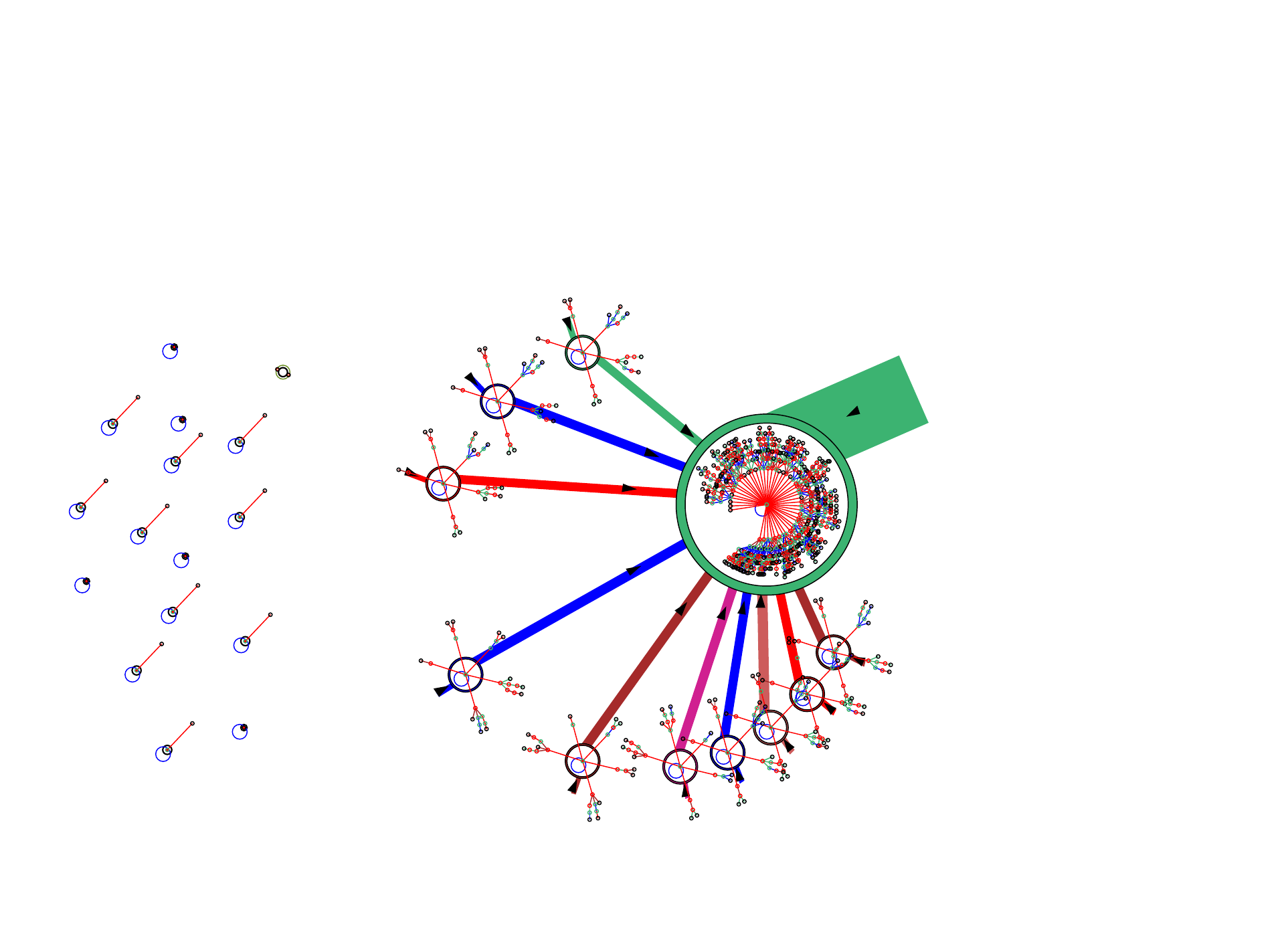}
             \hfill
             \begin{minipage}[b]{.1\linewidth}
{\tiny \baselineskip2ex               
\begin{verbatim}
1:90
2:14
3:14
4:14
5:14
6:14
7:0
8:14
9:0
10:0
11:14
12:0
13:0
14:0
15:14
16:0
17:14
18:0
19:14
10:0
11:14
12:0
13:0
14:0
15:14
16:0
17:14
18:0
19:14
20:0
21:0
22:0
23:0
24:0
25:0
26:0
27:0
\end{verbatim}}
\vspace{3ex}
\end{minipage}
\end{center}
    \vspace{-4ex}                  
       \caption [Disconnecting and isolating unreachable nodes]               
                {\textsf{Disconnecting and isolating unreachable nodes,
                    repositioned randomly on the far left, with the default
                unreachability threshold of zero. This example is a
                jump-graph of the 1d CA, $v2k3$ rcode 104, $n$=10.  The basins
                of attraction were redrawn at the nodes.
                \underline{\it Right}: The list of the number of jumps (inputs) to each basin (node) ---
                shown in the terminal by entering ``{\it xterm-}{\bf x}''.}}
       \label{unreach_BinG_r104_n10}
\end{figure}

\item[{\bf basins-i/I/s $\dots$}] ({\it f-jump-graph only}) enter ``{\it basins-}{\bf i}''
                             to redraw basins at jump-graph nodes but without the graph itself as in
                             figures~\ref{basin_r133_comp} and \ref{r110-f-ibaf}(b), or 
                             enter \mbox{``{\it basins-}{\bf I}''}  to keep the graph and draw basins
                             inside (or on top of) nodes as in
                             figures ~\ref{unreach_BinG_r104_n10} and \ref{r110-f-ibaf}(a).
                             Enter ``{\it basins-}{\bf s}'' to change the basin scale ---
                             suboptions appear\cite[\hspace{-1ex}\footnotesize{\#20.14}]{EDD}.
                             This is an alternative method for a flexible basin layout
                             with geometric possibilities.
                             Once drawn, basin state-space coordinates can be saved, to reload
                             into the ibaf-graph\cite[\hspace{-1ex}\footnotesize{\#20.14.1}]{EDD}.

\begin{figure}[H]
       \begin{center}
       \begin{minipage}[t]{1\linewidth}
         \includegraphics[bb=214 111 717 429, clip=, width=1\linewidth]{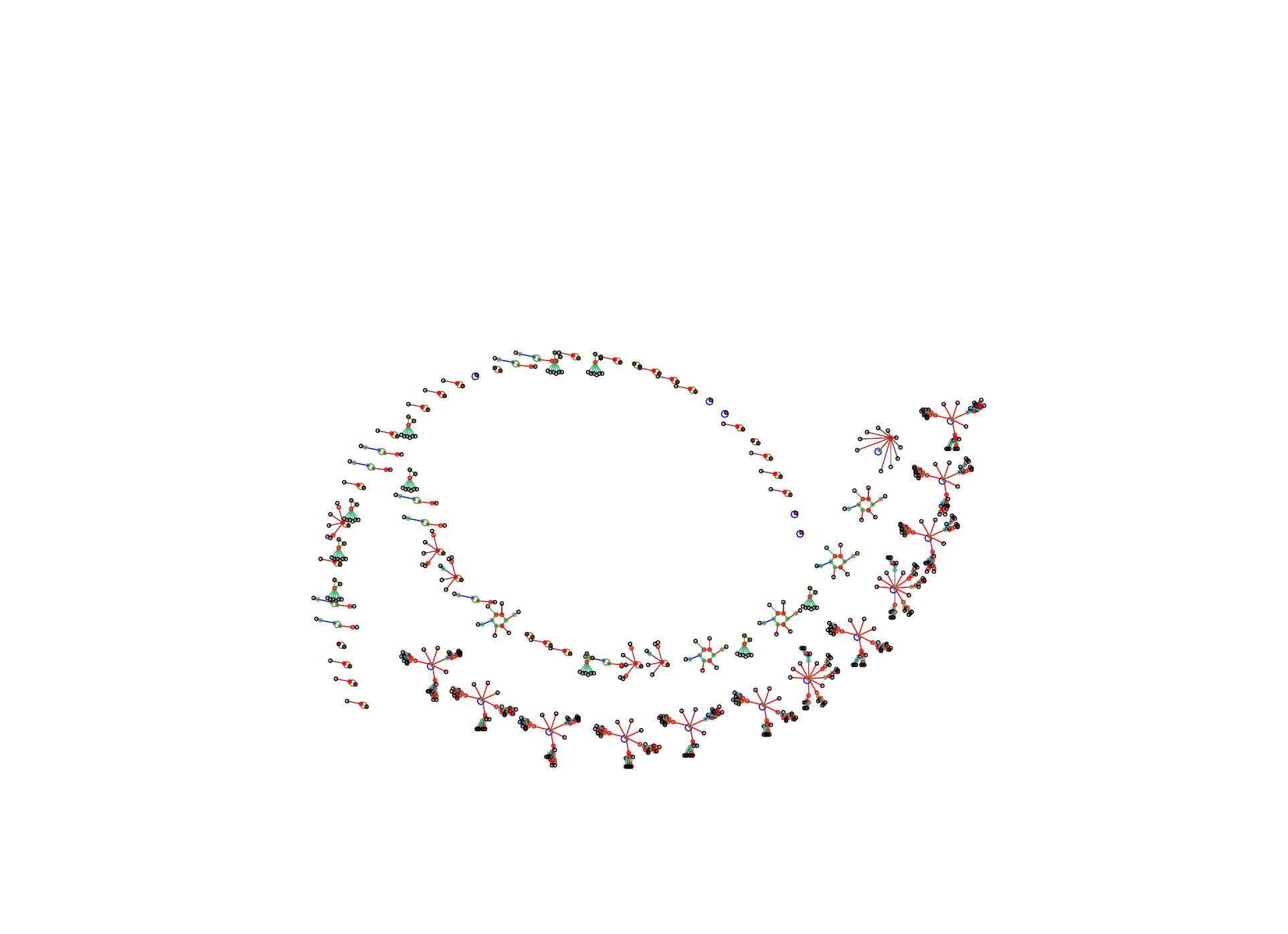}\\[-5ex]
        \begin{center}{\textsf{\small (a) all 73 basins --- nodes and fragments were dragged starting 
                         from a circle layout.}}\end{center}
       \end{minipage}
       \end{center}
      \vspace{1ex}
       \begin{minipage}[t]{.52\linewidth}
         \includegraphics[bb=285 110 610 291, clip=, width=1\linewidth]{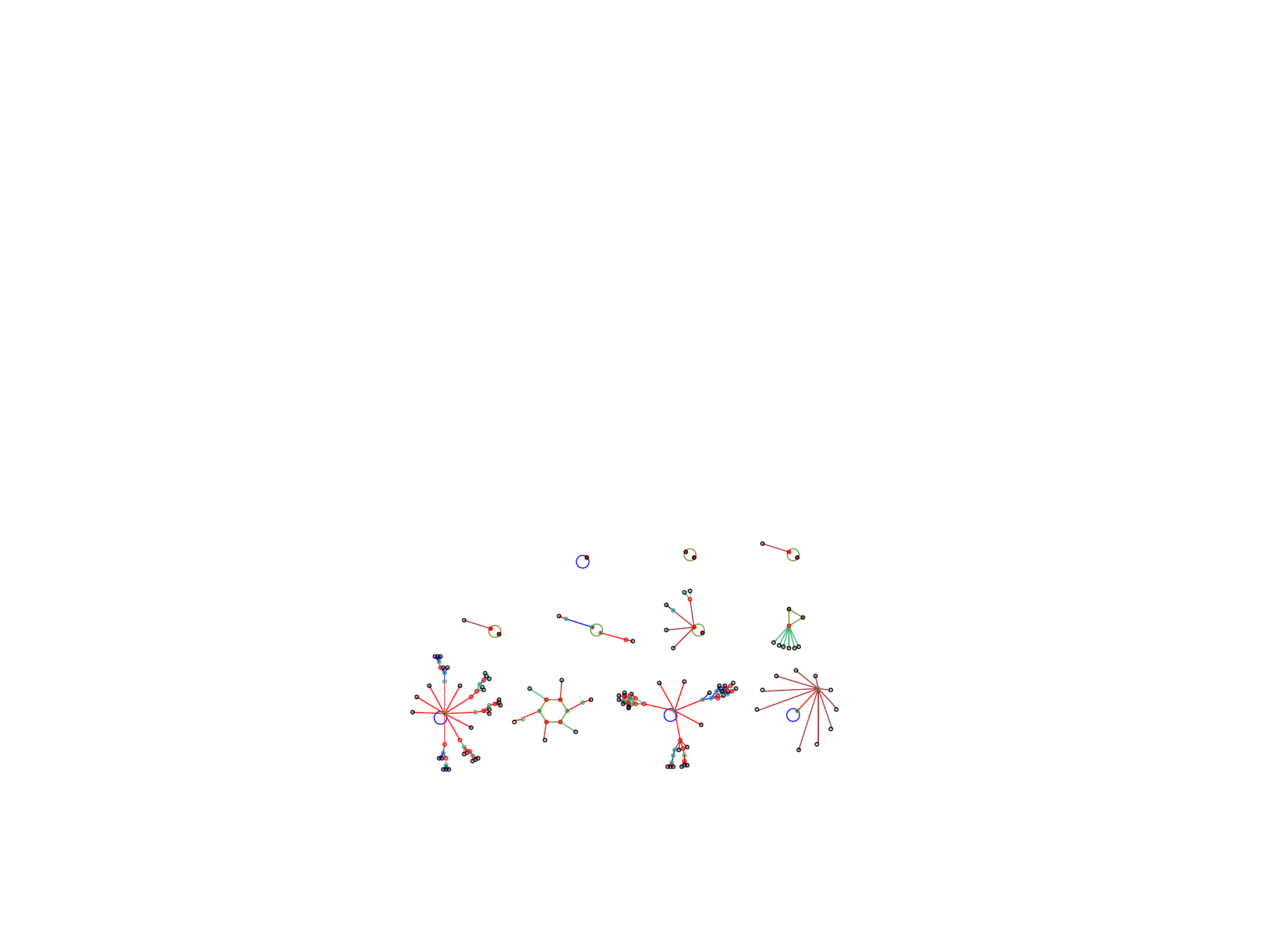}\\[-5ex]
        \begin{center}{\textsf{\small (b) 11 prototype basins from a square layout}}\end{center}             
       \end{minipage}
        \hfill
        \parbox[t]{.44\linewidth}{  \vspace{-27ex} 
       \caption [Inserting basins in the jump-graph]       
       {\textsf{Inserting basins in the \mbox{f-jump-graph} for a 1d CA 
           $v2k3$, $n$=10, rcode 133, (a)  uncompressed, (b) compressed to include just the prototype
           (non-equivalent) basins ---
        allowed if edges were omitted,  
        otherwise all basins are shown. This provides an alternative layout method to
        the ibaf-graph, which does not allow compression.}}
        \label{basin_r133_comp}}  
\end{figure}
 
\item[{\bf matrix-t/T $\dots$}] to toggle between the graph and its corresponding
  ``jump-table'' or ``adjacency-matrix'',
  suboptions appear\cite[\hspace{-1ex}\footnotesize{\#20.19}]{EDD}.
  For the network-graph and jump-graph,
  ``{\it matrix-}{\bf t}'' shows the numbers of links, ```{\it
    matrix-}{\bf T}'' the fractions of total links from each
  node.  For the ibaf-graph, ``{\it matrix-}{\bf t}'' shows the matrix
  as an unnumbered pattern, and ``{\it matrix-}{\bf T}'' shows the
  list of exhaustive pairs in the terminal.

\begin{figure}[H]
  \begin{minipage}[t]{.4\linewidth}
    \includegraphics[width=1\linewidth]{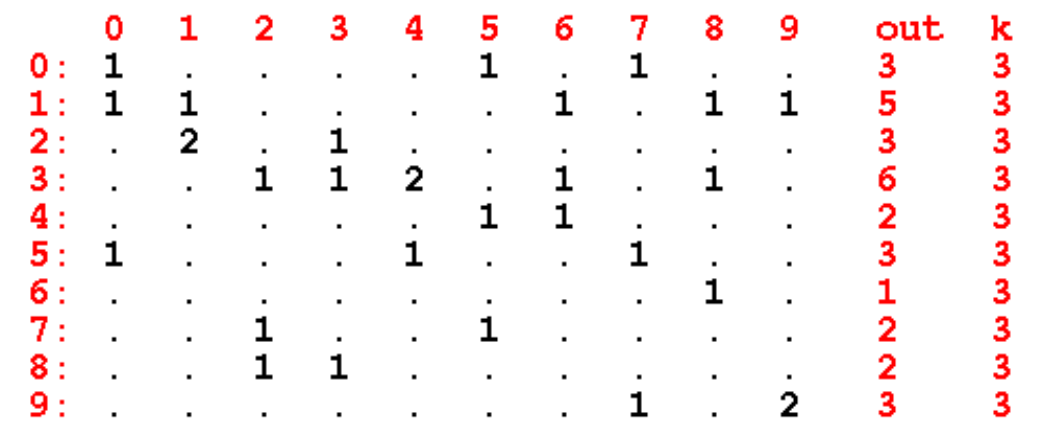}\\[-4ex]
   \begin{center}\textsf{\small (a) numbers of outputs}\end{center}
   \end{minipage}
  \hfill
  \begin{minipage}[t]{.53\linewidth}
    \includegraphics[width=1\linewidth]{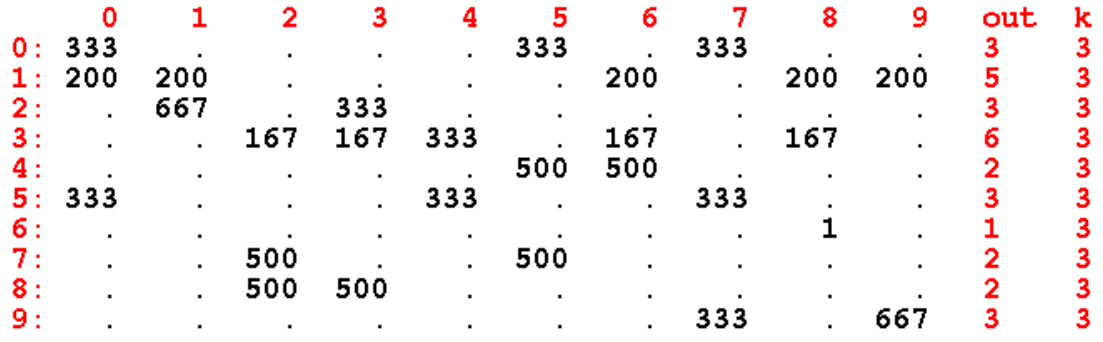}\\[-4ex]
   \begin{center}\textsf{\small (b) fraction of outputs}\end{center}
   \end{minipage}
       \caption [The adjacency-matrix of the network-graph]       
       {\textsf{
         The adjacency-matrix for a network-graph, this example for 
         the RBN $v2k3$, $n$=10 in figure~\ref{rbn10k3_network_graphs}.
         The rows \mbox{(0 to $n$-1)} show outputs to other nodes,
         columns \mbox{(0 to $n$-1)} show inputs from other nodes.
         Zero values are shown as dots. The adjacency-matrix (jump-table)
         for the jump-graph is similar.         
         (a) for numbers enter {\bf t}.
         (b) for fractions of each node's total outputs enter {\bf T}.        
         The last two red columns show the total outputs, and
         inputs ($k$) for each node.
         The jump-table provides more information in 4 extra columns\cite[\hspace{-1ex}\footnotesize{\#20.19.2}]{EDD}.
         \label{adjacency-matrix_RBN}
         }}
         \label{jump_matrix_RBN} 
\end{figure}

\begin{figure}[H]
  \begin{center}
    \begin{minipage}[t]{.45\linewidth}
      \includegraphics[width=1\linewidth]{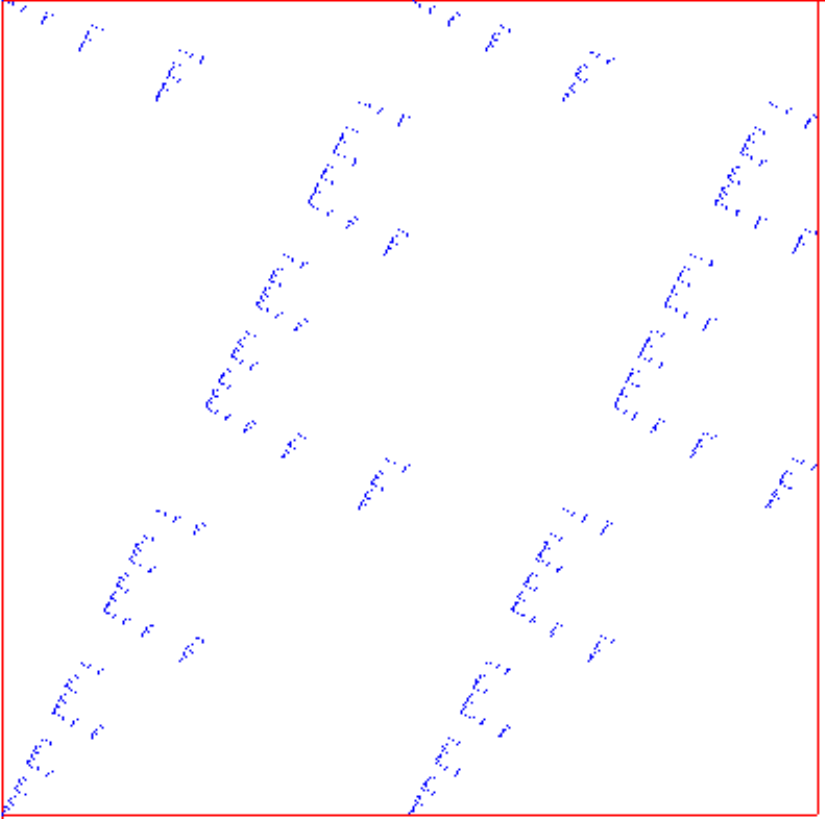}\\
   \textsf{\small (a) initial ibaf-matrix in a fixed frame.}
    \end{minipage}
   \begin{minipage}[b]{.38\linewidth} 
   \begin{minipage}[t]{1\linewidth}
     \includegraphics[width=1\linewidth]{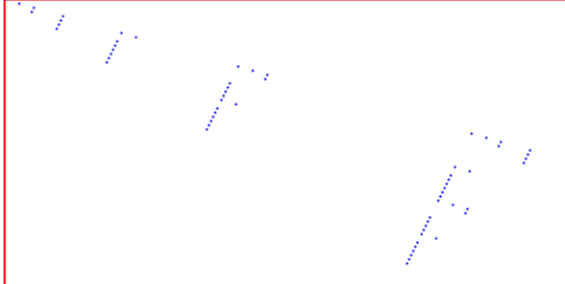}\\
   \textsf{\small (b) expanded +1 to one pixel}\\ 
   \end{minipage}\\
   \begin{minipage}[t]{1\linewidth}
      \includegraphics[width=1\linewidth]{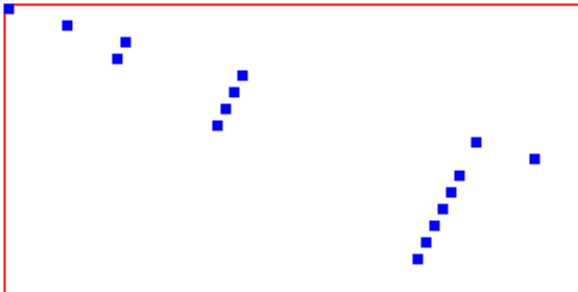}\\
   \textsf{\small (c) expanded +4 to 4 pixel block}   
   \end{minipage}
   \end{minipage}
   \vspace{-1ex}
  \end{center}
  \begin{center}
    \setlength{\fboxsep}{0pt}\fbox{
      \includegraphics[bb=188 242 1070 715,clip=,width=1\linewidth]{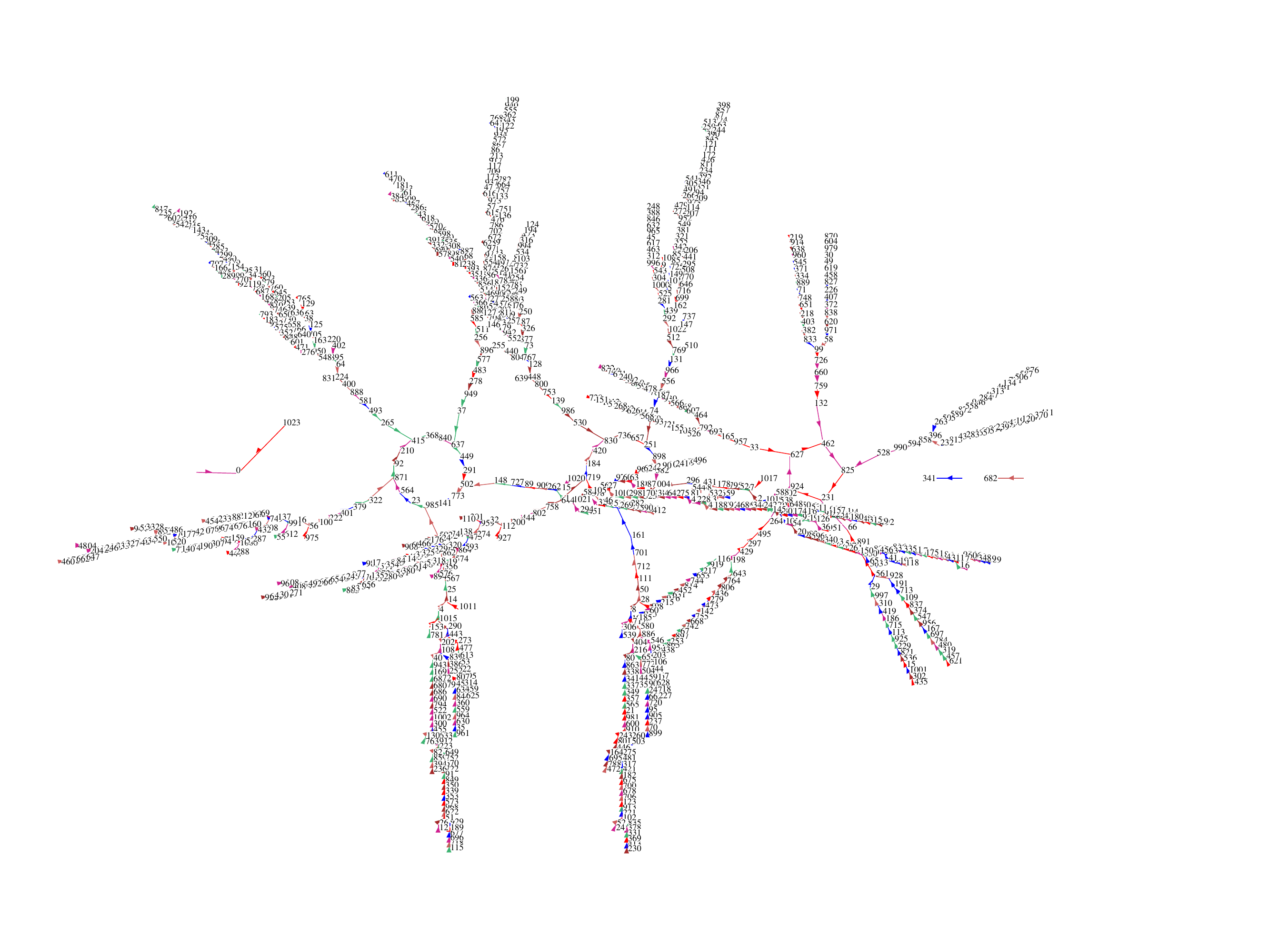}
    }
    \textsf{\small (d) The ibaf-graph showing numbers at nodes}
   \end{center}
   \vspace{-4ex}
       \caption[The ibaf-matrix]
               {\textsf{(a)-(c) The adjacency-matrix (ibaf-matrix) for the ibaf-graph (d)
                   for the $v2k3$ CA rcode 30 $n$=10 with
                   $S$=$2^n$=1024 states in 6 basins.
                   The ibaf-matrix is a pixel pattern where
                   rows \mbox{(0 to $v^n$-1)} show the single output of each node to which other
                   node, columns \mbox{(0 to $v^n$-1)} show inputs from other nodes.
                   (a) The initial $S$$\times$$S$ matrix is confined within a fixed size frame where pixels
                   may overlap, but
                  the matrix can be expanded to eliminate overlap showing as much as will fit. 
            (b) The expanded matrix, still with single pixels --- top left area.
           (c) Expanded to 4-pixel blocks --- top left area.
           The fractal pixel-pattern is equivalent to the return map
           by value\cite[\hspace{-1ex}\footnotesize{\#31.2.2.2}]{EDD}.}}
         \label{ibaf_matrix}  
\end{figure}
                            
\item[{\bf nodes-n $\dots$}] to toggle between scaled and unscaled nodes.
\item[{\bf nodes-N $\dots$}] to toggle node numbers on discs.
                            Numbers only appear on discs big enough to contain them.
\item[{\bf links-l $\dots$}] to toggle between scaled links/edges and 
                          thin lines (figure~\ref{rbn10k3_network_graphs}).
\item[{\bf Labels-+ $\dots$}] to toggle between existing ``Labels'' and the current node display.
                         Labels can be created/deleted in the drag-graph.
\item[{\bf arrows-A/}{$\boldsymbol <$}{\bf /}{$\boldsymbol >$} $\dots$] 
                            enter ``{\it arrows-}{\bf A}'' to toggle showing arrows.
                            Enter ``{\it arrows-}{$\boldsymbol <$}'' to decrease, or
                            ``{\it arrows-}{$\boldsymbol >$} to increase, the arrow size.
\item[{\bf in/out-z $\dots$}]  ({\it network-graph only})
                            to toggle scaling between inputs and outputs.
                            ``{\bf in/out}'' or ``{\bf out/in}'' shows which is active, though this should
                            be evident from the graph --- no change indicates that inputs=outputs.
                            
\item[{\bf file-f $\dots$}] to save node coordinates of the current graph layout,
  or load an existing file of the same system ---
             suboptions appear\cite[\hspace{-1ex}\footnotesize{\#35.3}]{EDD}.
\item[{\bf graph-g $\dots$}] ({\it ibaf-graph only}) to reset the default ibaf-graph layout
  of separate basin components, useful if this was rearranged with other layout options.
\item[{\bf circle/spiral-o/O $\dots$}]  enter ``{\it circle-}{\bf o}'' to show the graph
                           as a circle, or ``{\it spiral-}{\bf O}'' as a spiral. 
                           Circle layout is the default (except for the ibaf-graph and the 2d or 3d network-graph).
                           Spiral layout requires at least 30 nodes to look right.
\item[{\bf 1d/2d(tog)/3d-1/2/3 $\dots$}]  enter ``{\bf 1}'', ``{\bf 2}'' or ``{\bf 3}''
                              to show the graph arranged in 1d, 2d
                              or 3d (examples in figure~\ref{net_drag123}.
                              This is especially relevant
                              for a network-graph where the underlying network
                              is 2d or 3d --- then the graph in 2d or 3d is
                              the default. The initial 2d network-graph is either
                              square or hexagonal according to the underlying network ---
                              thereafter key ``{\it{2d(tog)-}{\bf 2}}'' 
                              toggles between square and hexagonal.
\item[{\bf rnd-r/R $\dots$}] enter ``{\it rnd-}{\bf r}'' to ``shake'' the layout,
                             repositioning nodes randomly nearby their current 
                             position.  Enter  ``{\it rnd-}{\bf R}'' for a completely random layout.
\item[{\bf quit-q} $\dots$]  to quit the initial-graph, returning to a point where the graph
                            was selected. For the ibaf-graph and f-jump-graph, {\bf return}
                            would generate the next mutant\cite[\hspace{-1ex}\footnotesize{\#28}]{EDD}.
\end{list}     


\subsection{drag-graph --- exclusive options}
\label{drag-graph  --- exclusive options}

\noindent Options in the drag-graph reminder apply to the ``active node''
(its  number is always shown in the title)
and (if active) --- its ``active fragment'' ---  which is dragged/dropped
with the pointer and left mouse button depressed.
As well as dragging, the options for display/rescale/rotate/flip
(section~\ref{initial and drag graphs --- shared options})
are confined just to  the active node+fragment, so the term
``drag/dragging'' also implies all these presentation options. There are four types of
interchangeable drag status that appear in the drag-options title,

\begin{list}{$\Box$}{\parsep 0ex \itemsep .6ex  
 \leftmargin 6ex  \rightmargin 0ex \labelwidth 5ex \labelsep 1ex}
\item[\underline{\it drag status in title} $\dots$]   \underline{\it functions and selection}
\item[{\color{BrickRed}{\bf single:}} $\dots$]
             to drag a single node --- enter ``{\it single-}{\bf s}'' at any time.
             Single node status allows a ``block'', and creating multi-line node
             ``labels''. Rotations and flips do not apply to a singe node.
\item[{\color{BrickRed}{\bf Block x-y:}} $\dots$]
            within single node status, enter ``{\it Block-}{\bf B}'' to define
            then drag a geometric block
            in 1d/2d/3d.
            The default is set by the last two active nodes. To leave block status
            and resume ``single node'' status, enter ``{\it exit-Block-}{\bf B}''
            or change status with one of the options ``{\bf s/i/o/e/a}''.
\item[{\color{BrickRed}{\bf inputs}}/{\color{BrickRed}{\bf outputs}}/{\color{BrickRed}{\bf either}}
              $\dots$] one of these is shown --- to drag a ``linked fragment'',
             a node and nodes linked to 
             it by inputs, outputs, or either (meaning irrespective of direction) ---
             enter one of the options \mbox{``{\it in/out/either-}{\bf i/o/e}''} at any time.
             An unlimited range ``{\color{BrickRed}{\bf step=nolimit:}}'', the default,
             appears in the title
             (or enter ``{\it nolimit-}{\bf 0}'') --- or set a shorter range by distance 
             measured in link-steps ``{\bf step-(1-9})'' --- for example if ``{\bf 2}'' is entered 
             ``{\color{BrickRed}{\bf step=2:}}'' appears in the title. 
             Linked fragment status allows link cuts/additions.
\item[{\color{BrickRed}{\bf allnodes:}} $\dots$]
             enter ``{\bf all-a}'' at any time to drag a node and the complete graph.
  
\end{list}

\noindent In addition to the shared options in
section~\ref{initial and drag graphs --- shared options}, a list of drag-options is set
out below in roughly the order they appear in the drag-reminder,
section~\ref{drag-reminder} The titles in red start with
``{\color{BrickRed}{\bf node x:}}'' giving the active node's number
followed by the drag status.
Note that the active node can be outside a ``block'' as well as inside.  As a check, drag
any node to see which nodes move --- these make up the currently
active fragment, or isolate the fragment with the toggle ``{\it just-}{\bf J}''\\

\begin{list}{$\Box$}{\parsep 0ex \itemsep .6ex  
    \leftmargin 6ex  \rightmargin 0ex \labelwidth 5ex \labelsep 1ex}
\item[\underline{\it options} $\dots$]   \underline{\it what they mean}
  
\item[{\bf drag-leftb} $\dots$] click the left mouse button on a node 
                        to activate it and hold down to drag the
                        node or fragment --- then release (drop) in a new
                        position. Initially, to activate a node it may
                        be necessary to click the right button first,
                        then the left, possibly a few times.  Hence
                        the reminder ``{\bf inactive?-rightb button-first}''.          

\item[{\bf elstc/snap-d} $\dots$] for a graph with many nodes (more likely for the
         ibaf-graph) dragging vibrates the
         graph. This is a consequence of the ``elastic'' option (the
         default) which continually redraws nodes/links for graphical
         animation, but animation/vibration can be suppressed by
         toggling to ``snap'' --- enter ``{\bf d}'' to toggle between
         the two --- in the drag-reminder ``{\bf elstc/snap-d}'' or
         ``{\bf snap/elstc-d}'' indicates which drag method is
         active. With ``snap'' just the active node is dragged leaving
         a trail as in figure~ \ref{snap/elastic}.  On release the node (+fragment)
          and links ``{\bf snap}'' into place and the trail
          disappears. For large graphs ``{\bf snap}'' is a more efficient method.
         
\begin{figure}[H]
   \begin{minipage}[t]{.5\linewidth}
     \includegraphics[height=.75\linewidth]{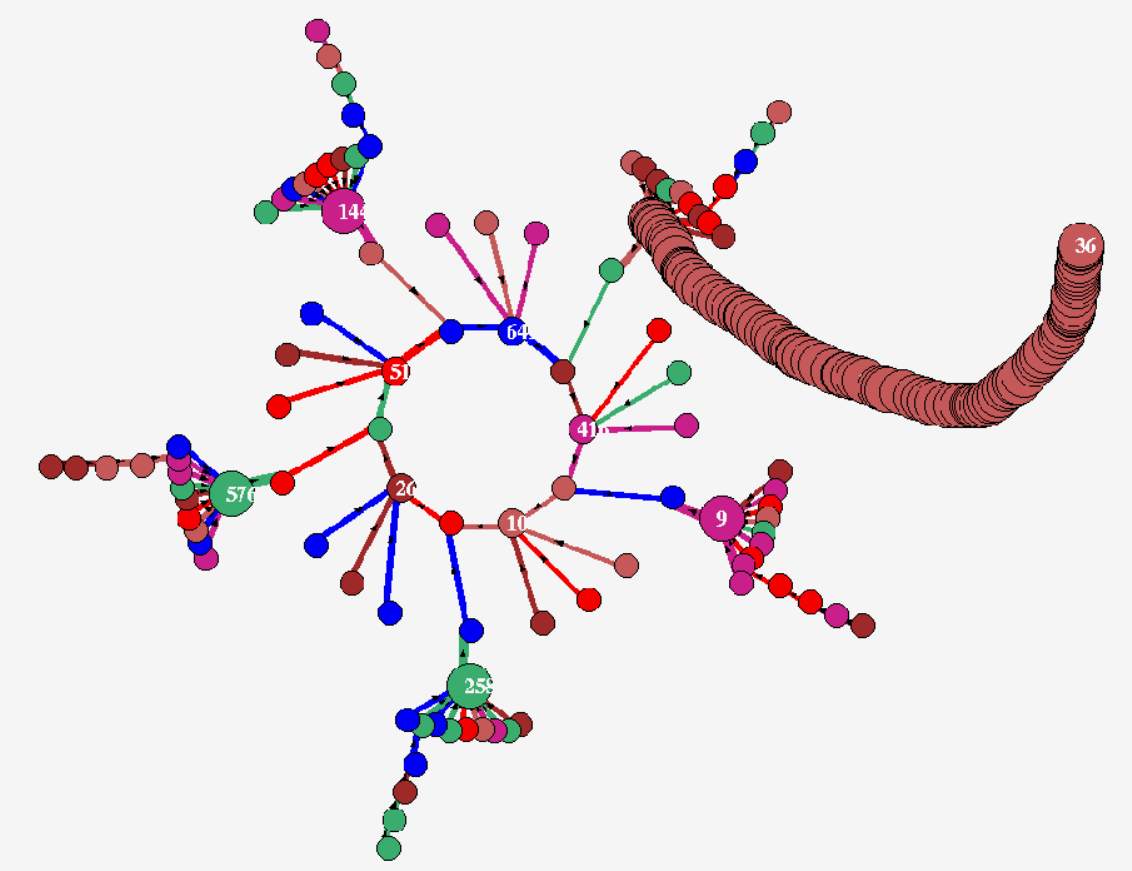}
         \vspace{-2ex} \textsf{\small (a) dragging with ``snap'' leaves a trail}
   \end{minipage}
   \begin{minipage}[t]{.5\linewidth}
     \includegraphics[bb=138 63 1152 810,  clip=,height=.75\linewidth]{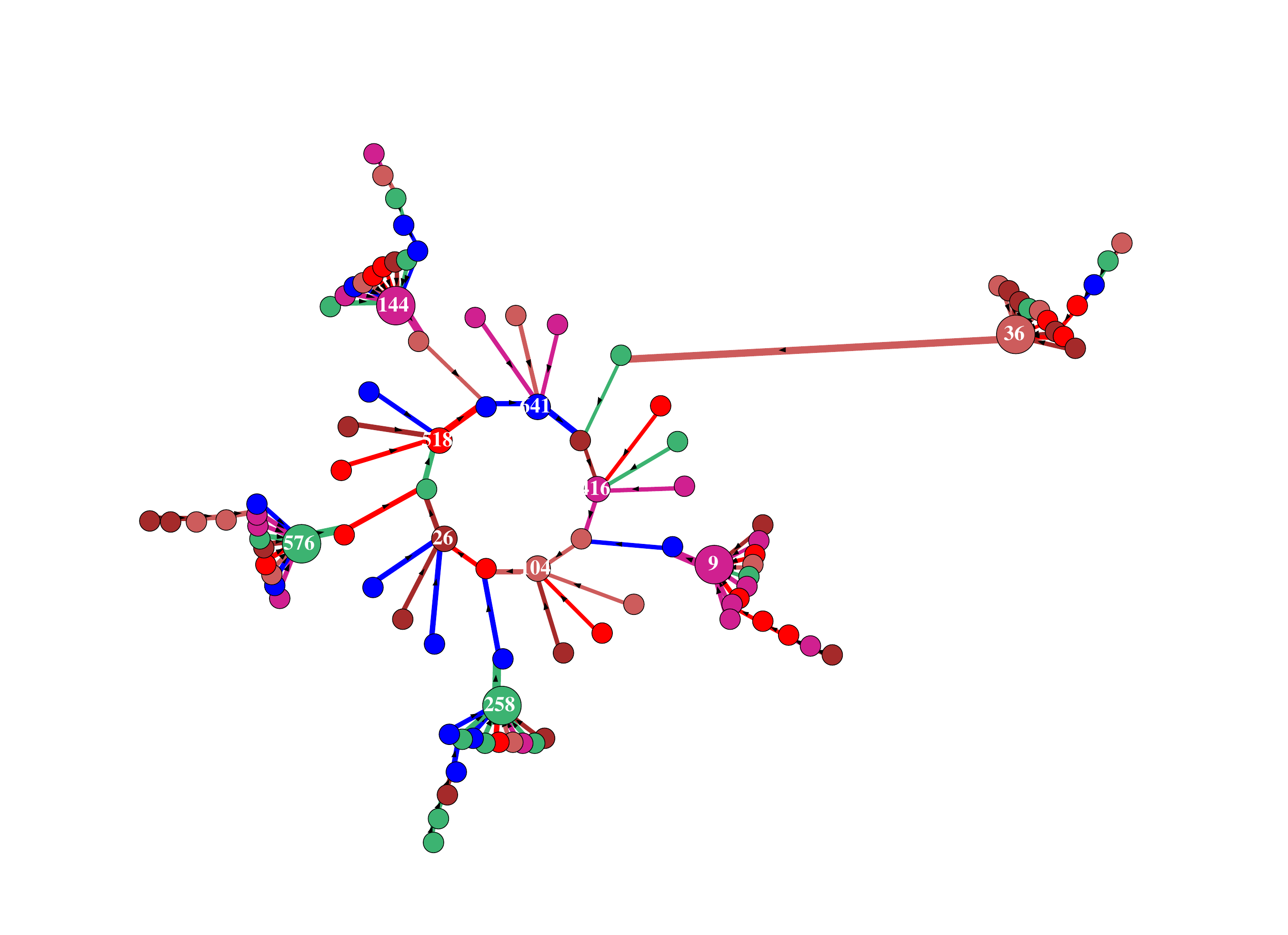}
         \vspace{-2ex} \textsf{\small (b) on release the new layout snaps into place}
   \end{minipage}\\[-0.5ex]
    \caption [dragging with ``snap'']
    {\textsf{Dragging with ``snap'' (instead of ``elastic'' --- toggle with ``{\bf d}'').
    The ibaf-graph of a basin of attraction isolated with
  ``{\it just-}{\bf j}''---
    the 4th basin in figure~\ref{r9-ibaf}.
    From the default layout (a) dragging node 38 to a new position
    with ``snap'' active and the left mouse button depressed.
    On left button release (b) node 38 and its linked fragment by
    inputs snaps into place.}}
     \label{snap/elastic}
\end{figure} 

\item[{\bf gap-g} $\dots$] ({\it applies to the whole graph in any
  status}) to recursively prune (disconnect) ``gap nodes'' which have
  only inputs or only outputs, so cannot transmit information. Applies
  to the whole graph in any drag status. Figure~\ref{prune-gap-nodes}
  shows an example for a single basin in the ibaf-graph isolated with
  ``{\it just-}{\bf j}''. Pruning breaks down components which can be dragged
  separately --- restore all connections with ``{\it net-}{\bf \#}''.
  
\begin{figure}[H]
  \begin{center}  
   \begin{tabular}{ |c|c|c| }
   \hline
     \includegraphics[bb=508 291 822 534,clip=, width=.2\linewidth]{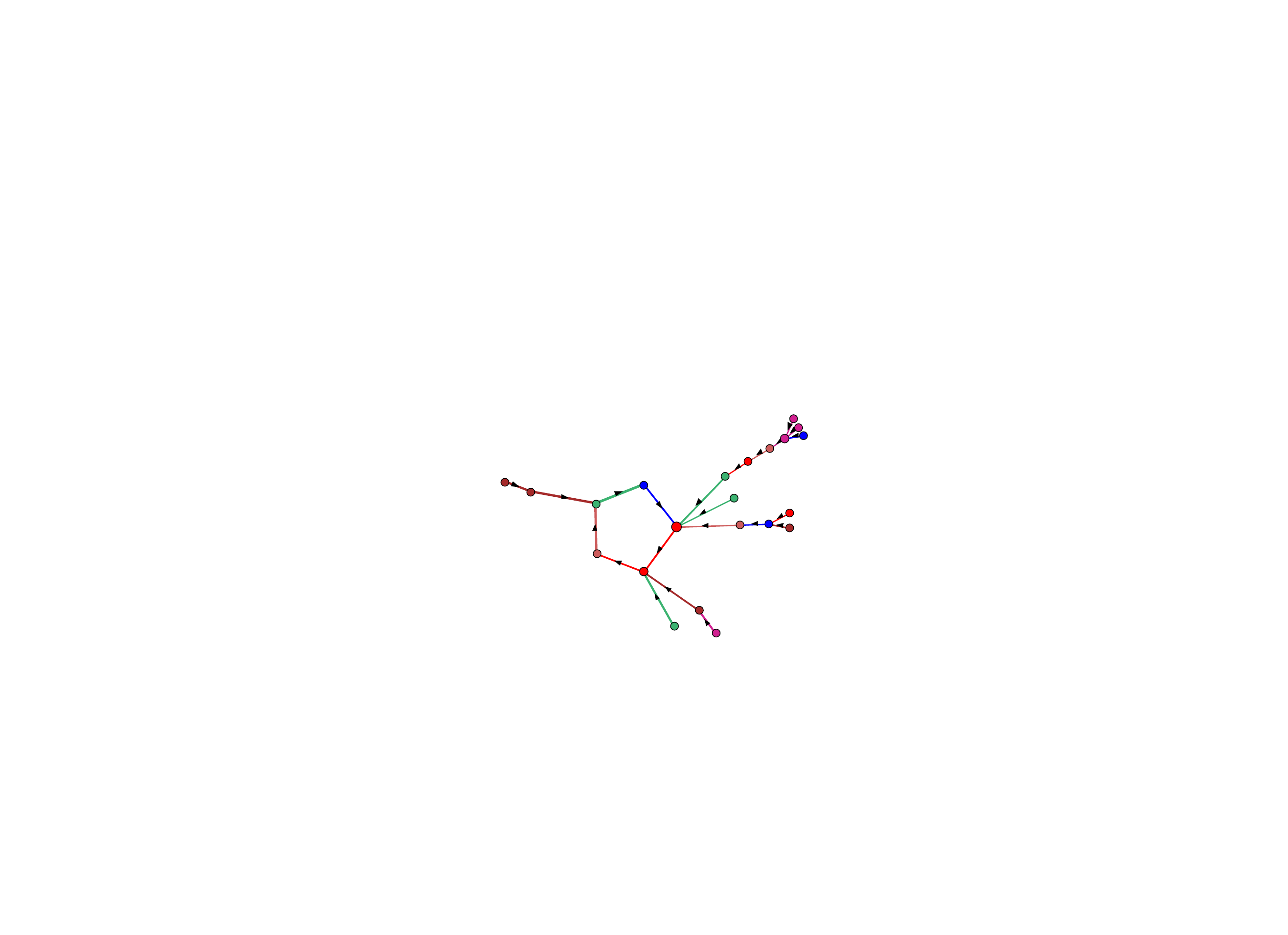}
     & \includegraphics[bb=508 291 822 534,clip=, width=.2\linewidth]{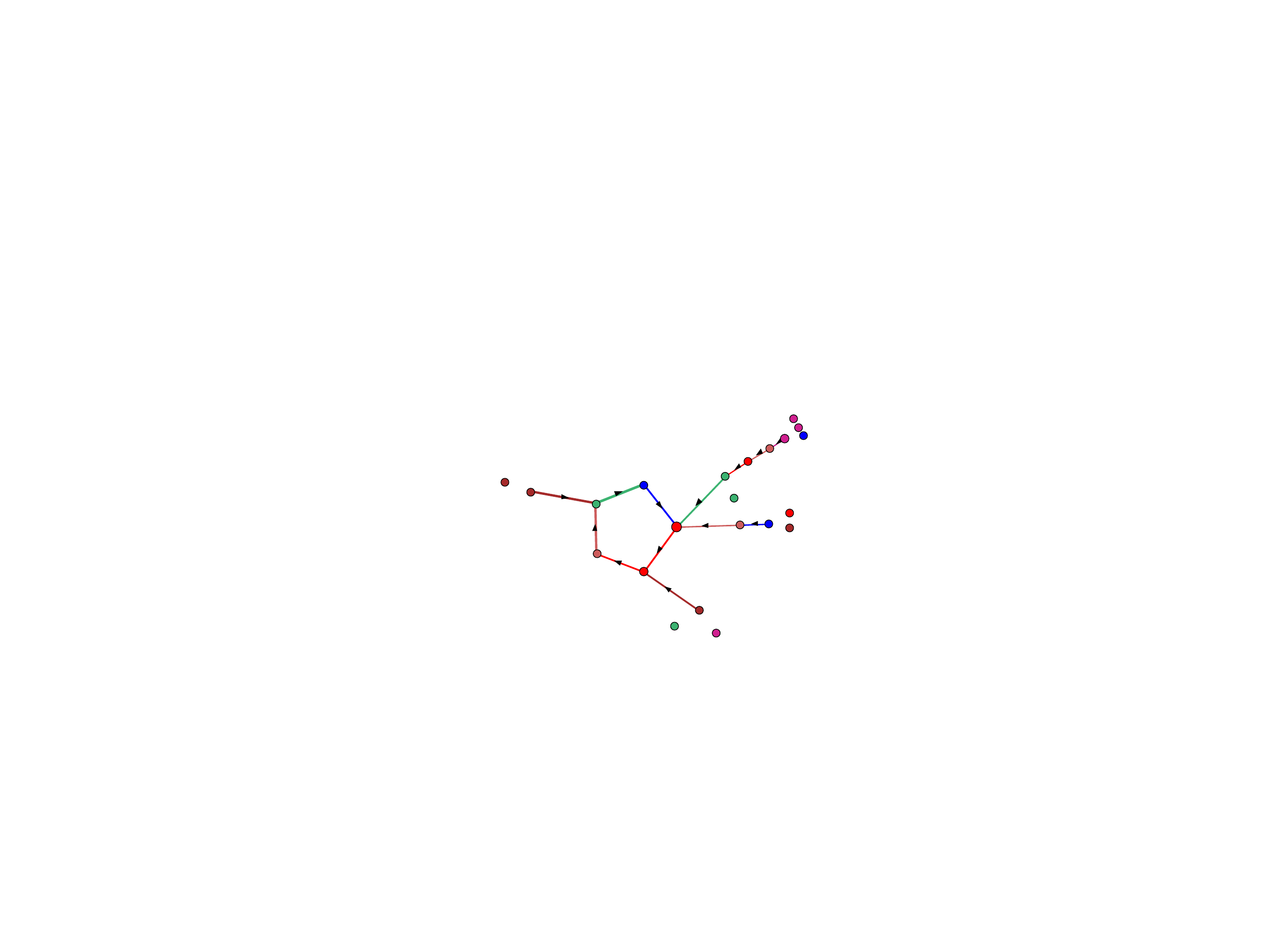}
     & \includegraphics[bb=508 291 822 534,clip=, width=.2\linewidth]{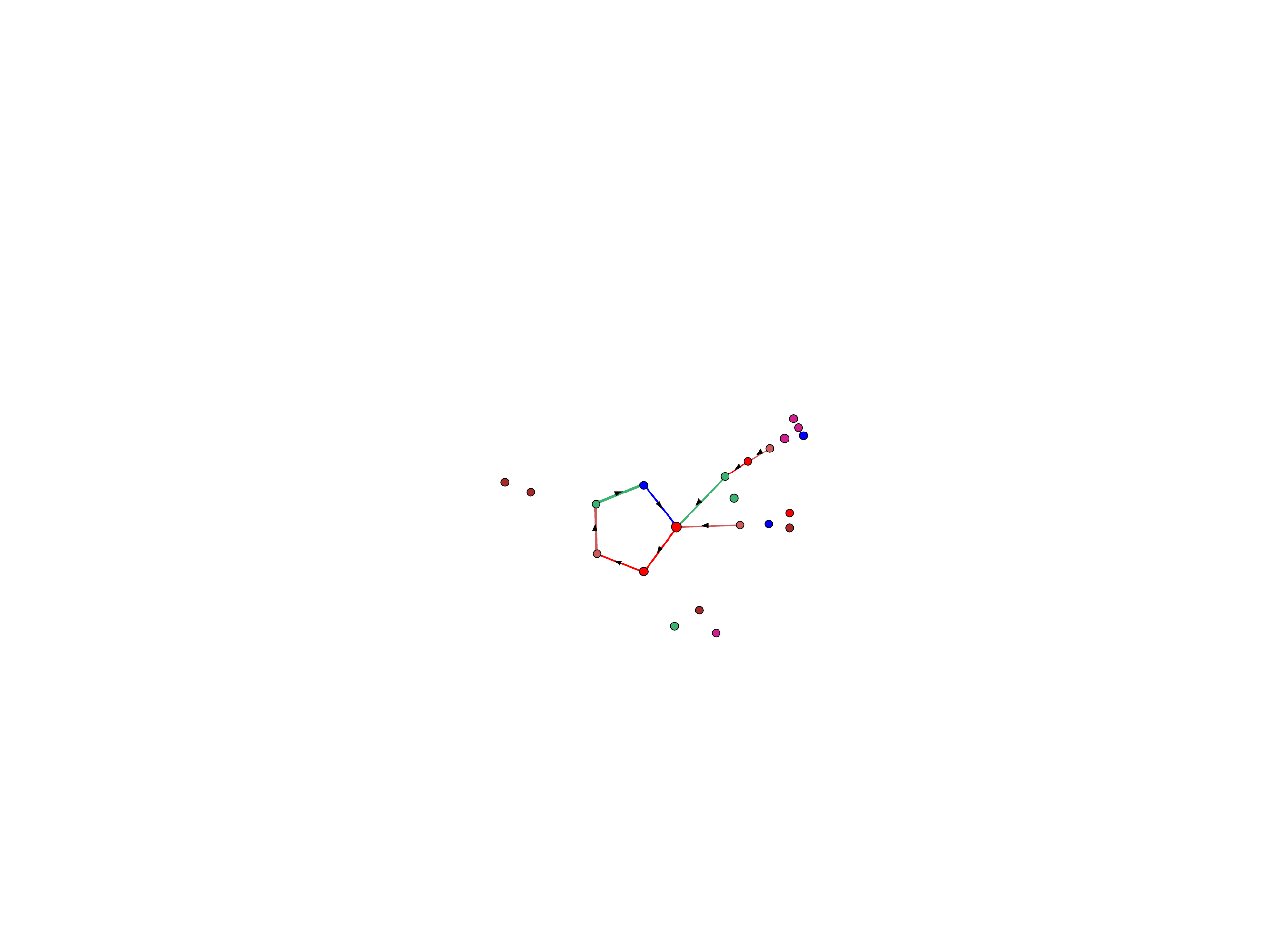}\\
     \hline
     \includegraphics[bb=508 291 822 534,clip=, width=.2\linewidth]{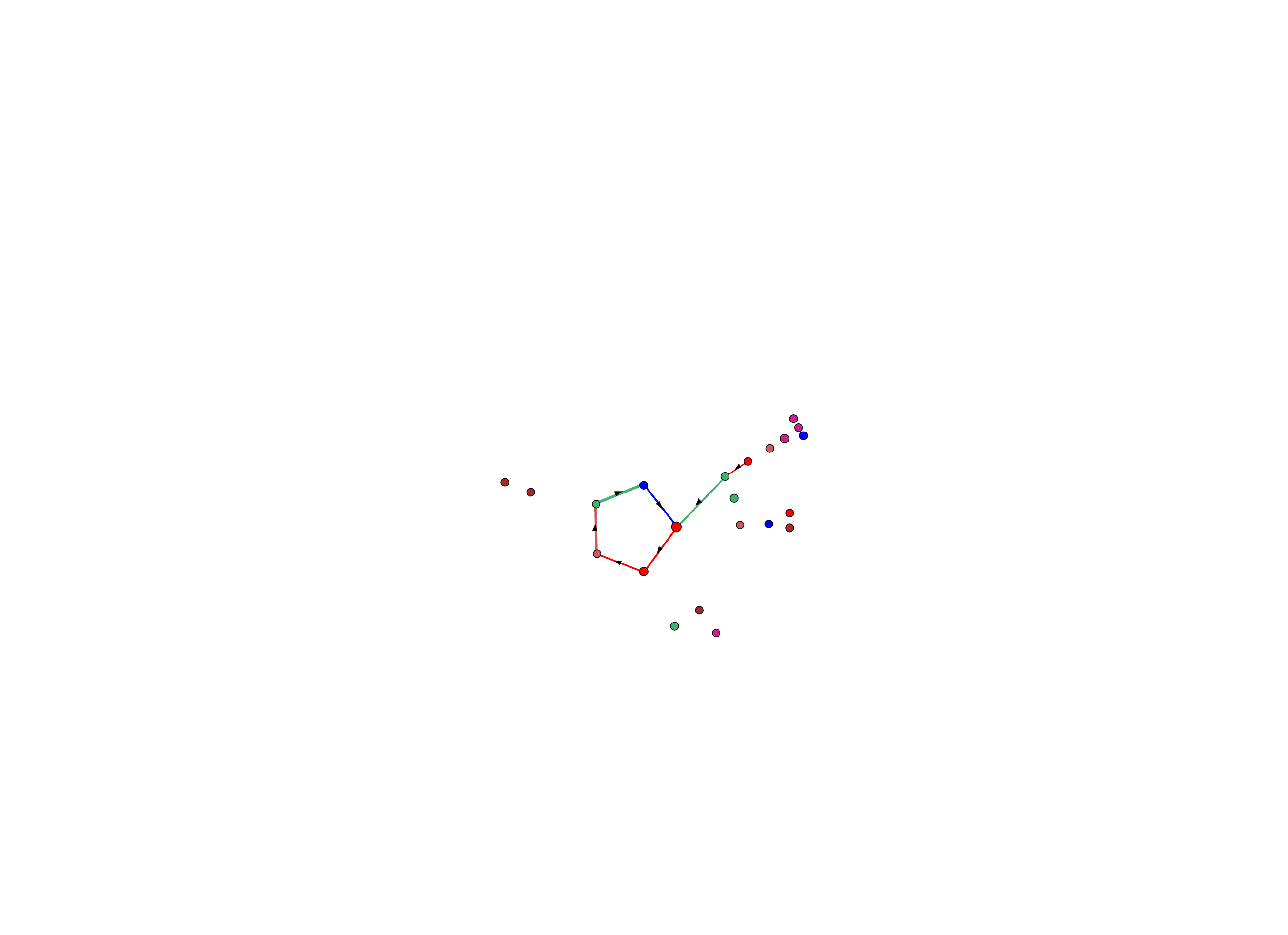}
     & \includegraphics[bb=508 291 822 534,clip=, width=.2\linewidth]{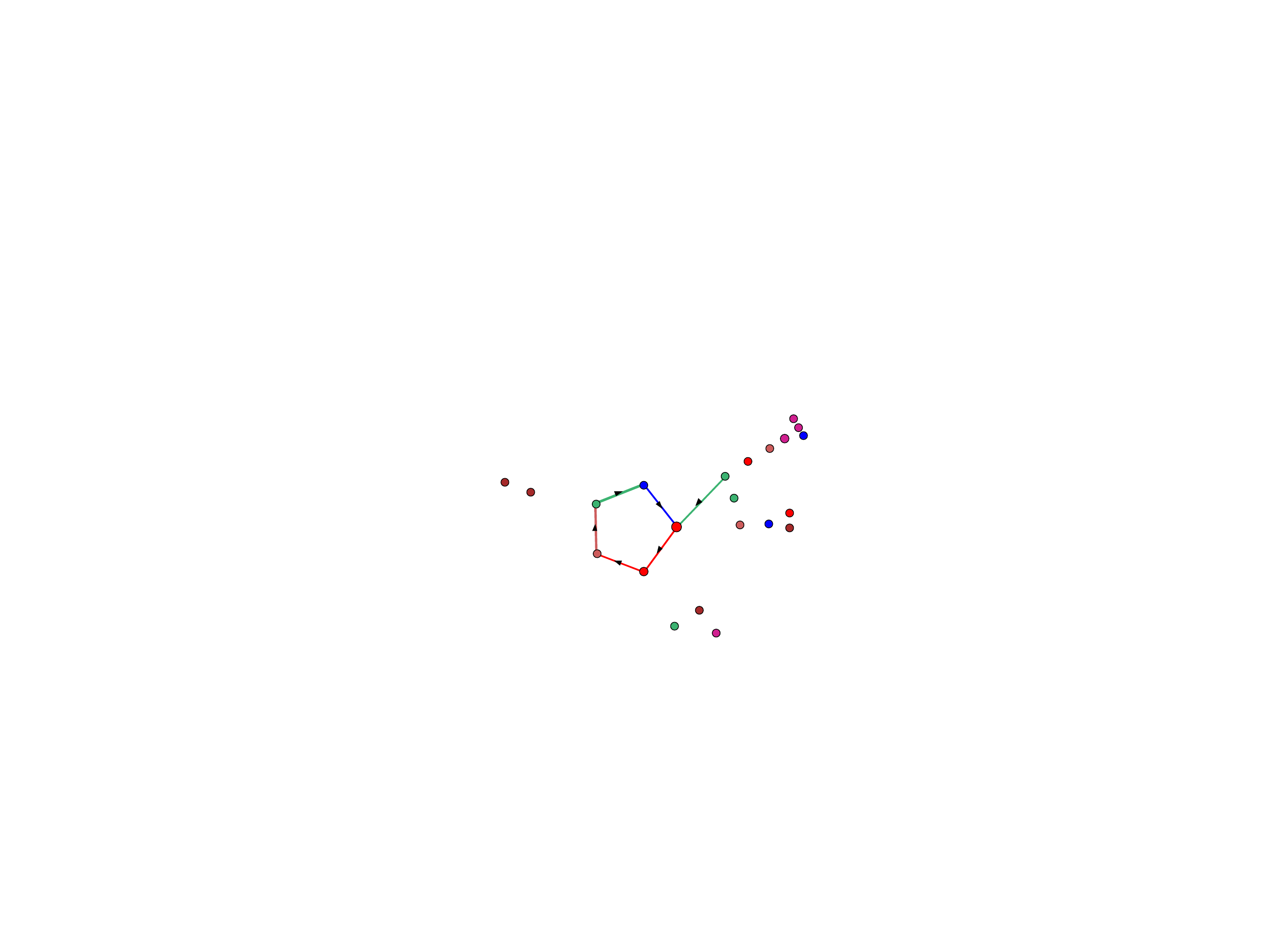}
     & \includegraphics[bb=508 291 822 534,clip=, width=.2\linewidth]{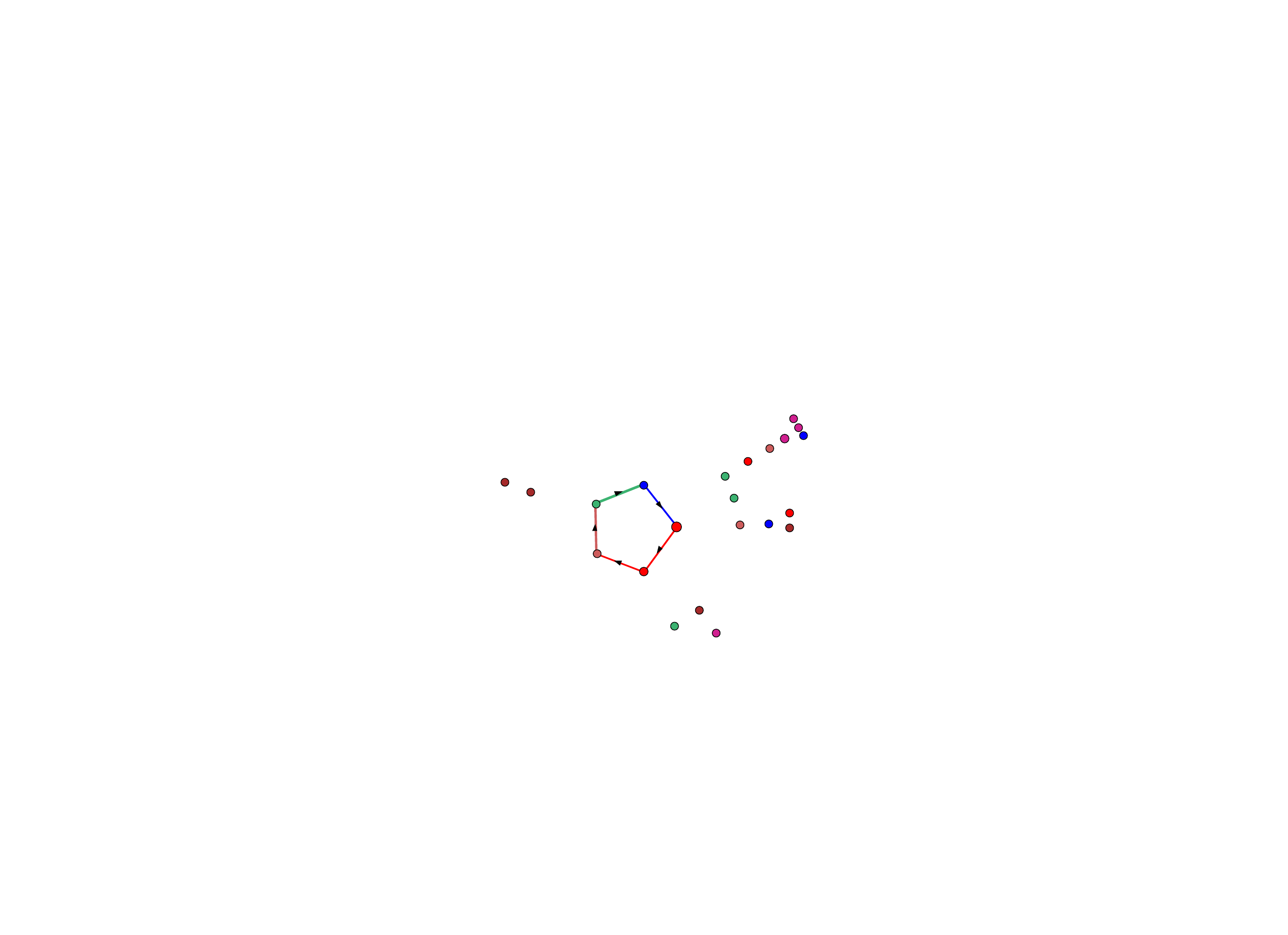}\\
    \hline
   \end{tabular}\\
   \end{center}  
   \vspace{-2ex}
    \caption [Pruning the graph]
       {\textsf{Pruning the ibaf-graph's gap-nodes with ``{\bf gap-g}''.
       In this example for the ibaf-graph, one small basin is isolated with ``{\bf just-j}''
       then recursively pruned of its leaf states until only the attractor remains
       --- rcode 110, $n$=10.}}
     \label{prune-gap-nodes}
\end{figure}

\item[{\bf nodes-E $\dots$}] ({\it not for ``single'' status}) 
  to equalise the display/size of all nodes in the fragment according
  to the active node, which can be outside a ``block'' as well as inside.

\item[{\bf block-B} $\dots$] ({\it ``single'' status only}) to define
  a block --- suboptions appear\cite[\hspace{-1ex}\footnotesize{\#20.11}]{EDD}.  A block is a
  sequence of node numbers (in 1d, 2d, or 3d) between two selected
  nodes.  The outer edges of the block can be set, or defaults
  accepted --- the last two nodes that were activated.  For a linked
  network-graph in 2d or 3d the outer edges can be the outer corners
  of a 2d or 3d block --- {\color{BrickRed}{\bf Block 6-21}} (for
  example) appears in the title.

\item[{\bf exit-block-B} $\dots$] ({\it in block status}) enter {\it exit-block-}{\bf B}
                         to exit a block and revert to ``single'' status, or exit
                         by changing status with ``{\it single-}{\bf s}'', ``{\it in/out/either-}{\bf i/o/e}'',
                         or ``{\it all-}{\bf a}''.

\begin{figure}[H]
   \begin{center}
     \includegraphics[ bb=431 242 960 675,clip=,height=.3\linewidth]{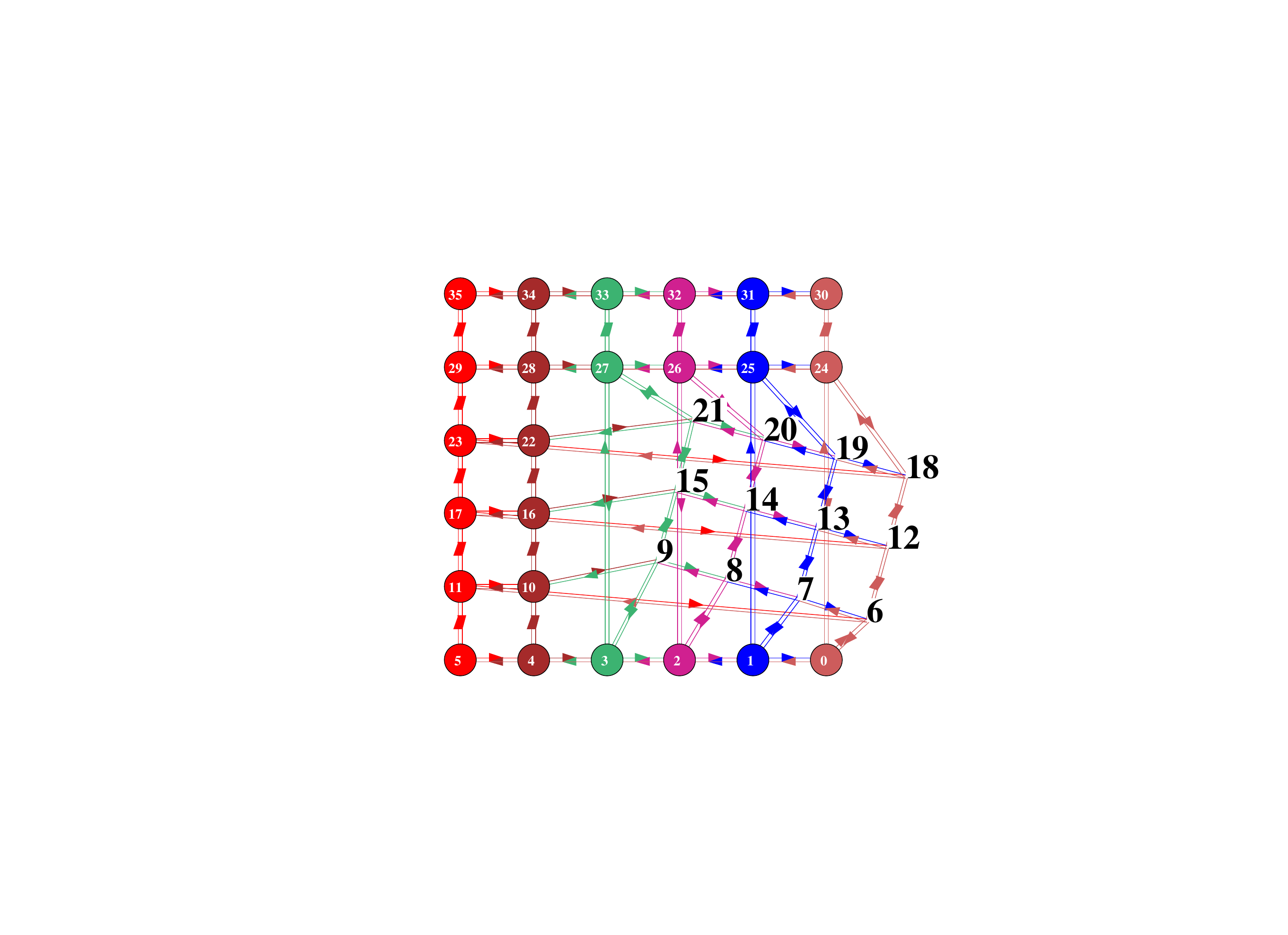}
     \includegraphics[bb=452 212 1106 690,clip=,height=.3\linewidth]{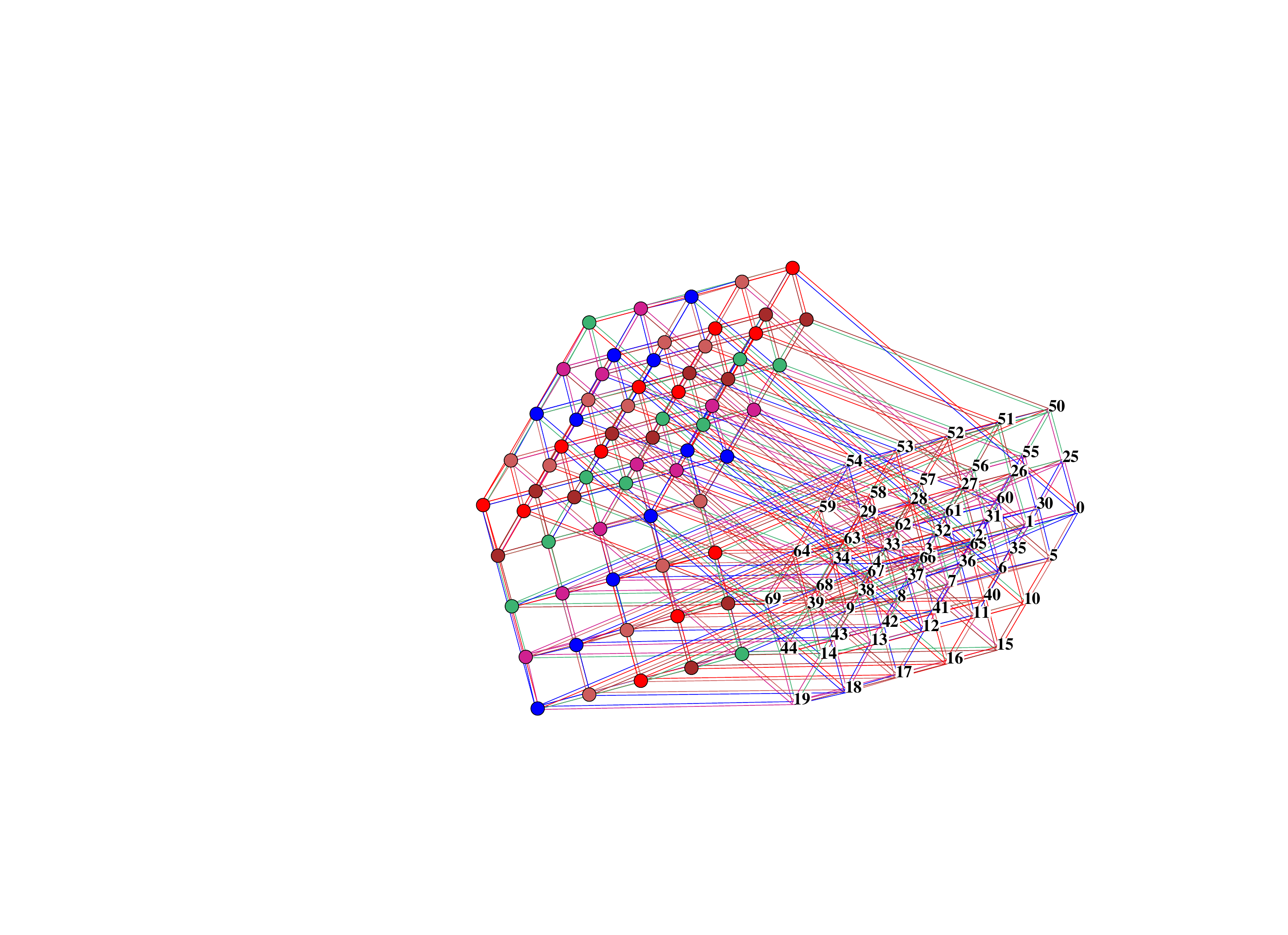}
  \end{center}  
   \vspace{-5ex}  
    \caption  [Dragging a 2d an 3d block]
       {\textsf{A regular network-graph with a block defined,
   dragged, rotated, nodes toggled to decimal, and font enlarged.    
 \underline{\it Left}: 2d $k$=4 6x6 with a 3x3 block.
  \underline{\it Right}: 3d $k$=6 5x5x5 with a 4x4x3 block.
       \label{2d an 3d block}}}
     \end{figure}

\item[{\bf just-j/J} $\dots$] Any active component or fragment can be show/printed in isolation.
  ``{\bf j}'' (lower case) will toggle showing a component with the
  active node, (for example: a basin in the ibaf-graph), to work on it
  in isolation as in figure~\ref{labels-sept}.  ``{\bf J}'' (capital)
  will toggle showing just the active frament linked to the active node
  (figures~\ref{frags-net}, \ref{frags-basin}).
  Note that other components/fragments
    are still there but invisible, so clicking in an empty
    space could switch to a different fragment.
                          
\begin{figure}[htb]
  \begin{minipage}[t]{1\linewidth} 
     \includegraphics[bb=333 207 977 672,clip=,height=.23\linewidth]{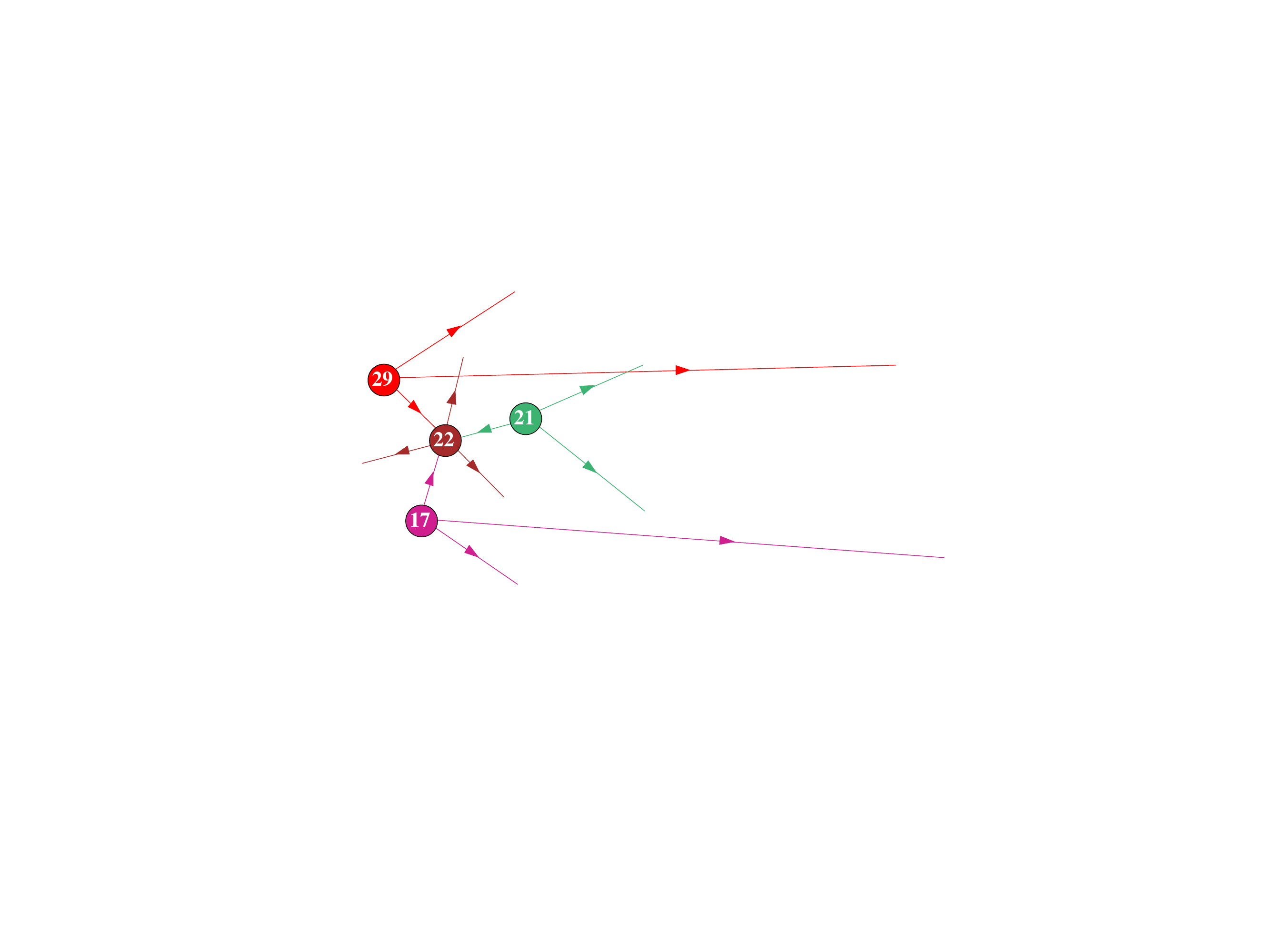}
     \hfill
     \includegraphics[bb=333 207 977 672,clip=,height=.23\linewidth]{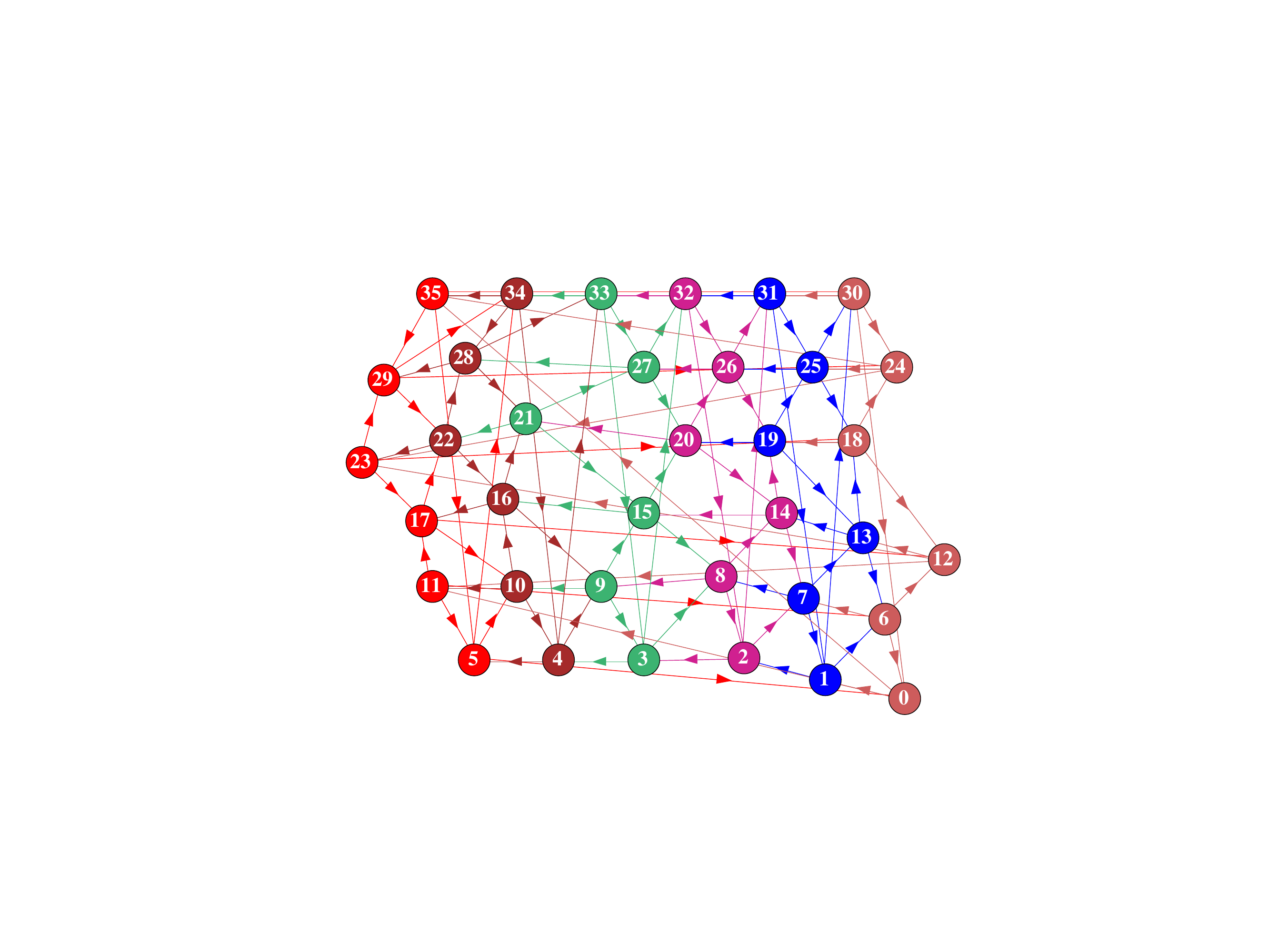}
     \hfill
     \includegraphics[bb=333 207 977 672,clip=,height=.23\linewidth]{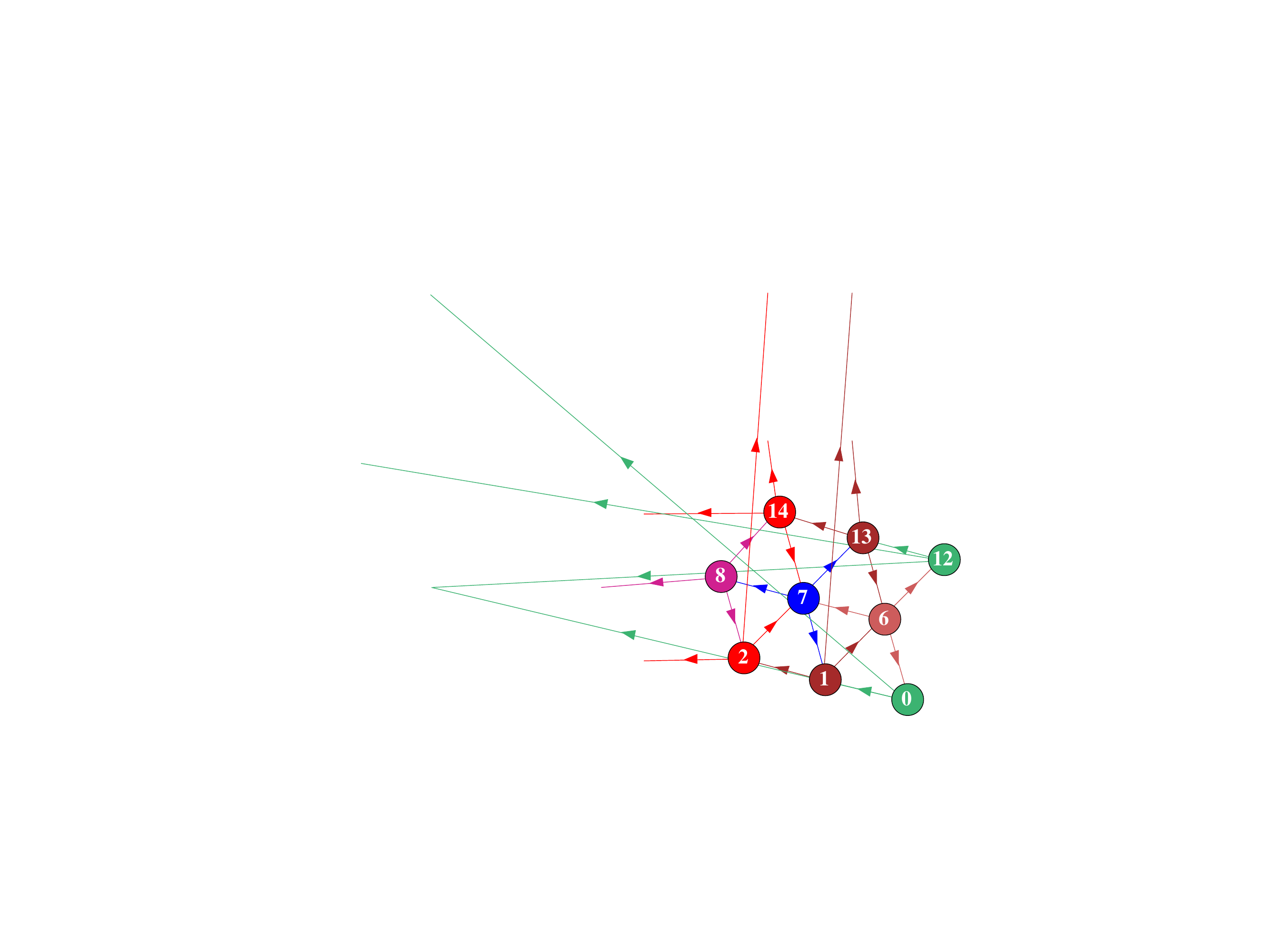}
 \end{minipage}     
   \vspace{-6ex}  
    \caption[Isolating fragments in a 2d network-graph] 
       {\textsf{Isolating fragments in a 6x6 2d network-graph with a $k$=3 triangular
       n-template\cite[\hspace{-1ex}\footnotesize{\#10.1.3}]{EDD}.
       \underline{\it Center}: The graph with a linked fragment and a block
       both slightly dragged and rotated.
       \underline{\it Left}: The isolated linked frament centered on node 22 according
       to ``inputs'' and ``step=1''.
       \underline{\it Right}: The isolated ``block'' defined between nodes 14 and 0.
       \label{frags-net}}}
       \vspace{-4ex}
\end{figure}    

\begin{figure}[htb]
  \begin{minipage}[c]{1\linewidth}
  \begin{center}
     \includegraphics[bb=413 391 1230 561,clip=,width=.8\linewidth]{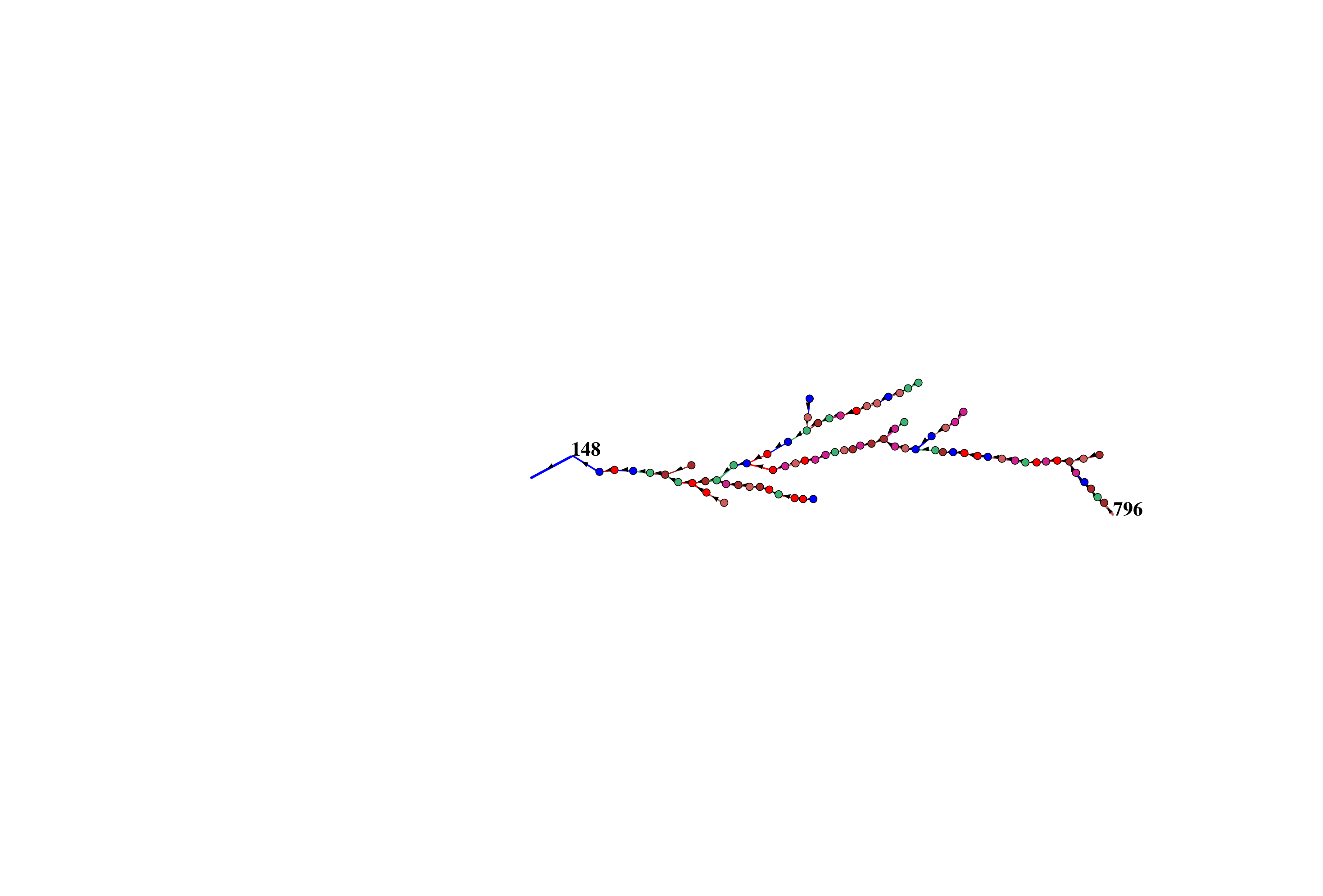}
                           \raisebox{2ex}{(a)}\\[-3ex]
     \includegraphics[bb=413 391 1230 561,clip=,width=.8\linewidth]{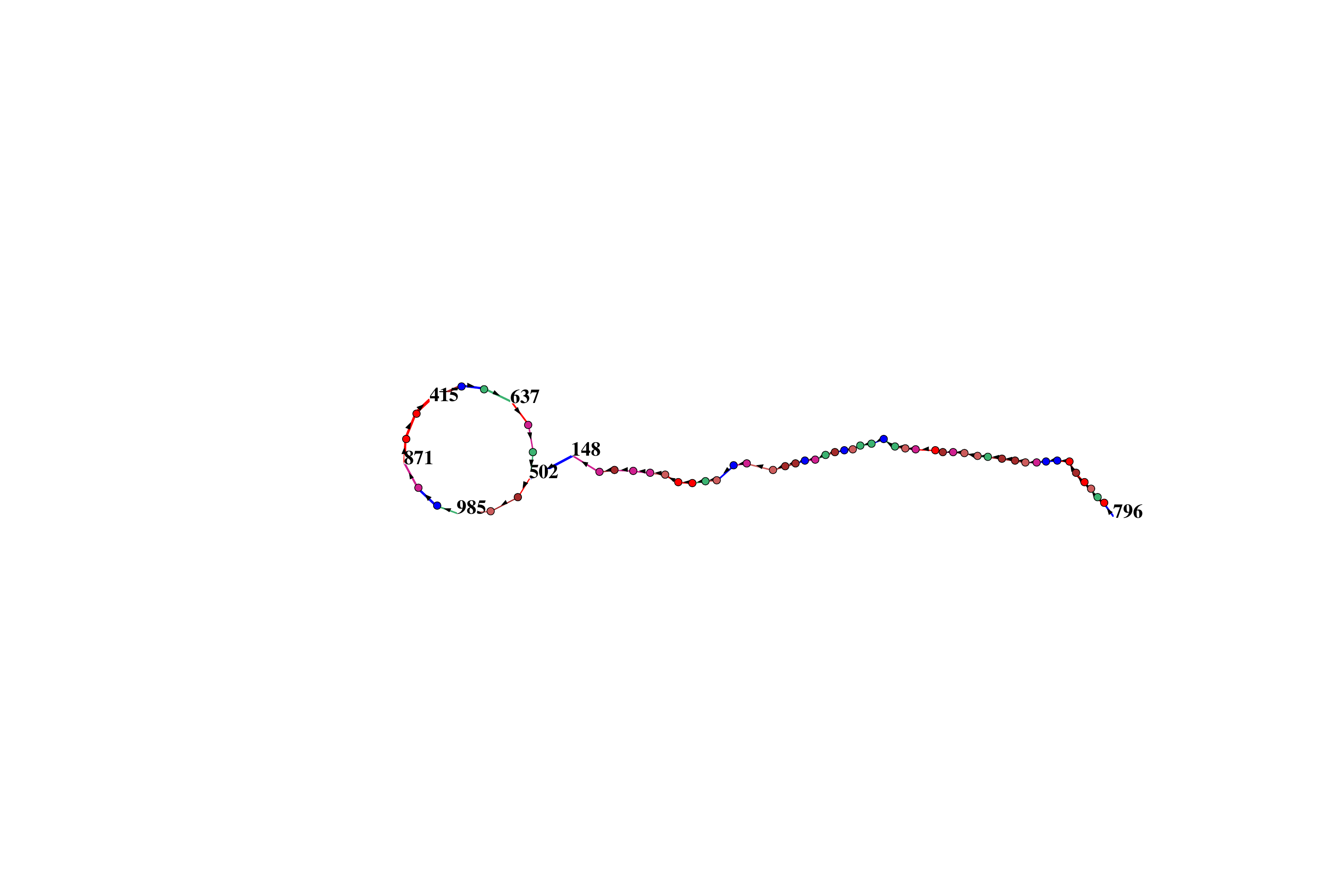}
                           \raisebox{2ex}{(b)}
  \end{center}
 \end{minipage}     
   \vspace{-3ex}  
    \caption[Isolating fragments in a basin of attraction] 
       {\textsf{Isolating linked fragments in the ibaf-graph
       of 1d CA rcode 110 k=3 n=10 (as in the second basin figure~\ref{ibaf_matrix}).
       (a) The fragment
       by ``inputs'' and ``step=nolimit'' from node 148 shows the entire incoming subtree.
       (b) The fragment
       by ``outputs'' and ``step=nolimit'' from the leaf node 796 shows the run-in
       to (and including) the attractor.  Numbered nodes on the attractor indicate
       the root of each equivalent subtree.       
       \label{frags-basin}}}
\end{figure}

\item[{\bf Label-L/+} $\dots$] ({\it ``single'' status only}) to create (or remove)
  a (multiple line) label locked onto the active node (figure~\ref{labels-sept}).
  Type the new label (max 90 characters)
  in suboptions\cite[\hspace{-1ex}\footnotesize{\#20.12}]{EDD},
  using the ``{\bf $\backslash$}'' (backslash)
  key for line breaks. A leading line break will to keep a numbered node visible.  
  Labels will move when nodes or fragments are dragged or the graph is rearranged,
  just like other node displays.
  Enter ``{\it Label-}{\bf +}''
  (plus sign) to toggle between showing all labels and current node
  displays. Labels persist if reverting to the initial-graph where all
  labels can also be toggled with ``{\it Label-}{\bf +}''.
                                              
  \hspace{3ex} The default label font size can contract/expand by
  applying ``{\bf (}'' or ``{\bf )}'' to the active node while in
  numbered display.  While dragging, all alphanumeric displays
  (numbered nodes and labels) revert to the default font size, but
  once dragging ends the correct font sizes reappear.
                     
\begin{figure}[H]
\begin{center}
  \includegraphics[bb=140 150 1112 708,clip=,width=.85\linewidth]{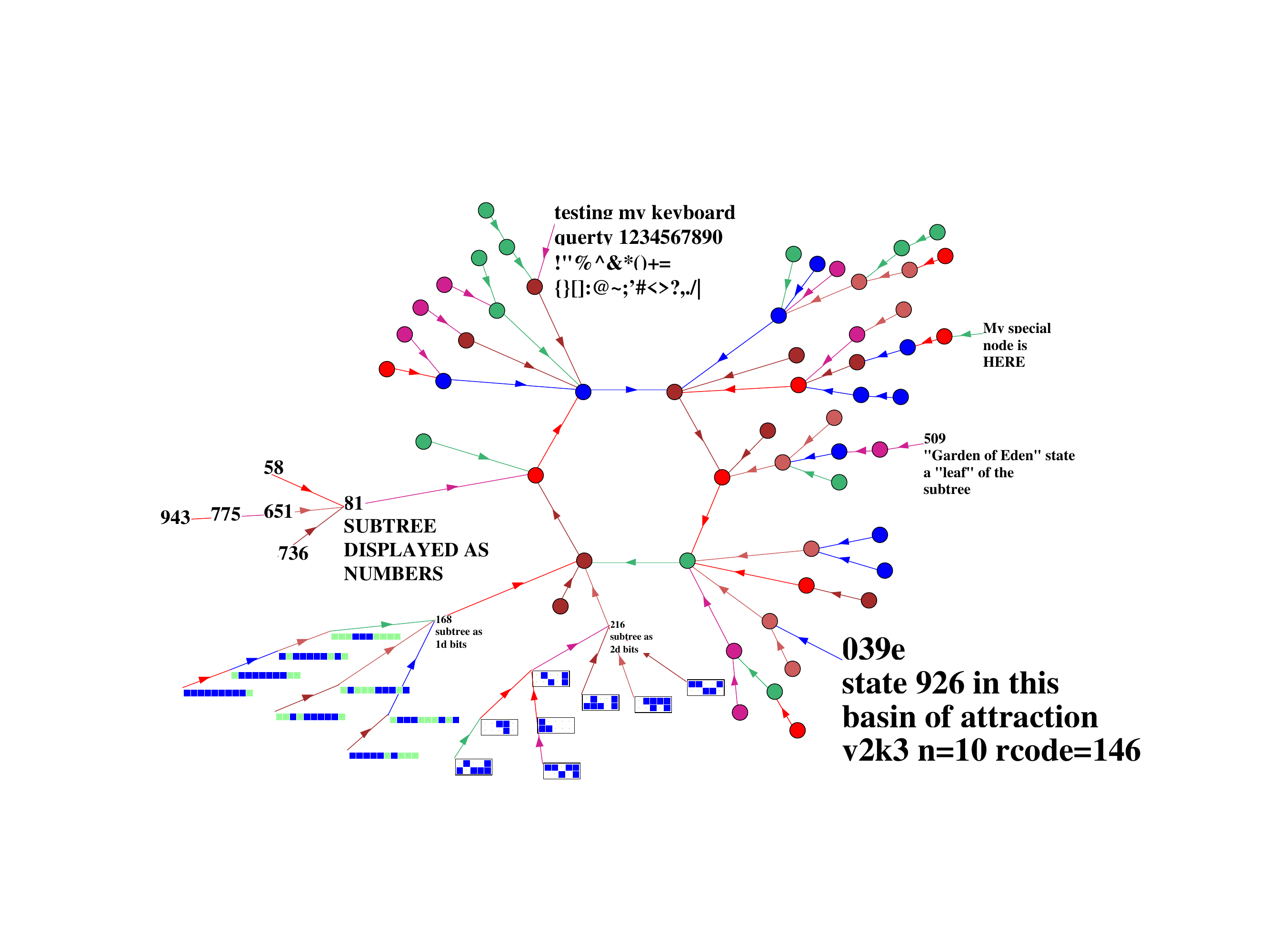}
\end{center} 
\vspace{-5ex}
        \caption [Example of arbitrary labels in ibaf]       
                 {\textsf{An example of a single ibaf-graph basin of attraction
                     (isolated with ``{\it just}-{\bf j}'')
                 with some arbitrary labels at various nodes, with various font sizes.
                 Without a leading line break a label will overwrite the current node display,
                 which can still be toggled (key ``{\bf =})''. 
                 For example the subtree to state 81 has been toggled to show all its
                 inputs as numbers, and state 216 as 2d patterns.}}
                 \label{labels-sept}             
\end{figure}

\item[{\bf Lnk23: cut/restore-c/r} $\dots$] ({\it linked-fragment only, for example})
                        enter ``{\it cut-}{\bf c}'' to cut (disconnect)
                        the active node (shown) from its immediate links, depending on
                        which option, {\color{BrickRed}{\bf inputs}}, {\color{BrickRed}{\bf outputs}}
                        or {\color{BrickRed}{\bf either}}
                        is active (figure~\ref{cut-links-to active}(a-c).
                        Enter ``{\it restore-}{\bf r}'' to undo the previous cut.
                        Enter ``{\it net-}{\bf \#}'' in any status (or in the initial-graph)
                        to restore links to the original.

\item[{\bf Lnk 23-19: cut/add/restore-C/A/R} $\dots$] ({\it linked fragment only, for example})
                        The active and previously active node numbers are shown 
                        in the prompt. Enter ``{\it cut-}{\bf C}'' to cut or \mbox{``{\it add-}{\bf A}''} to add,
                        or `{\it restore-}{\bf R}'' to restore 
                        a link between the pairs (figure~\ref{cut-links-to active}(d-f)),
                         depending on which option,
                        {\color{BrickRed}{\bf inputs}}, {\color{BrickRed}{\bf outputs}}
                        or {\color{BrickRed}{\bf either}} is active.
                        Enter ``{\it net-}{\bf \#}'' in any status (or in the initial-graph)
                        to restore links to the original.

\begin{figure}[H]
\textsf{\small
   \begin{minipage}{.25\linewidth} 
     \includegraphics[bb=416 119 911 685, clip=,height=1\linewidth]{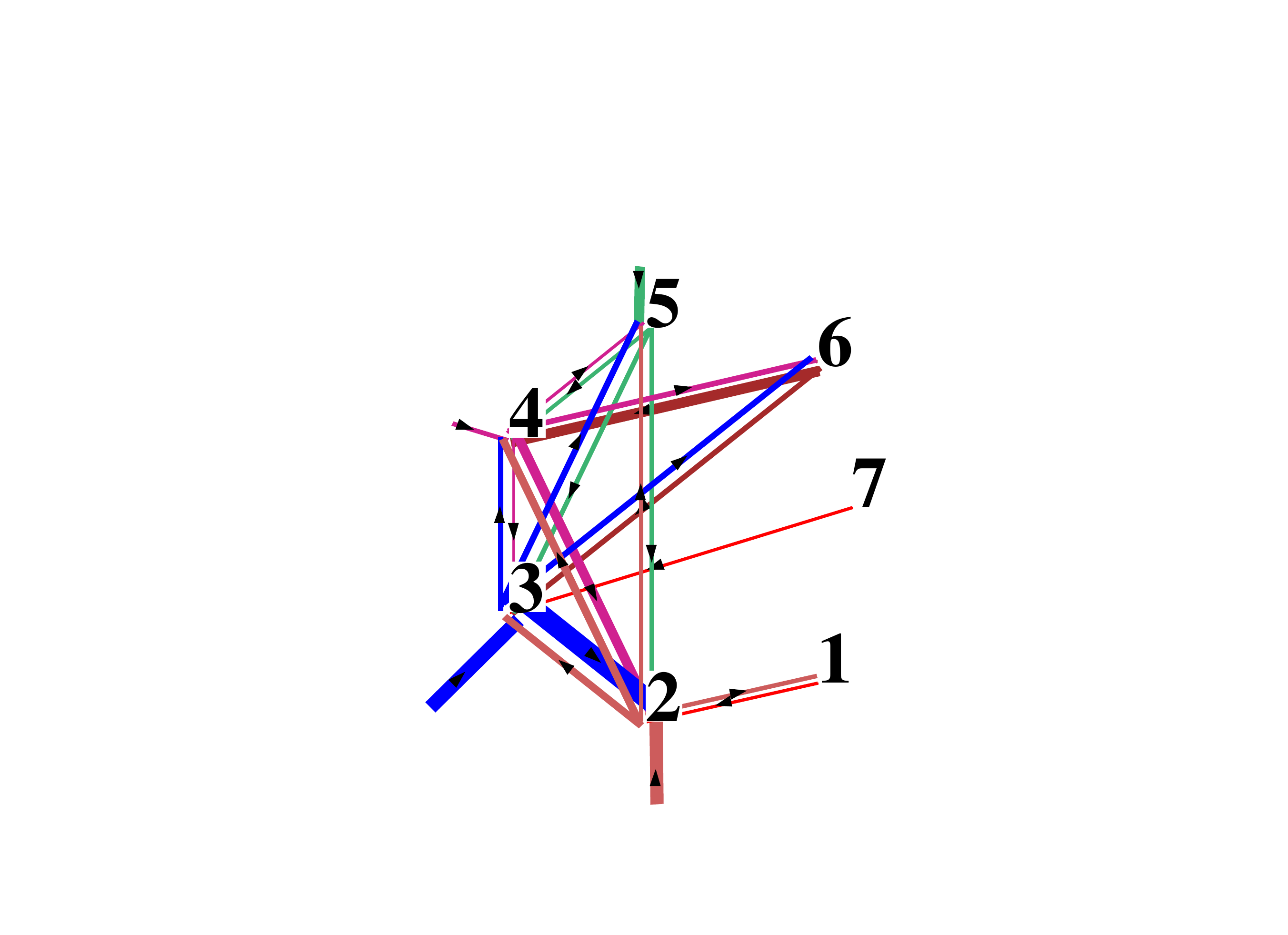}\\[-4ex]
         original
   \end{minipage}\\[1ex]
   \begin{minipage}{.25\linewidth}
     \includegraphics[bb=416 119 911 685, clip=,height=1\linewidth]{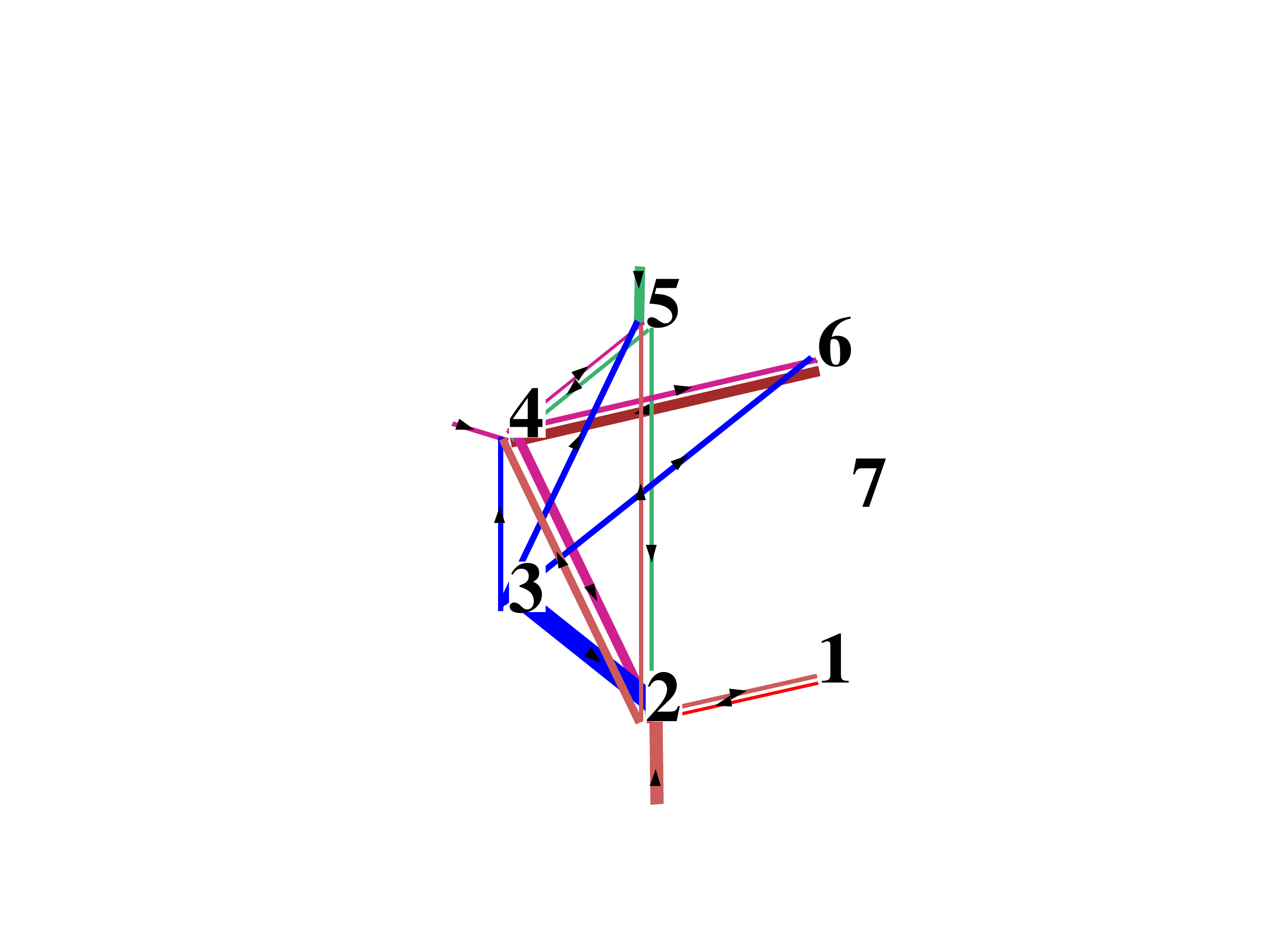}\\[-4ex]
         (a) in
   \end{minipage}
   \begin{minipage}{.25\linewidth}
     \includegraphics[ bb=416 119 911 685, clip=,height=1\linewidth]{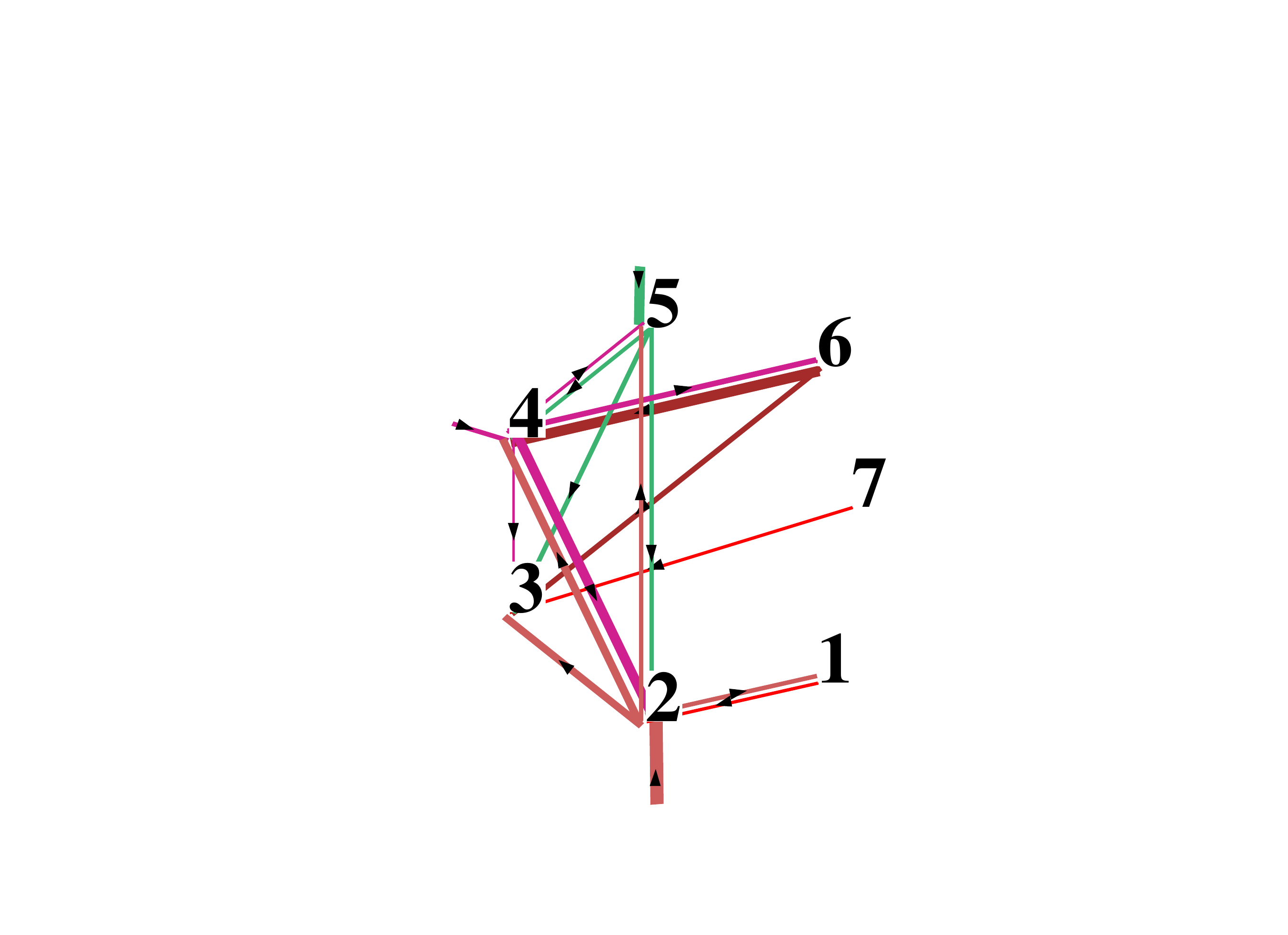}\\[-4ex]
         (b) out
   \end{minipage}
   \begin{minipage}{.25\linewidth}
     \includegraphics[bb=416 119 911 685, clip=,height=1\linewidth]{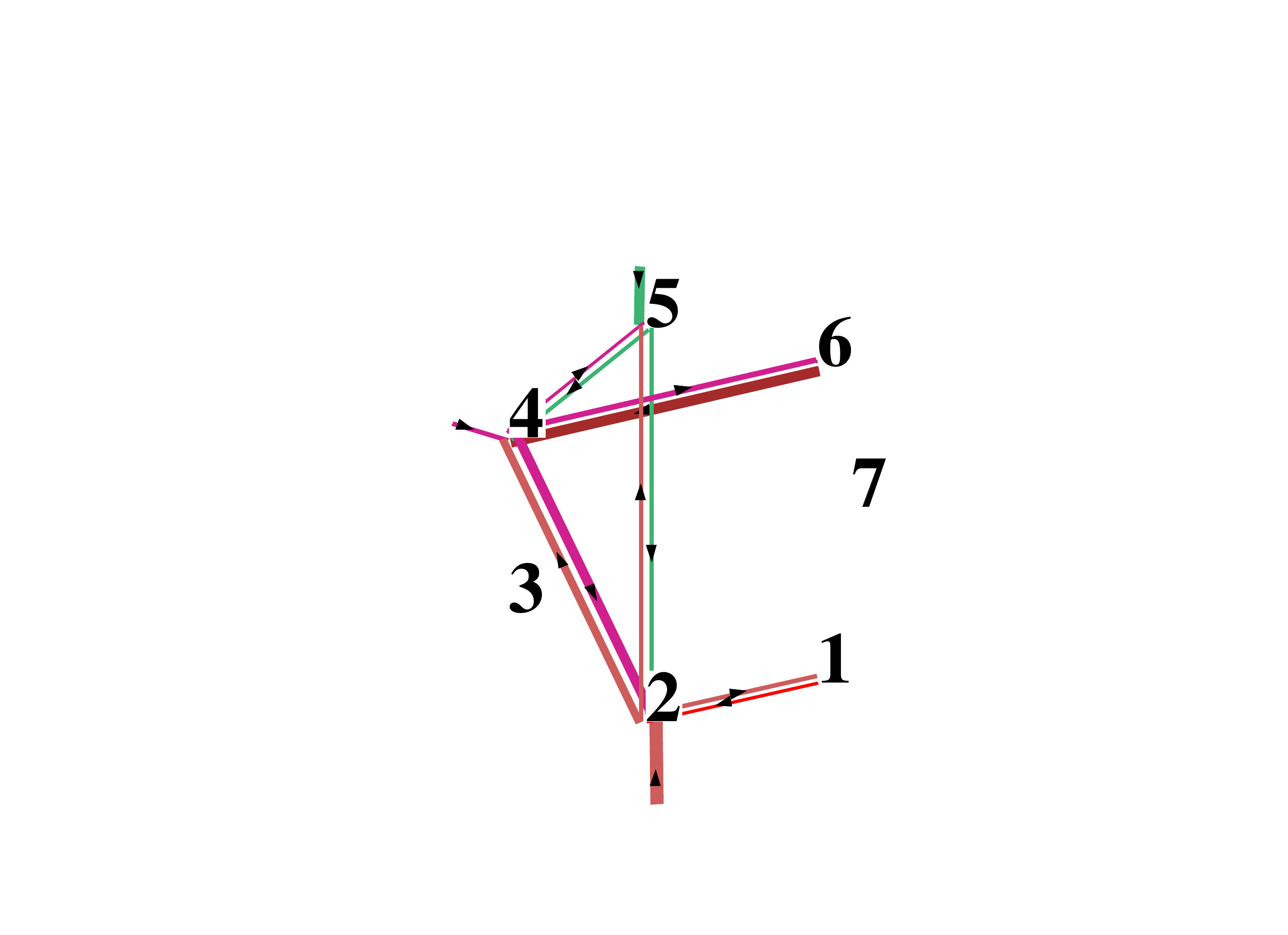}\\[-4ex]
         (c) either
   \end{minipage}
    \begin{minipage}{.2\linewidth} 
       cutting links to node 3
   \end{minipage}
   \\[1ex]
   \begin{minipage}{.25\linewidth}
     \includegraphics[bb=416 119 911 685, clip=,height=1\linewidth]{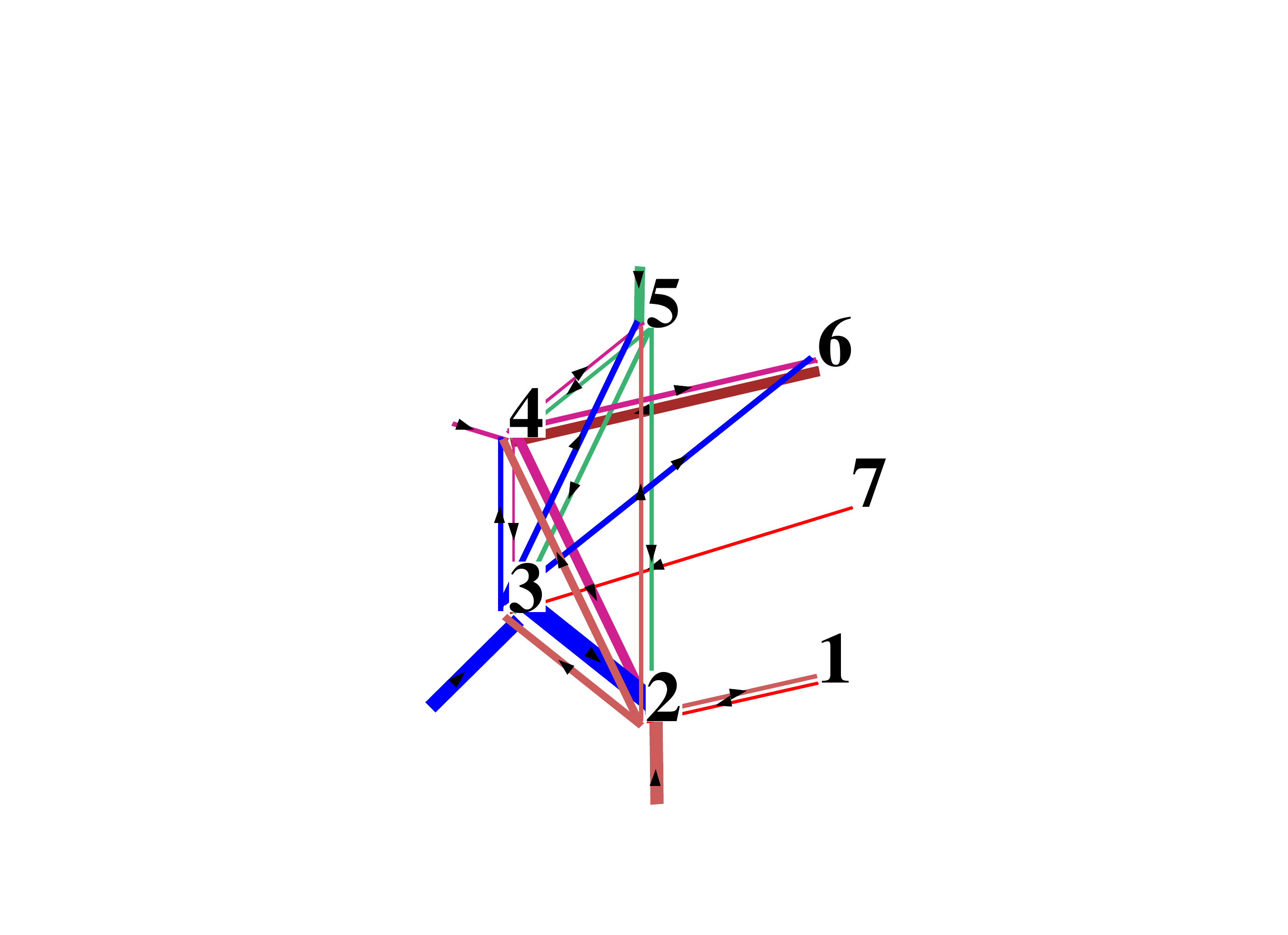}\\[-4ex]
         (d) in
   \end{minipage}
   \begin{minipage}{.25\linewidth}
     \includegraphics[bb=416 119 911 685, clip=,height=1\linewidth]{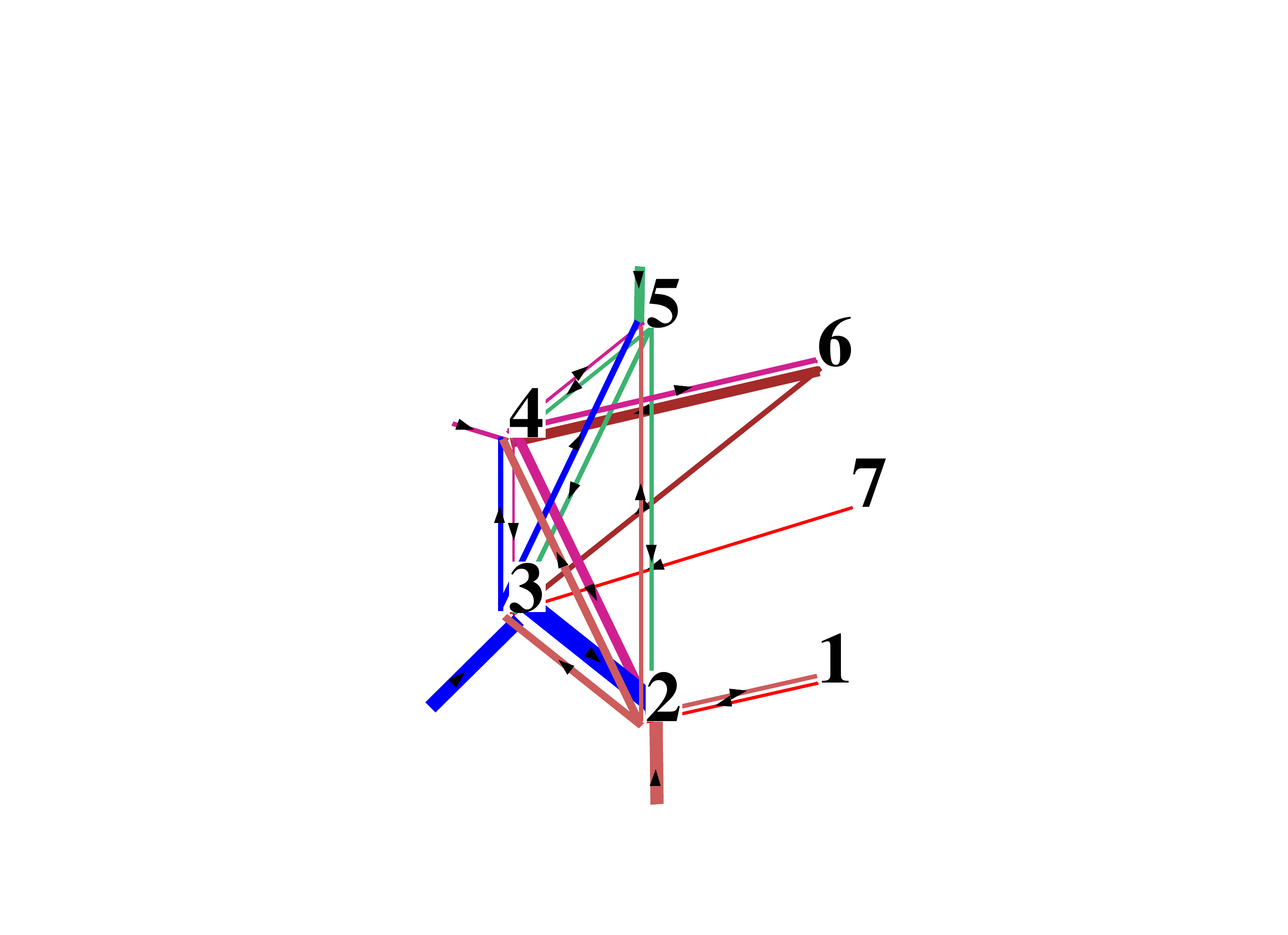}\\[-4ex]
         (e) out
   \end{minipage}
   \begin{minipage}{.25\linewidth}
     \includegraphics[ bb=416 119 911 685, clip=,height=1\linewidth]{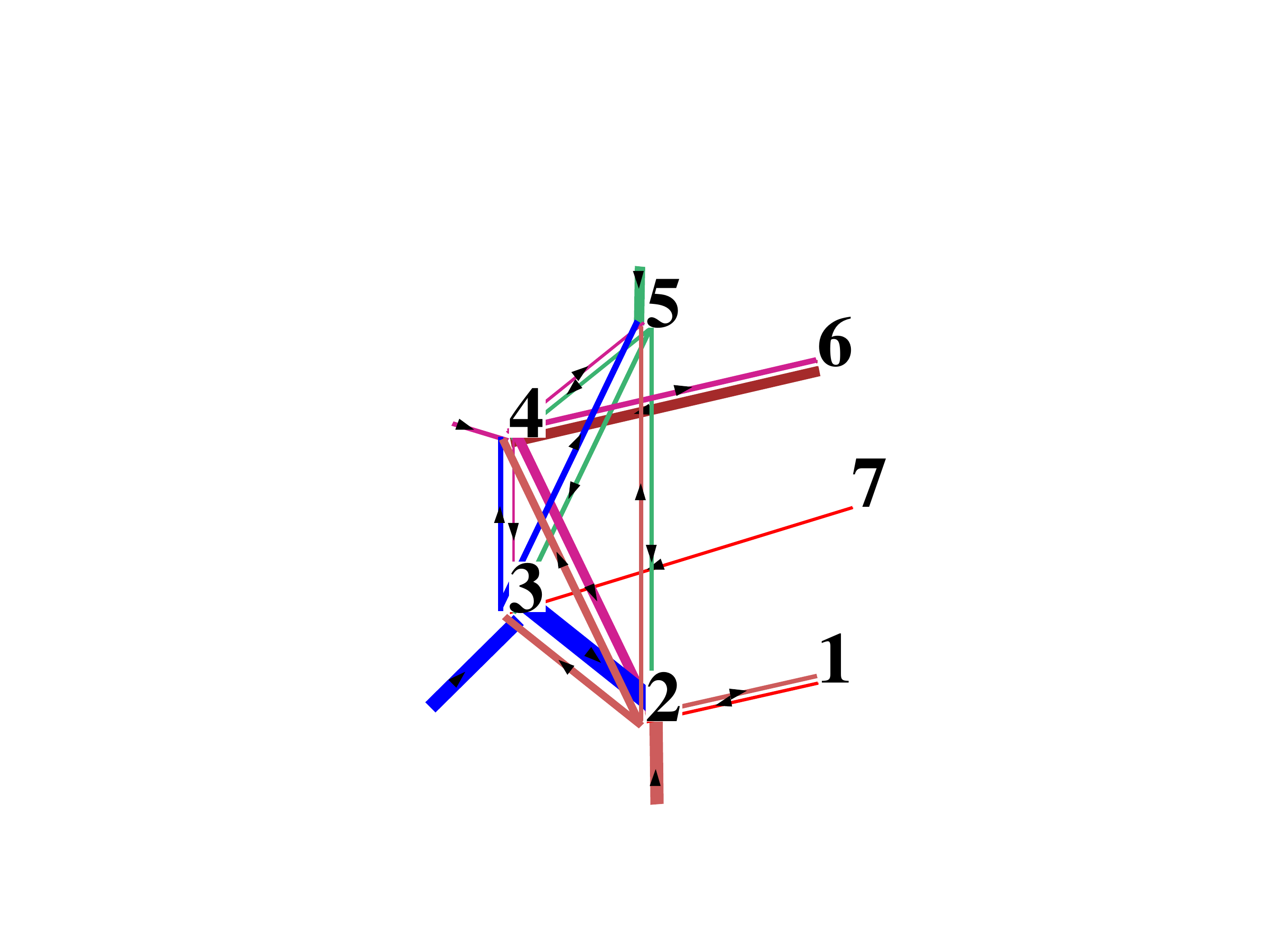}\\[-4ex]
         (f) either    
   \end{minipage}
   \begin{minipage}{.2\linewidth}
       cutting links between nodes 3 and 6
   \end{minipage}
   }\\[-73ex]
   \vspace{.1ex} \hspace{23ex}
   \begin{minipage}[t]{.73\linewidth}
    \caption [cutting links to the active node]
    {\textsf{Cutting links: 
\underline{\it Left}: the original jump-graph for $v2k3$ rcode 194 $n$=10 with
         nodes shown as decimal numbers
         and links scaled by jump probability. 
   \underline{\it Below}:  cutting links to active node 3 by
        (a) inputs, (b) outputs, (c) either.
        The self-link is cut in all cases.
   \underline{\it Bottom Row}:  cutting links between a pair of nodes;
              the active node 3 and node 6 by (d) inputs, (e)~outputs, (f)~either.
              The method to add links between a pair of nodes is equivalent.
     \label{cut-links-to active}}}
     \end{minipage}
\vspace{50ex}     
\end{figure}

\vspace{2ex}
\item[{\bf step-(1-9)} $\dots$] enter a number between {\bf 1} and {\bf 9} to
                        limit a linked fragment by the distance from the active node.
                        measured in link-steps (time-steps for the ibaf-graph) ---
                        a continuous chain of directed edges
                        depending on which status --- {\color{BrickRed}{\bf inputs}}, {\color{BrickRed}{\bf outputs}}
                        or {\color{BrickRed}{\bf either}} --- is active.                        
                        For example, enter ``{\bf 1}'' for immediate links,
                        ``{\bf 2}'' up to 2 steps away, etc., up to 9 steps. The current step distance is shown in the 
                        drag reminder, for example {\color{BrickRed}{\bf step=1}}.
                        For unlimited {\color{BrickRed}{\bf step=nolimit}} enter ``{\bf 0}'' (zero).
                        This can be set in any drag-status but applies for a linked fragment.

                        \hspace{3ex} The title of the drag-reminder shows the active node, current linked fragment type,
                        and step-limit, for example,

\begin{quote}
    {\color{BrickRed}{\bf node 14, inputs, step=nolimit:}}\\
    {\color{BrickRed}{\bf node 14, outputs, step=1:}}\\
    {\color{BrickRed}{\bf node 14, either, step=3:}}
\end{quote}

 \begin{figure}[H]
   \begin{center}
     \includegraphics[bb=434 243 1100 681,clip=,height=.3\linewidth]{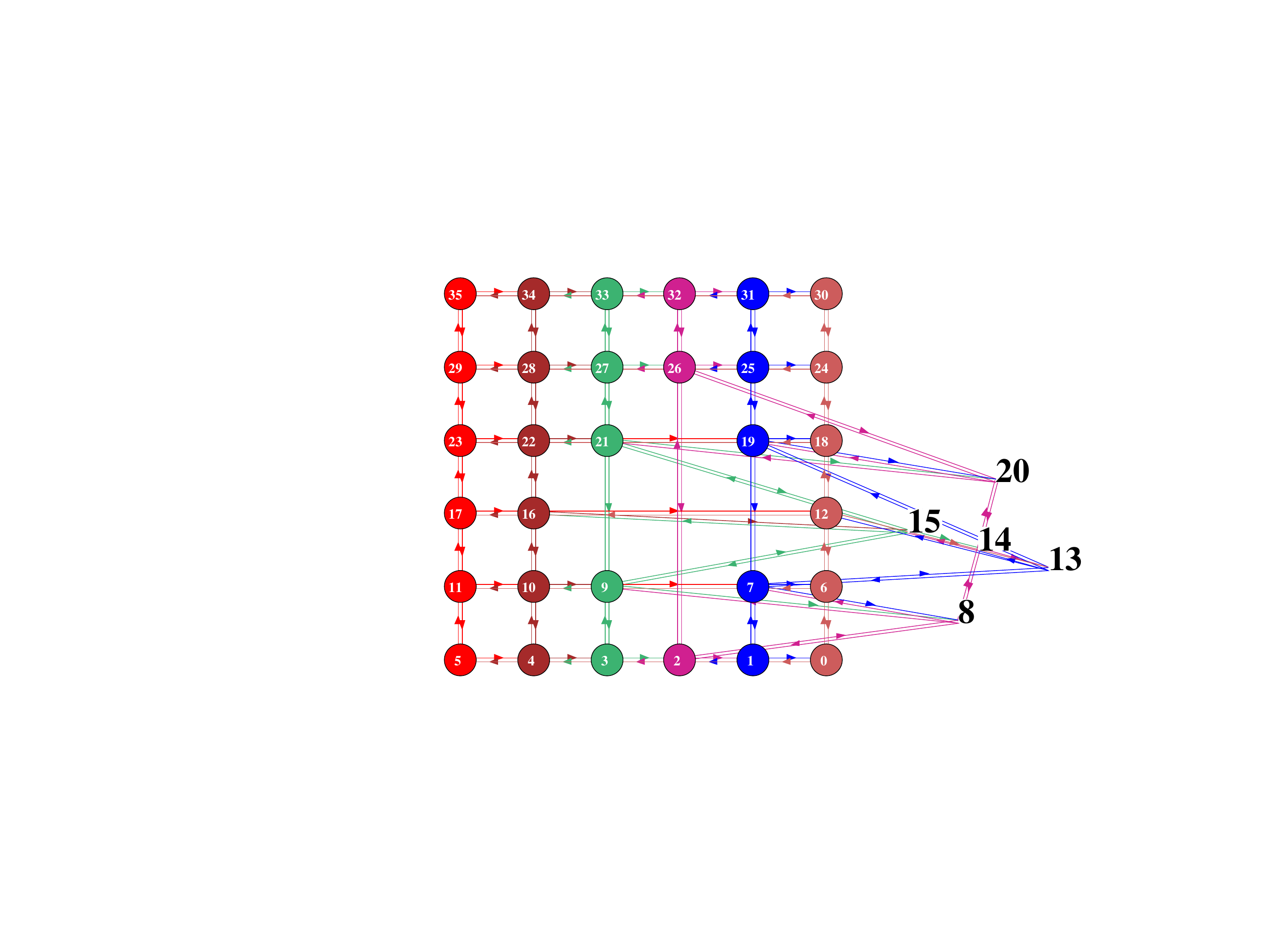}
     \includegraphics[bb=434 243 1206 681,clip=,height=.3\linewidth]{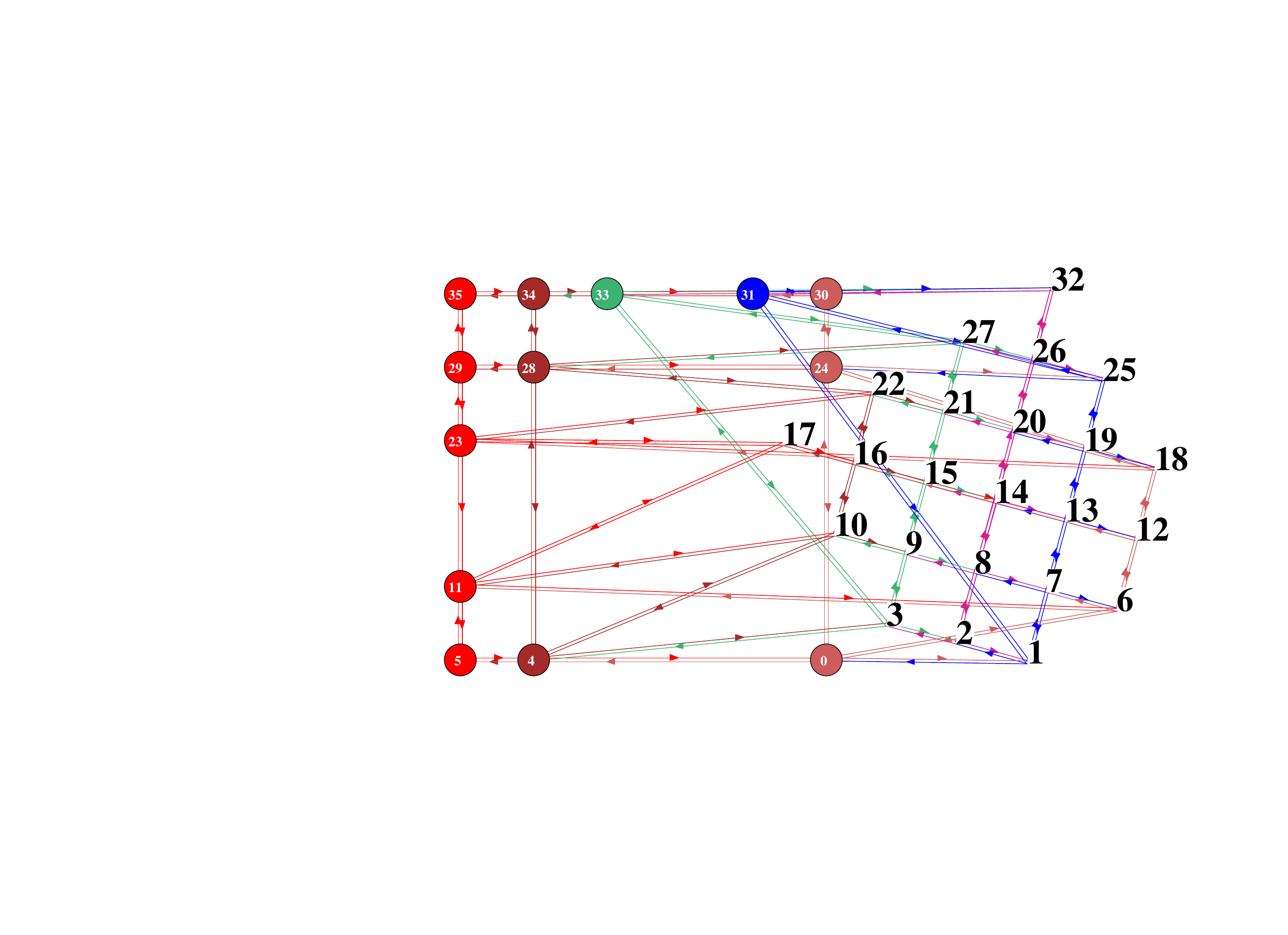}
  \end{center}  
   \vspace{-5ex}  
    \caption  [Dragging a 2d connected fragment]
       {\textsf{A regular network-graph 6x6x6 with a 2d ``either'' fragment on active node 14,
   dragged, rotated, nodes toggled to decimal, and font enlarged.    
 \underline{\it Left}: Step=1.
  \underline{\it Right}: Step=3.
       \label{step1-2d}}}
     \end{figure}

\item[{\bf nolimit-0} $\dots$] enter ``{\it nolimit-}{\bf 0}'' (zero) for an unlimited linked fragment,
                        a continuous chain of links relative to the active node,
                        depending on which option --- {\color{BrickRed}{\bf inputs}}, {\color{BrickRed}{\bf outputs}}
                        or {\color{BrickRed}{\bf either}} --- is active. Gap nodes would interrupt such
                        a chain.  {\color{BrickRed}{\bf step=nolimit}} is shown in the drag reminder.
                        This can be set in any drag-status but applies for a linked fragment.
\item[{\bf single-s} $\dots$]  to set ``single'' drag status for dragging just the active node.
                        The title changes to  {\color{BrickRed}{\bf single~node~23:}}
                        (for example). Single status also allows ``blocks'' and ``labels''. 
\item[{\bf in/out/either-i/o/e} $\dots$] 
                         enter ``{\it in-}{\bf i}'', ``{\it out-}{\bf o}'' or ``{\it either-}{\bf e}''
                         to define the type of link ---  {\color{BrickRed}{\bf inputs}}, {\color{BrickRed}{\bf outputs}}
                         or {\color{BrickRed}{\bf either}} (shown in the title) for dragging linked fragments.
                         
\item[{\bf all-a} $\dots$] enter ``{\it all-}{\bf a}'' to drag all nodes, the whole graph. 
                          The title changes to {\color{BrickRed}{\bf allnodes:}}. 
\item[{\bf exit-q} $\dots$]  Enter ``{\bf q}'' to exit the drag-graph and return to
                          initial-graph and reminder. Its easy to flip
                          between the two reminders to implement alternative functions. 
\end{list}

\section{Concluding remarks}
\label{Concluding remarks}

  \noindent This article has described three types of interactive
  graphs where all functions are driven by prompts from the keyboard
  and the pointer is used to drag/drop nodes and their connected
  fragments, to alter, unravel, and re-label the vizualization with
  continuous and immediate feedback.  These functions, in line with
  DDLab's GUI, do not require coding or scrips.

  The three graph types support each other and enhance
  DDLab's general functionality.  The ibaf-graph allows the user to
  investigate and reconfigure the classic basin of attraction field
  vizualization of the global dynamics of discrete dynamical networks.
  The jump-graph --- for visualizing the stability of basins of
  attraction --- is also used to set a flexible pre-layout for the
  ibaf-graph. The network-graph --- to visualize the connections of the
  discrete dynamical network --- also allows an alternative vizualization of
  space-time pattern dynamics, and can serve as an interactive
  vizualization tool for any arbitrary network.

  The snapshot examples in the figures presented give just a flavour
  of DDLab's interactive graph visualization repertoire.  For a
  deeper appreciation of the broad range of dynamic visualization
  functions, which may be refined and extended in future updates, its
  worthwhile to exercise the graphs hands-on within DDLab --- compiled
  versions, and the code itself in the c language, are available under
  the GNU General Public License for various platforms and operating
  systems\cite{Wuensche-DDLab} including Linux and Mac.


\end{document}